\documentclass[12pt]{article}
\usepackage{mathrsfs}
\usepackage{amsmath}
\usepackage{mathrsfs}
\usepackage{amssymb}
\usepackage{color}
\usepackage{psfrag}
\usepackage{graphicx}
\usepackage{subfigure}
\usepackage{cite}

\newcommand{\bea}{\begin{eqnarray}}
\newcommand{\eea}{\end{eqnarray}}
\newcommand{\be}{\begin{equation}}
\newcommand{\ee}{\end{equation}}
\newcommand{\vs}[1]{\vspace{#1 mm}}

\newcommand{\dsl}{\pa \kern-0.5em /}

\newcommand{\pa}{\partial}

\newcommand{\nn}{\nonumber\\}

\newcommand{\ba}{\begin{array}}
\newcommand{\ea}{\end{array}}
\newcommand{\bit}{\begin{itemize}}
\newcommand{\eit}{\end{itemize}}

\begin{document}
\topmargin 0mm
\oddsidemargin 0mm

\begin{flushright}

USTC-ICTS-19-11\\

\end{flushright}

\vspace{2mm}

\begin{center}

{\Large \bf On D-brane interaction \& its related properties}

\vs{10}

{\large Qiang Jia, J. X. Lu, Zihao Wu and Xiaoying Zhu}

\vspace{4mm}

{\em
Interdisciplinary Center for Theoretical Study\\
 University of Science and Technology of China, Hefei, Anhui
 230026, China\\
 and\\
  Peng Huanwu Center for Fundamental Theory, Hefei, Anhui 230026, China\\
 
}

\end{center}

\vs{10}

\begin{abstract}
We compute the closed-string cylinder amplitude between one Dp brane and the other Dp$\prime$ brane, placed parallel at a separation, with each carrying a general constant worldvolume flux and with $p - p' = 0, 2, 4, 6$ and $p \le 6$.  For the $p = p'$, we show that the main part of the amplitude for $p = p' < 5$ is a special case of that for $p = p' = 5$ or $6$ case. For all other $p - p' = 2, 4, 6$ cases, we show that the amplitude is just a special case of the corresponding one for $p = p'$ case. Combining both, we obtain the general formula for the amplitude, which is valid for each of the cases considered and for arbitrary constant worldvolume fluxes.  The corresponding general open string one-loop annulus amplitude is also obtained by a Jacobi transformation of the general cylinder one. We give also the general open string pair production rate. We study the properties of the amplitude  such as the nature of the interaction, the open string tachyonic instability, and the possible open string pair production and its potential enhancement. In particular, in the presence of pure magnetic fluxes or magnetic-like fluxes, we find that the nature of interaction is correlated with the existence of potential open string tachyonic instability. When the interaction is attractive,  there always exists an open string tachyonic instability when the brane separation reaches the minimum determined by the so-called tachyonic shift. When the interaction is repulsive,  there is no such instability for any brane separation.  We also find that the enhancement of open string pair production, in the presence of pure electric fluxes, can occur only for the $p - p' = 2$ case.   
\end{abstract}

\newpage
\section{Introduction}
Computing the interaction amplitude between one Dp and the other Dp$'$, placed parallel at a separation transverse to the Dp brane, with each carrying a general constant worldvolume flux (We consider only constant worldvolume flux(es) in this paper) and with\footnote{\label{fn1} For a system with $p - p' = \kappa = 0, 2, 4, 6$, we have as usual NN $= p' + 1$ for which the two ends of open string obey the Neumann boundary conditions, ND $= \kappa$ for which one end of the open string obeys the Dirichlet boundary conditions while the other the Neumann ones, and DD $= 9 - p' -  \kappa$ for which the two ends of open string obey the Dirichlet boundary conditions.} $p - p'  = 0, 2, 4, 6$ and\footnote{\label{fn2} In general, placing an infinitely extended Dp in spacetime will cause it to curve. For our purpose,  we try to avoid this to happen at least to the probe distance in which we are interested. For this, we need to limit our discussion in this paper to $p \le 6$ cases since these Dp branes have well-behaved supergravity configurations which are all asymptotically flat. Moreover, when the string coupling is small, i.e. $g_{s} \ll 1$, placing one such Dp in spacetime will keep the spacetime flat even for a probe distance to the brane in the substringy scale $\alpha'^{1/2} \gg r \gg g^{1/(7 - p)}_{s} \alpha'^{1/2}$   as discussed in section 2 of \cite{Lu:2007kv}. } $p \le 6$, has its own interest by itself.  As we will see, the amplitude itself exhibits many interesting properties. For example, the contribution from the so-called  NS-NS sector or R-R sector has a nice form, determined by the certain properties of the worldvolume background fluxes relevant to the amplitude, and can be expressed in terms of  certain $\theta$-functions and the Dedekind $\eta$-function.  The total amplitude can also be expressed in terms of a certain 
$\theta$-function, usually the $\theta_{1}$-function,  using a special form of the more general identity relating various different $\theta$-functions obtained from the contributions of the NS-NS and R-R sectors after the so-called Gliozzi-Scherk-Olive (GSO) projection, and the Dedekind $\eta$-function,   so exhibiting the expected modular property of the amplitude.

A Dp brane carrying no worldvolume flux is a non-perturbative stable Bogomol'ny-Prasad-Sommereld (BPS) solitonic extended object in superstring theories (for example, see\cite{Duff:1994an}), preserving one half of the spacetime supersymmetries.  It has its tension and carries the so-called RR charge.  When we place two such Dp branes parallel at a separation, the net interaction between the two actually vanishes due to the 1/2 BPS nature of this system. We can check this explicitly by computing the lowest order stringy interaction amplitude in terms of the closed string tree-level cylinder diagram. We have here the so-called  NS-NS contribution, due to the brane tension, which is attractive, and the so-called R-R contribution, due to the RR charges, which is repulsive.  The BPS nature of each Dp brane identifies its tension with its RR charge in certain units and as such the sum of the two gives an expected zero net interaction by making use of the  usual `abstruse identity' \cite{Polchinski:1995mt}.
This same amplitude can also be computed  via the so-called  open string one-loop annulus diagram.  The same conclusion can be reached.  

When one of the above two Dp branes is replaced by a Dp$'$ with $p > p'$ and $p \le 6$,  we have only the NS-NS contribution since different brane RR charge does not interact\footnote{\label{fn3} For the $p - p' =2$ case, the long-range interaction is attractive since the contribution from either the exchange of massless dilaton or the exchange of massless graviton is attractive while for the $p - p' = 6$ case the long-range interaction is repulsive since the contribution from the exchange of massless dilaton is repulsive and exceeds the attractive contribution from the exchange of massless graviton. However, for the $p - p' =4$ case, the repulsive contribution from the exchange of massless dilaton just cancels the attractive one from the exchange of massless graviton and this gives a net vanishing interaction.  Each of these can be checked explicitly, for example, see \cite{Ouyang:2014bha}.
Each of these remains to be true for any brane separation as we will show later in this paper.}. For the $p -p'  = 2$ case, we have an attractive interaction while for the $p - p' = 6$ case we have a repulsive one.  As such, the underlying spacetime supersymmetries are all broken.  However, for the $p - p' = 4$ case, the net interaction vanishes and the underlying system is still BPS, preserving 1/4 of spacetime supersymmetries.  Each of these, regarding the supersymmetry breaking or preservation, can be checked explicitly following \cite{polbooktwo,Lu:2009au}, for example. 

When the brane worldvolume fluxes are present, we now expect in general a non-vanishing interaction. Except for the ($p= 6, p' = 0$) case mentioned above and the ($p = 6, p' \le 6$) cases to be considered later in this paper,   the long-range interaction between the Dp and Dp$'$ for other cases is in general attractive when the electric and/or magnetic fluxes\footnote{\label{fn4}The electric flux on a Dp-brane stands for the presence of F-strings, forming the so-called (F, Dp) non-threshold bound state\cite{Witten:1995im, Schmidhuber:1996fy, Arfaei:1997hb,Lu:1999qia, Lu:1999uca, Lu:1999uv, Hashimoto:1997vz,DiVecchia:1999uf}, while a magnetic flux stands for that of co-dimension 2 D-branes inside the original Dp brane, forming the so-called (D(p-2), Dp) non-threshold bound state\cite{Breckenridge:1996tt, Costa:1996zd, Di Vecchia:1997pr},  from the spacetime perspective. These fluxes are in general quantized.  We will not discuss their quantizations in the text for simplicity due to their irrelevance for the purpose of this paper.} on the Dp$'$ are parallel to the corresponding ones on the Dp, respectively.  The reason for this is simple since only different constituent branes contribute to this long-range interaction and each contribution is from the respective NS-NS sector and is attractive.    For example, if we have both electric and magnetic fluxes present on Dp and Dp$'$, the F-strings (see footnote (\ref{fn4})) within Dp$'$ have no interaction with their parallel F-strings within Dp but have a long-range attractive interaction with D(p - 2) branes (see footnote (\ref{fn4})) within Dp.  However, as indicated above for $p = 6$ and $p' \le 6$,  the long-range interaction can be repulsive in the presence of certain types of fluxes. This has been demonstrated in the simplest possible cases in \cite{Lu:2009pe} when $p - p' = 2$.  We will spell out the condition in general for this to be true later in this paper.

For certain type of fluxes (to be specified later on), the nature of the interaction at small brane separation (attractive or repulsive) remains unclear if it is computed in terms of the closed string tree-level cylinder amplitude.  In general, this implies new physics to appear.  The best description is now in terms of the open string one-loop annulus amplitude for which many interesting properties such as certain instabilities become manifest. 

When only magnetic fluxes are present (or even no fluxes are present with $p - p' = 2$),  we find that there is a correlation between the nature of the interaction between the Dp and the Dp$'$ and the potential open string tachyonic instability.  If the interaction is attractive,  the open string connecting the two D-branes has a tachyonic shift to its spectrum\cite{Bolognesi:2012gr, Ferrara:1993sq}.  We have then the onset of tachyonic instability when the brane separation reaches the minimum determined by the shift.  Once this instability develops, the attractive brane interaction diverges. We have then the so-called tachyon condensation and as such a phase transition occurs, releasing the excess energy of this system. For example, for $p = p'$, this process restores not only the gauge symmetry from $U (1) \times U(1) \to U(2) $ but also the supersymmetry from none to half of the spacetime supersymmetries \cite{Lu:2018nsc}.  In the so-called weak field limit, the corresponding instability is just the analog of the Nielsen-Olesen one\cite{Nielsen:1978rm} of non-abelian gauge theories such as the electroweak one and the gauge symmetry restoration was considered in \cite{Ambjorn:1988tm, Ambjorn:1989bd}. On the other hand, if the interaction is repulsive, we don't have this tachyonic shift and therefore have no tachyonic instability to begin with.

 When we have only worldvolume electric fluxes present, the underlying system is in general no longer 1/2 BPS and breaks all its supersymmetries, therefore unstable.  This manifests itself by the appearance of an infinite number of simple poles in the integrand of the integral representation of the open string one-loop annulus amplitude, implying that the interaction amplitude has an imaginary part. Each of these simple poles actually indicates the corresponding open string pair production under the action of the applied electric fluxes\cite{Lu:2009yx,  Lu:2018suj, Lu:2018nsc}.  The imaginary amplitude just reflects the decay of the underlying system via the so-called open string pair production, releasing the excess energy of the system until it reaches the corresponding 1/2 BPS stable one.  This is the analog of the Schwinger pair production in quantum electrodynamics (QED) \cite{Schwinger:1951nm}.  For unoriented bosonic string and type I superstring, this was pursued a while ago in \cite{Bachas:1992bh, Porrati:1993qd}.  
 When the applied electric flux reaches its so-called critical field determined by the fundamental string tension, the open string pairs are produced cascadingly and there is also an instability developed.
 
 When both electric and magnetic fluxes are present in a certain way, the open string pair production has an enhancement, uncovered recently in  \cite{Lu:2009au, Lu:2009pe, Lu:2017tnm, Lu:2018suj, Lu:2018nsc, Jia:2018mlr}.	
	
As explained in \cite{Lu:2018nsc, Jia:2018mlr}, there is no open string pair production for a single Dp brane in Type II string theories even if we apply an electric flux on the brane, unless it reaches its critical value\footnote{When the applied electric field reaches its critical value, it will break the virtual open strings from the vacuum and the pair production is due to the breaking of each of these open strings, not to the separation of the pair of the virtual open string and the virtual anti open string under the electric field.}. The simple reason for this is that  each of the virtual open strings in the pairs from the vacuum has its two ends carrying charge $+$ and $-$, respectively, giving a net zero-charge of the open string, called the charge-neutral open string, and the net force acting on the string vanishes under the applied constant electric field. So the electric field can only stretch each open string but cannot separate the virtual open string and the virtual anti-open string in each pair to give rise to the pair production.  This can also be explained by the fact that  a Dp brane carrying a constant worldvolume electric flux is actually a 1/2 BPS non-threshold bound state \cite{Lu:1999uca,DiVecchia:1999uf}, therefore it is stable and cannot decay via the open string pair production.		
	
In order to produce the pair production in Type II string theories, a possible choice is to let the two ends of the open string experience different electric fields since the charge-neutral nature of the open strings cannot be altered. The above mentioned two Dp-brane system is probably the simplest one for this purpose. We compute this pair production rate \cite{Lu:2009yx, Lu:2009au, Lu:2017tnm, Lu:2018suj} and find it indeed non-vanishing. However, for any realistic electric field applied, the rate is in general vanishingly small and  it has no practical use. But when an additional magnetic flux is added in a certain way, the rate can be enhanced enormously\cite{Lu:2009au,Lu:2017tnm} and the largest possible rate is for the system of two D3 branes when the electric and magnetic fields are collinear \cite{Lu:2017tnm,Jia:2018mlr}.  This enhanced pair production may have the potential to be detected\cite{Lu:2018nsc}. 
	
	For this to occur in practice, we need to assume one of the D3 branes to be relevant to our 4-dimensional world and the other D3 to be invisible (hidden or dark)  to us. For this simple system, it appears that there is a possibility for the detection of the pair production but there is an issue if one carefully examines the underlying physics as discussed in detail in \cite{Lu:2018nsc}.  The mass spectrum of the open string connecting the two D3 at a separation $y$ has a mass shift $m = y/(2\pi \alpha')$ at each mass-level. That the corresponding modes at each given mass-level all have this same shift is due to the underlying supersymmetry in the absence of worldvolume fluxes. For example, the lowest-mass eight bosons and eight fermions all have the same mass $y/(2\pi\alpha')$ which becomes massless at $y = 0$ and the underlying system is 1/2 BPS in the absence of worldvolume fluxes.  In general, the laboratory electric and magnetic fields are much smaller than the string scale and the weak field limit holds.  So the contribution to the pair production is due to the above 8 bosonic ($8_{B}$) and 8 fermionic ($8_{F}$) lowest-mass charged modes of the open string.  From the worldvolume viewpoint,  these $8_{B} + 8_{F}$ massive charged modes, in the absence of worldvolume fluxes, are due to the symmetry breaking of $U(2) \to U(1) \times U(1)$ with one of scalars taking its expectation value $\sim y$ but the underlying 16 supersymmetries remain intact, giving rise to the 4-dimensional N = 4 massive gauge theory with one massive charged vector (W-boson), 5 massive charged scalars and their corresponding fermionic super partners, all with mass $m = y/(2\pi \alpha')$.  In the presence of worldvolume fluxes and if the brane interaction is non-vanishing, the underlying supersymmetries are all broken. The presence of practical magnetic fluxes can also give a tiny mass shift to the massive charged vector\cite{Ferrara:1993sq}.    
	
	We therefore naturally expect the mass scale $m = y/(2\pi\alpha')$ no less than a few TeV (since no supersymmetry has been found in LHC yet) and this requires the electric and magnetic fields 21 orders of magnitude larger than the current laboratory limit to make the detection possible. The other choice is to take the other D3 as a dark one and for this, we don't have a priori knowledge of the mass scale $m$. If it happens to be no larger than the electron mass, we may have an opportunity to detect the open string rate if the QED Schwinger pair production becomes feasible. Even so, we still have to explain why the other charged fermions other than the one identified with the electron, the charged scalars and the charged vector, all having the same mass $m \sim m_{e} = 0.51 \, {\rm MeV}$, are not the Standard model particles. 
	  
	All these issues, one way or the other, are due to that the 16 ($8_{B} + 8_{F}$) relevant modes all have  the same mass $m = y/(2\pi \alpha')$ from the underlying supersymmetry.  In addition, the currently available laboratory electric and magnetic fields are too small.  The natural question is: does there exist a possibility that we can get around these issues in practice? 

A while ago, one of the present authors along with his collaborator considered a system of one Dp and one Dp$'$, placed parallel at a separation transverse to the Dp brane, with $p - p' = 2$ and with each  carrying only one flux \cite{Lu:2009pe}, and found that whenever there is an electric flux present along the common direction shared by both the Dp and Dp$'$,  there is an open string pair production enhancement even in the absence of a magnetic flux.  The novel feature found in \cite{Lu:2009pe} is that the Dp$'$-brane plays effectively as a magnetic flux of stringy order (see footnote (\ref{fn4})).  In other words,  if our D3 brane has a nearby D-string, for example as a cosmic string, this D-string appears effectively as a stringy magnetic field.   This  field can give rise to the pair production enhancement, which can hardly possible with a laboratory magnetic field, if our D3 carries an electric flux along the D-string direction. In addition, the underlying system breaks all supersymmetries intrinsically.  So this consideration may provide a solution to the above question raised. This is the other line motivating us to consider the brane interaction in general between one Dp and the other Dp$'$ with $p - p' = k =0, 2, 4, 6$ as specified at the outset of this introduction.   

Without further ado, we in this paper compute the lowest-order stringy interaction amplitude between one Dp and the other Dp$'$, placed parallel at a separation transverse to the Dp,  with each carrying a general worldvolume flux and with $p - p' = 0, 2, 4, 6$ and $p \le 6$. We will show that the key part of the amplitude in terms of the $\theta_{1}$-functions and the Dedekind $\eta$-function for each of the $p = p' < 5$ cases is  a special case of that for the $p = p' = 5$ or $6$ case.   We further demonstrate that the amplitude for $p - p' = 2, 4, 6$ can be obtained, respectively,  from the corresponding $p = p'$ case by choosing specific magnetic fluxes along the 2, 4, 6 ND-directions, a trick greatly simplifying the computations\footnote{We will also provide the physical rationale for this.}. We  compute first the closed-string cylinder amplitude using the closed string boundary state representation of D-brane\cite{Callan:1986bc, Green:1994, Billo:1998vr, Di Vecchia:1999rh, Di Vecchia:1999fx}, which has the advantage of holding true for a general worldvolume constant flux\cite{DiVecchia:1999uf}.  By a Jacobi transformation of this, we can obtain the corresponding open-string annulus amplitude. We will also compute the open string pair production rate if any and discuss the relevant analytical structures of the amplitude.   We will explore the nature (attractive or repulsive) of the
interaction at large brane separation and small brane separation, respectively, and study
various instabilities such as the onset of tachyonic one at small brane separation. In particular, we find that there is a correlation between the nature of interaction being attractive and the existence of tachyonic shift, which can give rise to the onset of tachyonic instability when the brane separation reaches the minimum determined by the tachyonic shift. We will
determine at which conditions there exists the open string pair production and its possible
enhancement. We will also speculate possible applications of the enhanced open string pair production for practical purpose.

Before we move to discuss the plan of this paper, we would like to point out a few things which are relevant and will be mentioned only briefly here due to the scope of this paper\footnote{This paragraph is added  to address the comments/questions by one of the referees.  We thank the anonymous referee for these which help to clarify certain issues not addressed in the early version of this paper.}. Note that the added fluxes on either Dp or Dp$'$, considered in this paper,  are general, with each electric one standing for F-strings \cite{Witten:1995im, Schmidhuber:1996fy, Arfaei:1997hb,Lu:1999qia, Lu:1999uca, Lu:1999uv, Hashimoto:1997vz,DiVecchia:1999uf} and each magnetic one standing for co-dimension 2 D-branes \cite{Breckenridge:1996tt, Costa:1996zd, Di Vecchia:1997pr} inside the Dp or Dp$'$,  so these fluxes represent different intersecting D-branes and/or F-strings. We know that the magnetized D-branes are related to the corresponding intersecting D-branes via T-dualities\cite{Berkooz:1996km, Balasubramanian:1996uc, Breckenridge:1996tt, Breckenridge:1997ar,Rabadan:2001mt, Di Vecchia:1999fx} (see \cite{Blumenhagen:2006ci} for a more complete list of references, in particular for phenomenological applications). We also know that the electrified D-branes are related to moving D-branes again via T-dualities \cite{Callan:1995xx, Di Vecchia:1999fx}.  We limit ourselves in this paper to consider the Dp and the Dp$'$ to be placed parallel, not oblique, at a separation only for simplicity and for the purpose of seeking the open string pair production enhancement.  The extension of the present consideration to the oblique Dp and Dp$'$ may provide more feasibilities in giving rise to the open string pair production in practice. For example, we can use the intersecting branes to give a lower string scale and/or to give light stringy states\cite{Anastasopoulos:2011hj}.  In this paper we only use the so-called no-force condition to discuss whether the underlying system preserves certain supersymmetry but this can also analyzed in detail following that for intersecting branes \cite{polbooktwo, Behrndt:1996pm, Berkooz:1996km, Balasubramanian:1996uc,
Bergshoeff:1996rn, Breckenridge:1996tt, Breckenridge:1997ar, SheikhJabbari:1997cv, Rabadan:2001mt} (see \cite{Blumenhagen:2006ci} for a rather complete list of references on phenomenological applications to supersymmetry breaking for intersecting branes).  One of primary purposes of this paper is to seek a D-brane system containing a D3 as our own world which can give rise to the earthbound laboratory testable open string pair production without the need of string compactifications.  We also know that with string compactfications the intersecting D-branes in Type II string theory, for example the intersecting D6 branes, provide phenomenologically interesting models\cite{Uranga:2003pz, Lust:2004ks, Blumenhagen:2005mu, Blumenhagen:2006ci}.  In this paper, we consider only the system of a single Dp and a single Dp$'$ with each carrying a general flux. The extension of this to N and N$'$ branes with each stack carrying the respective overall U(1) flux in a similar setting looks straightforward. For example, the two corresponding amplitudes differ by an overall factor $N N'$.  We can understand this by noting that the closed string cylinder or the open string annulus interaction between each given brane in the first stack and that in the second stack is the same as we compute in this paper and total counting factor is $N N'$. Also note that the closed string cylinder or the open string annulus amplitude is the lowest order stringy interaction for the system considered.  This multiplicity factor $N N'$ can also be understood from the long-range interaction between the two stacks computed from the effective field theory, for example, see \cite{Ouyang:2014bha}.   

The  paper is organized as follows. In section 2, we give a brief review of the closed-string boundary state representation of Dp-brane carrying a general constant worldvolume flux and set up conventions for latter sections.  We give also a general discussion on 
computing the closed-string cylinder amplitude between a Dp and a Dp$'$, placed parallel at a separation transverse to the Dp,  with each carrying a general constant worldvolume flux and with $p - p' = 0, 2, 4, 6$ and $p \le 6$.   In section 3,  
we first compute the closed-string cylinder amplitude for each of the $ 0 \le p = p' \le 6$ cases. In this computation, we will make use of certain tricks and simplifications in evaluating this amplitude.  Once this is done, we will see the expected nice structure of the amplitude in terms of
certain $\theta$-functions and the Dedekind $\eta$-function. We study the nature of interaction and find that the repulsive interaction can only be possible for $p = p' = 6$ and for certain purely magnetic fluxes present.  For all other cases, the long-range interaction is attractive. We also find the correlation between the nature of interaction being attractive and the existence of tachyonic shift, which will give rise to the onset of tachyonic instability when the brane separation reaches the minimum determined by the tachyonic shift.  We  compute the decay rate of the underlying system and the corresponding open string pair production rate when they exist and discuss their potential enhancement.  In section 4, we  move to compute the amplitude for each of $p - p' = 2, 4, 6$ and $p \le 6$, respectively, using the known $p = p'$ one with a specific choice of the magnetic fluxes on the Dp$'$, along now the 2, 4, 6 ND-directions. Basically, the amplitude for each $p - p' = 2, 4, 6$ and $p \le 6$ can simply be obtained from the corresponding $p = p'$ one computed in section 3 by a special choice of magnetic fluxes along the 2, 4, 6 ND-directions on the Dp$'$ brane.  We also provide the underlying physical reason for this.   Similar properties of the amplitude as discussed in section 3 are also given. We discuss and conclude this paper in section 5.

\section{The Dp brane boundary state}
In this section, we give a brief review of Dp brane boundary state carrying a general constant worldvolume flux, following \cite{Di Vecchia:1999fx}. 	We also give a general discussion in computing the closed-string cylinder amplitude between a Dp brane and a Dp$'$ brane, placed parallel at a separation transverse to the Dp, with each carrying a general constant worldvolume flux and with $p - p' = 0, 2, 4, 6$ and $p \le 6$. 

For such a description of Dp-brane, there are two sectors, namely NS-NS and R-R sectors. In either sector, we have two implementations for the boundary conditions of a Dp brane, giving two boundary states $|B, \eta \rangle$, with $\eta = \pm$. However, only the combinations 
\bea\label{gsopbs}
 &&|B \rangle_{\rm NS} = \frac{1}{2} \left[|B, + \rangle_{\rm NS} - |B, - \rangle_{\rm NS}\right],\nn
 &&|B \rangle_{\rm R} = \frac{1}{2} \left[|B, + \rangle_{\rm R} + |B, - \rangle_{\rm R}\right],
 \eea
  are selected by the  Gliozzi-Scherk-Olive (GSO) projection in the NS-NS and R-R sectors, respectively.   The boundary state $|B, \eta\rangle$ for a Dp-brane can be expressed as the product of a matter part and a ghost part \cite{Billo:1998vr, Di Vecchia:1999rh, Di Vecchia:1999fx}, i.e. 
 \be\label{bs}
  |B, \eta\rangle = \frac{c_p}{2} |B_{\rm mat}, \eta\rangle |B_{\rm g}, \eta\rangle,
 \ee
 where 
 \be\label{mgbs}
 |B_{\rm mat}, \eta\rangle = |B_X \rangle|B_\psi, \eta\rangle,  \quad |B_{\rm g},\eta\rangle = |B_{\rm gh}\rangle |B_{\rm sgh}, \eta\rangle
 \ee
  and the overall normalization
$c_p = \sqrt{\pi}\left(2\pi \sqrt{\alpha'}\right)^{3 - p}. $  

As discussed in \cite{DiVecchia:1999uf, Di Vecchia:1999fx}, the operator structure of the
boundary state holds true even with a general constant  worldvolume flux and is always of the form 
\be\label{xbs}
 |B_X\rangle ={\rm exp} (-\sum_{n =1}^\infty \frac{1}{n} \alpha_{- n} \cdot M
\cdot {\tilde \alpha}_{ - n}) |B_X\rangle_0,
\ee
 and 
 \be\label{bspsinsns}
 |B_\psi, \eta\rangle_{\rm NS} = - {\rm i}~ {\rm exp} (i \eta
\sum_{m = 1/2}^\infty \psi_{- m} \cdot M \cdot {\tilde \psi}_{- m})|0\rangle
\ee for the NS-NS sector and 
\be\label{bspsirr}
 |B_\psi,\eta\rangle_{\rm R} = - {\rm exp} (i \eta \sum_{m = 1}^\infty
\psi_{- m} \cdot M \cdot {\tilde \psi}_{- m}) |B
\eta\rangle_{0\rm R}
\ee for the R-R sector. The ghost boundary states are the standard ones as given in \cite{Billo:1998vr}, independent of the fluxes, which we will not present here.   The M-matrix\footnote{We have changed the previously often used symbol $S$ to the current $M$ to avoid a possible confusion with the S-matrix in scattering amplitude.}, 
the zero-modes  $|B_X\rangle_0$ and $|B,
\eta\rangle_{0\rm R}$ encode all information about the overlap
equations that the string coordinates have to satisfy. They can be determined respectively
\cite{Callan:1986bc,DiVecchia:1999uf, Di Vecchia:1999fx} as 
\be\label{mmatrix}
M = ([(\eta -
\hat{F})(\eta + \hat{F})^{-1}]_{\alpha\beta},  -
\delta_{ij}),
\ee
\be\label{bzm}
|B_X\rangle_0 = [- \det
(\eta + \hat F)]^{1/2} \,\delta^{9 - p} (q^i - y^i) \prod_{\mu
= 0}^9 |k^\mu = 0\rangle,
\ee 
for the bosonic sector and 
\be\label{rrzm}
|B_\psi, \eta\rangle_{0\rm R} = (C \Gamma^0 \Gamma^1
\cdots \Gamma^p \frac{1 + {\rm i} \eta \Gamma_{11}}{1 + {\rm i} \eta
} U )_{AB} |A\rangle |\tilde B\rangle,
\ee
 for the R-R sector. In
the above, the Greek indices $\alpha, \beta, \cdots$ label the
world-volume directions $0, 1, \cdots, p$ along which the Dp
brane extends, while the Latin indices $i, j, \cdots$ label the
directions transverse to the brane, i.e., $p + 1, \cdots, 9$. We define $\hat F = 2\pi \alpha' F$ with $F$ the external
worldvolume field. We also have denoted by $y^i$ the
positions of the D-brane along the transverse directions, by $C$ the
charge conjugation matrix and by $U$ the matrix 
\be\label{umatrix}
 U (\hat F) = \frac{1}{\sqrt{- \det (\eta + \hat F)}} ; {\rm exp} \left(- \frac{1}{2} {\hat
F}_{\alpha\beta}\Gamma^\alpha\Gamma^\beta\right); 
\ee
with the symbol $;\,\, ;$ denoting the indices of the $\Gamma$-matrices completely anti-symmetrized  in each term of the exponential expansion.
$|A \rangle |\tilde B\rangle$ stands for the spinor vacuum of the R-R
sector. Note that the $\eta$ in the above
denotes either sign $\pm$ or the worldvolume Minkowski flat metric and should be clear from the content.

We now come to compute the closed-string tree-level cylinder amplitude between a Dp and a Dp$'$ as stated earlier via
 \be\label{ampli}
 \Gamma = \langle B_{p'} (\hat F') | D |B_{p} (\hat F) \rangle,
 \ee
  where $D$ is the closed string propagator defined as 
\be\label{prog}
 D =
\frac{\alpha'}{4 \pi} \int_{|z| \le 1} \frac{d^2 z}{|z|^2} z^{L_0}
 {\bar z}^{{\tilde L}_0}.
 \ee 
 Here $L_0$ and ${\tilde L}_0$ are
the respective left and right mover total zero-mode Virasoro
generators of matter fields, ghosts and superghosts. For example,
$L_0 = L^X_0 + L_0^\psi + L_0^{\rm gh} + L_0^{\rm sgh}$ where $L_0^X,
L_0^\psi, L_0^{\rm gh}$ and $L_0^{\rm sgh}$ are the respective ones from
matter fields $X^\mu$, matter fields $\psi^\mu$, ghosts $b$ and $c$,
and superghosts $\beta$ and $\gamma$, and their
explicit expressions can be found in any standard discussion of
superstring theories, for example in \cite{Di Vecchia:1999rh},
therefore will not be presented here. The above total  amplitude has
contributions from both NS-NS and R-R sectors, respectively, and can
be written as $\Gamma_{p, p'} = \Gamma_{\rm NSNS} + \Gamma_{\rm RR}$. In
calculating either $\Gamma_{\rm NSNS}$ or $\Gamma_{\rm RR}$, we need to
keep in mind that the boundary state used should be the GSO
projected one as given in (\ref{gsopbs}). 

For this, we
need to calculate first the amplitude $ \Gamma (\eta',
\eta) = \langle B', \eta'| D |B, \eta\rangle $ in each sector
with $\eta' \eta = +\, {\rm or} - $, $B' = B_{p'} (\hat F')$ and $B = B_{p}(\hat F)$.  Actually, $\Gamma(\eta', \eta)$ depends only on the product of $\eta'$ and $\eta$, i.e., $\Gamma(\eta', \eta) = \Gamma (\eta' \eta)$. In the NS-NS sector, this gives $\Gamma_{\rm NSNS} (\pm) \equiv \Gamma (\eta', \eta)$ when $\eta' \eta = \pm$, respectively.  Similarly we have $\Gamma_{\rm RR} (\pm) \equiv \Gamma (\eta', \eta)$ when $\eta'\eta = \pm$ in the R-R sector. We then have
\be\label{ns-r}
\Gamma_{\rm NSNS} = \frac{1}{2} \left[\Gamma_{\rm NSNS} (+) - \Gamma_{\rm NSNS} (-)\right], \quad \Gamma_{\rm RR} = \frac{1}{2}\left[\Gamma_{\rm RR} (+) + \Gamma_{\rm RR} (-) \right].
\ee
 Given the structure of the
boundary state, the amplitude $\Gamma (\eta'
\eta)$ can be factorized as 
\be\label{amplitude}
\Gamma (\eta' \eta) = \frac{
c_{p'} c_{p}}{4} \frac{\alpha'}{4 \pi} \int_{|z| \le 1} \frac{d^2 z}{|z|^2}
A^X \, A^{\rm bc}\, A^\psi (\eta'\eta)\, A^{\beta\gamma} (\eta'
\eta).
\ee 
In the above, we have 
\bea\label{me}
&&A^X = \langle B'_X | |z|^{2 L^X_0} |B_X \rangle,\quad
A^\psi (\eta' \eta) = \langle B'_\psi, \eta'| |z|^{2 L_0^\psi}
|B_\psi, \eta \rangle, \nn
&&A^{\rm bc} = \langle B_{\rm gh} | |z|^{2
L_0^{\rm gh}} | B_{\rm gh}\rangle,\quad A^{\beta\gamma} (\eta' \eta) =
\langle B_{\rm sgh}, \eta'| |z|^{2 L_0^{\rm sgh}} |B_{\rm sgh}, \eta\rangle.
\eea
The above ghost and superghost matrix elements $A^{\rm bc} $ and $A^{\beta\gamma} (\eta' \eta)$, both independent of the fluxes and the dimensionalities of the branes involved, can be calculated to give,
\be\label{ghme}
A^{\rm bc} = |z|^{- 2} \prod_{n = 1}^\infty \left(1 - |z|^{2n}\right)^2,
\ee 
and in the NS-NS sector
\be\label{sghmensns}
A_{\rm NSNS}^{\beta\gamma} (\eta' \eta) = |z| \prod_{n = 1}^\infty \left(1 + \eta'\eta |z|^{2n - 1}\right)^{-2},
\ee
while in the R-R sector
\be\label{sghmerr}
 A_{\rm RR}^{\beta\gamma} (\eta' \eta) =  {}_{\rm R0}\langle B_{\rm sgh}, \eta'|B_{\rm sgh},\eta\rangle_{\rm 0R}\, |z|^{\frac{3}{4}} \prod_{n = 1}^\infty \left(1 + \eta'\eta |z|^{2n}\right)^{-2},
\ee
where ${}_{\rm R0}\langle B_{\rm sgh}, \eta'|B_{\rm sgh},\eta\rangle_{\rm 0R}$ denotes the superghost zero-mode contribution which requires a regularization along with the zero-mode contribution of matter field $\psi$ in this sector.  We will discuss this regularization later on. 

With the above preparation, we are ready to compute the closed string tree-level cylinder amplitudes for the systems under consideration.  We first compute the closed-string tree-level cylinder amplitude for the case of $p = p'$.  This serves as the basis for computing the amplitude for each of the $p \neq p'$ cases. The general steps follow those given in section 2 of \cite{Lu:2018suj} but with a few refinements. Once the closed string tree-level cylinder amplitude is obtained, we use a Jacobi transformation to obtain the corresponding open string one-loop  annulus amplitude.  We will have these done in the following two sections one by one.  We also discuss the properties of the respective amplitude in each case.

\section{ The amplitude and its properties: the $p = p'$ case\label{section3}}

 As indicated already in the previous section,  the computations of the amplitude boil down to computing the matrix elements of matter part, i.e, $A^{X}$ and $A^{\psi}$ given in (\ref{me}).  For this, the following property of the matrix M given in (\ref{mmatrix}) can be used to simplify their computations greatly,
 \be\label{matrixmp}
 M_\mu\,^\rho (M^T)_\rho\,^\nu = (M^T)_\mu\,^\rho M_\rho\,^\nu = \delta_\mu\,^\nu,
 \ee
 where $T$ denotes the transpose of matrix. Following \cite{Lu:2018suj}, for a system of two Dp branes, placed parallel at a separation $y$, with one carrying flux $\hat F'$ and the other carrying flux $\hat F$, we can then have, 
 \be\label{matrixx}
 A^X = V_{p + 1} \frac{\left[\det(\eta + \hat F') \det(\eta + \hat F)\right]^{\frac{1}{2}}} { \left(2 \pi^2 \alpha' t\right)^{ \frac{9 - p}{2}}}\, e^{- \frac{y^2}{2\pi \alpha' t}} \prod_{n = 1}^\infty \left(\frac{1}{1 - |z|^{2n}}\right)^{9 - p} \prod_{\alpha = 0}^p \frac{1}{1 - \lambda_\alpha |z|^{2n}},
 \ee
and in the NS-NS sector
\be\label{matrixpsinsns}
A^\psi_{\rm NSNS} (\eta'\eta) = \prod_{n = 1}^\infty \left(1 + \eta'\eta |z|^{2n - 1}\right)^{9 - p} \prod_{\alpha = 0}^p \left(1 + \eta'\eta \lambda_\alpha |z|^{2n - 1}\right),
\ee
while in the R-R sector
\be\label{matrixpsirr} 
A^\psi_{\rm RR} (\eta'\eta) = {}_{R0}\langle B'_\psi, \eta' |B_\psi, \eta\rangle_{\rm 0R}\, |z|^{\frac{5}{4}} \prod_{n = 1}^\infty \left(1 + \eta'\eta |z|^{2n}\right)^{9 - p} \prod_{\alpha =0}^p \left(1 + \eta' \eta \lambda_\alpha |z|^{2n}\right),
\ee  
 where $ {}_{R0}\langle B'_\psi, \eta' |B_\psi, \eta\rangle_{\rm 0R}$ denotes the zero-mode contribution in this sector which, when combined with the zero-mode contribution from the superghost, needs a regularization mentioned earlier.  We will present the result of this regularization later on.  In the above, $|z| = e^{- \pi t}$, $V_{p + 1}$ denotes the volume of the Dp brane worldvolume, $\lambda_\alpha$ are the eigenvalues of the matrix $w_{(1 + p) \times (1 + p)}$ defined in the following 
 \be\label{matrixw}
 W = M M'^T = \left(\begin{array}{cc}
 w_{(1 + p)\times(1 + p)} & 0\\
 0 & \mathbb{I}_{(9 - p)\times (9 - p)}\end{array}\right),
 \ee
 where the matrix $M$ or $M'$ is the one given in \eqref{mmatrix} when the corresponding flux is $\hat F$ or $\hat F'$, respectively, and $\mathbb{I}$ stands for the unit matrix. We can also express the matrix $w$ in terms of matrix $s$ and $s'$ as
 \be\label{matrixlw} 
 w_{\alpha}\,^{\beta} = (s s'^{T})_{\alpha}\,^{\beta} = \left[(\mathbb{I} - \hat F) (\mathbb{I}+ \hat F)^{-1} (\mathbb{I}+ \hat F')(\mathbb{I}- \hat F')^{-1}\right]_{\alpha}\,^{\beta},
 \ee
 where
 \be\label{matrixls}
 s_{\alpha}\,^{\beta} = [(\mathbb{I}- \hat F) ( \mathbb{I} + \hat F)^{-1}]_{\alpha}\,^{\beta},
 \ee
 and similarly for $s'$ but with $\hat F$ replaced by $\hat F'$.  Note that the two factors $(\mathbb{I} - \hat F)$ and $(\mathbb{I} + \hat F)^{-1}$ in $s$ are inter-changeable and this remains also true for the $s'$.   In the above `$\mathbb{I}$' stands for the $(1 + p) \times (1 + p)$ unit matrix.  For matrix $s$, we have $s_{\alpha}\,^{\gamma} (s^{T})_{\gamma}\,^{\beta} = \delta_{\alpha}\,^{\beta}$. This holds also for the matrix $s'$ and matrix $w$.  The above orthogonal matrix $W$, satisfying
 \be\label{orthogonalw}
 W W^T = W^T W = \mathbb{I}_{10 \times 10},
 \ee
 can be obtained from a redefinition of the certain oscillator modes, say $\tilde a_{n \nu}$, which is a trick used in simplifying the evaluation of the matrix elements of matter part from the contribution of oscillator modes.   Let us take the following as a simple illustration for obtaining the matrix $W$. In obtaining $A^X$,  we need to evaluate, for given $n > 0$, the following matrix element,
 \be\label{illustration}
 \langle 0| e^{- \frac{1}{n} \alpha_n^\mu (M')_\mu\,^\nu \tilde\alpha_{n \nu}} |z|^{ 2\alpha_{- n}^\tau \alpha_{n \tau}} e^{- \frac{1}{n}  \alpha_{-n}^\rho (M)_\rho\,^\sigma \tilde\alpha_{-n\sigma} }| 0\rangle = \langle 0| e^{- \frac{1}{n} \alpha_n^\mu (M')_\mu\,^\nu \tilde\alpha_{n\nu}} e^{- \frac{|z|^{2n}}{\, \,n}  \alpha_{-n}^\rho (M)_\rho\,^\sigma \tilde\alpha_{-n\sigma} }| 0\rangle,
 \ee
 where $|0\rangle$ stands for the vacuum.  Purely for simplifying the evaluation of the matrix element on the right of the above equality, we first define $\tilde{\alpha}'_\mu = (M' )_{\mu}\,^{\rho}\tilde{\alpha}_\rho$ where we have omitted the index $n$ since this works for both $n > 0$ and $n < 0$, due to the matrix $M'$ being real.  Note that the commutation relation $[\tilde{\alpha}'_{n\,\mu}, \tilde{\alpha}'_{m\, \nu}] = \eta_{\mu\nu} \delta_{n + m, 0}$ continues to hold,  using the property of matrix $M'$ as given in \eqref{matrixmp}.  With this property of matrix $M'$, we can have $\tilde \alpha_{\mu} =  (M'^T)_{\mu}\,^{\nu} \tilde\alpha'_{\nu}$. Substituting this into \eqref{illustration} for $n < 0$ and also dropping the prime on $\tilde \alpha'$, we have \eqref{illustration} as
 \be\label{illustrationone}
 \langle 0| e^{- \frac{1}{n}  \tilde \alpha_{n}^{\mu}  \alpha_{n \mu} }e^{- \frac{|z|^{2n}}{\,\,n}  \alpha_{-n}^\rho W_\rho\,^\sigma \tilde\alpha_{-n\sigma}} | 0\rangle,
 \ee  
where $W$ is precisely the one given in \eqref{matrixw}.  Since $W$ is an unit matrix in the absence of fluxes, we expect that it can be diagonalized with the deformation of adding fluxes using the following non-singular matrix $V$,
\be\label{unitaryz}
V = \left(\begin{array}{cc}
v _{(1 + p)\times(1 + p)} & 0\\
  & \mathbb{I}_{(9 - p)\times (9 - p)}\end{array}\right),
 \ee
such that 
\be\label{diagonalw}
W = V W_0 V^{-1}.
\ee
In the above,
\be\label{w0}
W_0 = \left(\begin{array}{ccccc}
 \lambda_0 &&&&\\
 &\lambda_1&&&\\
 &&\ddots&&\\
 &&&\lambda_p&\\
 &&&&\mathbb{I}_{(9 - p)\times(9 - p)}
 \end{array}\right),
 \ee
and $v$ is a $(1 + p)\times (1 + p)$ non-singular matrix. We  prove (\ref{unitaryz}), (\ref{diagonalw}) and (\ref{w0}) to hold true in general in Appendix A.  We now further define\footnote{This purely serves the purpose of  simplifying  the evaluation of the matrix element \eqref{illustrationone}. For this, we keep the annihilation operator $\alpha'_{n \mu}$ with a lower Lorentz index $\mu$ while the creation operator $\alpha'^{\nu}_{-n}$ with an upper Lorentz index $\nu$.  It will be opposite for the corresponding oscillators with tilde.}, for $n > 0$, $\alpha'_{n\mu} =  (V^{-1})_{\mu}\,^{\nu}\alpha_{n \nu}$ and $\alpha'^{\mu}_{- n} =  \alpha^{\nu}_{-n}\, V_{\nu}\,^{\mu}$, and  $\tilde\alpha'_{-n\mu} =  (V^{-1})_{\mu}\,^{\nu}\tilde \alpha_{- n \nu}$ and $\tilde \alpha'^{\mu}_{n} =  \tilde \alpha^{\nu}_{n}\, V_{\nu}\,^{\mu}$. Note that  $ \tilde {\alpha}'^{\mu}_{n}\alpha'_{n \mu}=  \tilde\alpha^{\mu}_{n} \alpha_{n \mu}$. The matrix element  \eqref{illustrationone} becomes  
\be\label{illustwo}
\langle 0| e^{- \frac{1}{n}  \tilde \alpha'^{\mu}_{n} \alpha'_{n \mu} }e^{- \frac{|z|^{2n}}{\,\,n}  \lambda_{\rho}\, \alpha'^{\rho}_{-n} \tilde\alpha'_{-n\rho}} | 0\rangle.
\ee 
We have now the commutator relations $[\alpha'_{n \mu}, \alpha'^{\nu}_{- m}] = n \delta^{\nu}_{\mu} \delta_{n, m}$ and $[\tilde\alpha'^{\mu}_n, \tilde\alpha'_{- m \nu}]  = n \delta^\mu_\nu \delta_{n, m}$ when $n, m > 0$.  We still have $\alpha'_{n \mu} | 0\rangle = \tilde\alpha'^{\mu}_{n} | 0\rangle = 0$ and $ \langle 0| \alpha'^{\mu}_{- n}  = \langle 0| \tilde\alpha'_{-n \mu} = 0$. The evaluation of \eqref{illustwo} becomes then as easy as the case without the presence of fluxes, giving the results of \eqref{matrixx} to \eqref{matrixpsirr}, respectively.    

The $p + 1$ eigenvalues $\lambda_{\alpha}$ with $\alpha = 0, \cdots p$ are not all independent and can actually be determined by the given worldvolume fluxes.  First from the given property of $w$, we have $\det w = 1$, which gives 
\be\label{eigen1}
\prod_{\alpha = 0}^{p}  \lambda_{\alpha} = 1.
\ee
 The eigenvalue $\lambda$ satisfies the following equation
 \be\label{eigeneq}
  \det \left(\lambda\, \delta_{\alpha}\,^{\beta} - w_{\alpha}\,^{\beta}\right) = 0,
 \ee   
  as well as the equation 
  \be\label{inverseeigeneq}
    \det \left(\lambda^{-1}\, \delta_{\alpha}\,^{\beta} - (w^{-1})_{\alpha}\,^{\beta}\right) = 0.
  \ee 
 The last one can also be written as
    \be\label{neweigeneq}
    \det \left(\lambda^{-1}\, \delta_{\alpha}\,^{\beta} - w_{\alpha}\,^{\beta}\right) = 0,
  \ee    
  where we have used $(w^{-1})_{\alpha}\,^{\beta} = (w^{T})_{\alpha}\,^{\beta} = \eta^{\beta\beta'} w_{\beta'}\,^{\alpha'} \eta_{\alpha\alpha'}$.  In other words, for every eigenvalue $\lambda$ of $w$, its inverse $\lambda^{-1}$ is also an eigenvalue. So the $p + 1$
  eigenvalues $\lambda_{\alpha}$ are pairwise.  When $p =$ even, this must imply that one of the eigenvalues is unity.   Given this property of $\lambda_{\alpha}$, the equation (\ref{eigen1}) satisfies automatically.   For convenience, we now relabel the eigenvalues pairwise as $\lambda_{\alpha}$ and $\lambda^{-1}_{\alpha}$ with $\alpha = 0, 1\cdots [(p - 1)/2] $ and keep in mind that there is one additional unity eigenvalue, i.e., $\lambda = 1$, when $p =$ even. Here $ [(p - 1)/2]$ denotes the corresponding integral part of $(p - 1)/2$. For example, for $p = 6$, it gives an integer $2$.   
  
  For a general $p \le 6$, we need at most the following three equations to determine the corresponding eigenvalues $\lambda_{\alpha}, \lambda^{-1}_{\alpha}$ with $\alpha = 0, 1, \cdots, [(p - 1)/2]$ plus $\lambda = 1$ if $p = $ even. For $p  = $ even, we 
  have $\lambda = 1$ and 
 \be\label{d-eigenvalue-evenp}
 1 + \sum_{\alpha = 0}^{\left[\frac{p - 1}{2}\right]} \left(\lambda_{\alpha} + \lambda^{-1}_{\alpha}\right) = {\rm tr} w, \,1 + \sum_{\alpha = 0}^{\left[\frac{p - 1}{2}\right]} \left(\lambda^{2}_{\alpha} + \lambda^{-2}_{\alpha} \right)= {\rm tr} w^{2}, \,
 1 + \sum_{\alpha = 0}^{\left[\frac{p - 1}{2}\right]} \left(\lambda^{3}_{\alpha} + \lambda^{- 3}_{\alpha} \right) ={\rm tr} w^{3},
 \ee
 while for $p = $ odd, we have instead 
 \be\label{d-eigenvalue-oddp}
 \sum_{\alpha = 0}^{\left[\frac{p - 1}{2}\right]} \left(\lambda_{\alpha} + \lambda^{-1}_{\alpha}\right)= {\rm tr} w, \quad  \sum_{\alpha = 0}^{\left[\frac{p - 1}{2}\right]} \left(\lambda^{2}_{\alpha} + \lambda^{-2}_{\alpha}\right) ={\rm tr} w^{2}, \quad
 \sum_{\alpha = 0}^{\left[\frac{p - 1}{2}\right]} \left(\lambda^{3}_{\alpha} + \lambda^{- 3}_{\alpha} \right)={\rm tr} w^{3}.
 \ee
  In the above, the $w$ is given in (\ref{matrixlw}) in terms of fluxes $\hat F$ and $\hat F'$.  Concretely, we list the respective equations needed to determine the corresponding eigenvalues in Table \ref{eigeneq} for $p \le 6$.
 \begin{table}
\begin{center}
 \begin{tabular}{|c|c|}   
\hline
 p & Equation(s) for eigenvalue(s)\\
 \hline
 0& $\lambda = 1$\\
 \hline
 1 & $\lambda_{0} + \lambda^{-1}_{0} = {\rm tr} w$\\
 \hline
 2 &$\lambda_{0} + \lambda^{-1}_{0} = {\rm tr} w - 1$,\, $\lambda = 1$\\
 \hline
 3 & $\sum_{\alpha =0}^{1} (\lambda_{\alpha} + \lambda^{-1}_{\alpha}) = {\rm tr} w, \,   \sum_{\alpha =0}^{1} (\lambda^{2}_{\alpha} + \lambda^{-2}_{\alpha}) = {\rm tr} w^{2}$\\
\hline
 4 & $\sum_{\alpha =0}^{1} (\lambda_{\alpha} + \lambda^{-1}_{\alpha}) = {\rm tr} w - 1, \,  \sum_{\alpha =0}^{1} (\lambda^{2}_{\alpha} + \lambda^{-2}_{\alpha}) = {\rm tr} w^{2} - 1, \, \lambda  = 1$\\
 \hline 
 5& $\sum_{\alpha =0}^{2} (\lambda_{\alpha} + \lambda^{-1}_{\alpha}) = {\rm tr} w, \,   \sum_{\alpha =0}^{2} (\lambda^{2}_{\alpha} + \lambda^{-2}_{\alpha}) = {\rm tr} w^{2}, \, \sum_{\alpha =0}^{2} (\lambda^{3}_{\alpha} + \lambda^{-3}_{\alpha}) = {\rm tr} 
 w^{3}$\\
 \hline
 6& $\sum_{\alpha =0}^{2} (\lambda_{\alpha} + \lambda^{-1}_{\alpha}) = {\rm tr} w - 1, \,   \sum_{\alpha =0}^{2} (\lambda^{2}_{\alpha} + \lambda^{-2}_{\alpha}) = {\rm tr} w^{2} - 1, \, \sum_{\alpha =0}^{2} (\lambda^{3}_{\alpha} + \lambda^{-3}_{\alpha}) = {\rm tr} 
 w^{3} - 1$,\\
 &$ \lambda  = 1$\\
 \hline
 \end{tabular}
\caption{The equations needed to determine the corresponding eigenvalues for $p \le 6$.}\label{eigeneq}
\end{center}
 \end{table}
We actually don't need to solve the eigenvalues from the equations given in Table \ref{eigeneq} for the matrix elements given in (\ref{matrixx}), (\ref{matrixpsinsns}) and (\ref{matrixpsirr}),  respectively, for each case. 
Let us use one particular example for $p = 3$ to illustrate this.  For example, the following product  in (\ref{matrixx}) can be expressed in terms of ${\rm tr} w, ({\rm tr} w)^{2}$ and ${\rm tr} w^{2}$ as  
\bea\label{example}
  &&\left(1 - \lambda_{0} |z|^{2n}\right) \left(1 - \lambda^{-1}_{0} |z|^{2n}\right) \left(1 - \lambda_{1} |z|^{2n}\right)\left(1 - \lambda^{-1}_{1} |z|^{2n}\right)\qquad\qquad \qquad\qquad\nn
 &\,& \qquad = 1 - \sum_{\alpha = 0}^{1} \left(\lambda_{\alpha} + \lambda^{-1}_{\alpha}\right) |z|^{2n} + \left[2 + \left(\lambda_{0} + \lambda^{-1}_{0}\right)\left(\lambda_{1} + \lambda^{-1}_{1}\right)\right] |z|^{4n}  \nn
 &\,& \qquad\qquad -  \sum_{\alpha = 0}^{1} \left(\lambda_{\alpha} + \lambda^{-1}_{\alpha}\right) |z|^{6n} + |z|^{8n} \nn
 &\,&\qquad = 1 - {\rm tr} w\, |z|^{2n} + \frac{1}{2} \left[\left({\rm tr} w\right)^{2} - {\rm tr} w^{2}\right] |z|^{4n} - {\rm tr} w \, |z|^{6n} + |z|^{8n} .
 \eea 
 We are now ready to express the amplitude (\ref{amplitude}) given in the previous section in the NS-NS or R-R sector in a more compact form.  For the NS-NS sector, using (\ref{ghme}), (\ref{sghmensns}), (\ref{matrixx}) and (\ref{matrixpsinsns}) for 
 the contributions from the ghost $bc$, superghost $\beta\gamma$, the matter $X$ and $\psi$, respectively, we have the NSNS-amplitude as
 \bea\label{amplitudensns-eta}
 && \Gamma_{\rm NSNS} (\eta'\eta) = \frac{ V_{p + 1} \sqrt{\det (\eta + \hat F')\det(\eta + \hat F)}}{(8 \pi^2 \alpha')^{\frac{1 + p}{2}}} \int_0^\infty \frac{d t} {t^{\frac{9 - p}{2}}} e^{- \frac{y^2}{2\pi\alpha' t}}\, |z|^{-1}\nn
&&\times\prod_{n = 1}^\infty \left(\frac{1  + \eta'\eta |z|^{2n - 1}}{1 - |z|^{2n}}\right)^{7 + \delta_{p, {\rm even}} - p} \prod_{\alpha = 0}^{[\frac{p - 1}{2}]} \,
\frac{\left(1 + \eta'\eta \lambda_\alpha  |z|^{2 n - 1}\right)\left(1 +\eta'\eta \lambda^{-1}_\alpha  |z|^{2 n - 1}\right)}{\left(1 - \lambda_\alpha |z|^{2n}\right) \left(1 - \lambda^{-1}_\alpha |z|^{2n}\right)},
\eea  
and similarly for the R-R sector, using (\ref{ghme}), (\ref{sghmerr}), (\ref{matrixx}) and (\ref{matrixpsirr}), we have  
\bea\label{amplituderr-eta}
&&\Gamma_{\rm RR} (\eta'\eta) = \frac{V_{p + 1} \sqrt{\det(\eta + \hat F')\det(\eta + \hat F)}}{(8\pi^2\alpha')^{\frac{1 + p}{2}}} {}_{\rm 0R}\langle B', \eta'| B, \eta\rangle_{\rm 0R}
\int_0^\infty \frac{dt}{t^{\frac{9 - p}{2}}} \, e^{- \frac{y^2}{2\pi \alpha' t}} \nn
&& \times \prod_{n = 1}^\infty \left(\frac{1 + \eta' \eta |z|^{2n}}{1 - |z|^{2n}}\right)^{7 + \delta_{p, {\rm even}} - p} \, \prod_{\alpha = 0}^{[\frac{p - 1}{2}]} \,
\frac{\left(1 + \eta'\eta \lambda_\alpha  |z|^{2 n}\right)\left(1 +\eta'\eta \lambda^{-1}_\alpha  |z|^{2 n}\right)}{\left(1 - \lambda_\alpha |z|^{2n}\right) \left(1 - \lambda^{-1}_\alpha |z|^{2n}\right)}.
\eea
 In obtaining the above,  we have used the following relations
\be\label{relations}
 \frac{c^2_p}{16 \pi (2 \pi^2 \alpha')^{\frac{7 - p}{2}} }= \frac{1}{(8\pi^2\alpha')^{\frac{1 + p}{2}}}, \qquad \int_{|z|\le 1}\frac{ d^2 z}{|z|^2} = 2\pi^2 \int_0^\infty dt, 
 \ee
 where the explicit expression for $c_p$ as given right after \eqref{mgbs} has been used and $|z| = e^{- \pi t}$ as given earlier.  The above zero-mode contribution 
\be\label{0mme}
{}_{\rm 0R}\langle B', \eta'| B, \eta\rangle_{\rm 0R} \equiv {}_{\rm 0R}\langle B_{\rm sgh}, \eta'| B_{\rm sgh}, \eta\rangle_{\rm 0R} \times {}_{\rm 0R}\langle B'_\psi, \eta'| B_\psi, \eta\rangle_{\rm 0R} 
\ee
 can be computed for the general fluxes $\hat F$ and $\hat F'$, using the expression for the R-R sector zero-mode 
\eqref{rrzm} along with  \eqref{umatrix} and following the regularization scheme given in \cite{Yost,Billo:1998vr},  as 
 \bea\label{0mme-explicit}
 {}_{\rm 0R}\langle B', \eta'| B, \eta\rangle_{\rm 0R} &=& - \frac{ 2^{4} \,\delta_{\eta\eta', +}}{\sqrt{\det(\eta + \hat F)\det(\eta + \hat F')}} \nn
 &\,&\times \sum^{\left[\frac{p + 1}{2}\right]}_{n = 0} \left(\frac{1}{2^{n} n!}\right)^{2}  \left(2 n\right)! \hat F_{\left[\alpha_{1}\beta_{1}\right. \cdots}  \hat F_{\left.\alpha_{n}\beta_{n}\right]} \hat F'^{\left[\alpha_{1}\beta_{1}\right. \cdots}  \hat F'^{\left.\alpha_{n}\beta_{n}\right]}. 
 \eea
 From (\ref{amplitudensns-eta}) and (\ref{amplituderr-eta}) as well as (\ref{ns-r}), we have the GSO projected NSNS-amplitude,  
 \bea\label{amplitudensns}
 \Gamma_{\rm NSNS} &=& \frac{1}{2} \left[ \Gamma_{\rm NSNS} (+) - \Gamma_{\rm NSNS} (-)\right] \nn
 &=& \frac{ V_{p + 1} \sqrt{\det (\eta + \hat F')\det(\eta + \hat F)}}{2\,(8 \pi^2 \alpha')^{\frac{1 + p}{2}}} \int_0^\infty \frac{d t} {t^{\frac{9 - p}{2}}} e^{- \frac{y^2}{2\pi\alpha' t}}\, |z|^{-1} \left[\prod_{n = 1}^{\infty} A_{n} (+) -  \prod_{n = 1}^{\infty} A_{n} (-) \right],\nn
 \eea 
 and the GSO projected RR-amplitude 
 \bea\label{amplituderr}
 \Gamma_{\rm RR} &=& \frac{1}{2} \left[ \Gamma_{\rm RR} (+) + \Gamma_{\rm RR} (-)\right]\nn
 &=&\frac{V_{p + 1} \sqrt{\det(\eta + \hat F')\det(\eta + \hat F)}}{2\,(8\pi^2\alpha')^{\frac{1 + p}{2}}} {}_{\rm 0R}\langle B', \eta'| B, \eta\rangle_{\rm 0R}
\int_0^\infty \frac{dt}{t^{\frac{9 - p}{2}}} \, e^{- \frac{y^2}{2\pi \alpha' t}} \prod_{n = 1}^{\infty} B_{n} (+),\quad
\eea
 where 
 \bea\label{AB}
 A_{n} (\pm) &=&  \left(\frac{1   \pm |z|^{2n - 1}}{1 - |z|^{2n}}\right)^{7 + \delta_{p, {\rm even}} - p} \prod_{\alpha = 0}^{[\frac{p - 1}{2}]} \,
\frac{\left(1 \pm \lambda_\alpha  |z|^{2 n - 1}\right)\left(1 \pm \lambda^{-1}_\alpha  |z|^{2 n - 1}\right)}{\left(1 - \lambda_\alpha |z|^{2n}\right) \left(1 - \lambda^{-1}_\alpha |z|^{2n}\right)}, \nn
B_{n} (+) &=& \left(\frac{1 +  |z|^{2n}}{1 - |z|^{2n}}\right)^{7 + \delta_{p, {\rm even}} - p} \, \prod_{\alpha = 0}^{[\frac{p - 1}{2}]} \,
\frac{\left(1 +  \lambda_\alpha  |z|^{2 n}\right)\left(1 + \lambda^{-1}_\alpha  |z|^{2 n}\right)}{\left(1 - \lambda_\alpha |z|^{2n}\right) \left(1 - \lambda^{-1}_\alpha |z|^{2n}\right)}.
\eea
In obtaining (\ref{amplituderr}), we have used the property of the zero-mode (\ref{0mme-explicit}) which has the only contribution from $\eta\eta' = +$.

To proceed, we express the eigenvalues $\lambda_{\alpha} = e^{2\pi i \,\nu_{\alpha}}$ with $\alpha = 0, \cdots [(p - 1)/2]$. $\nu_{\alpha}$ takes either real or purely imaginary value. In the former
case, it looks that we could take $\nu_{\alpha} \in [0, 1/2]$ since only the combination of $\lambda_{\alpha} + \lambda^{-1}_{\alpha} = 2 \cos 2\pi \nu_{\alpha}$ appears in the amplitude and $\nu_{\alpha} = 0$ corresponds to the absence of fluxes. However,  taking account of the zero-mode contribution in the R-R sector as discussed in Appendix B (see (\ref{0mme-nu}) given below), we still need to take $\nu_{\alpha} \in [0,1)$.  In the latter case,
one can show that at most  one of the $\nu_{\alpha}$'s can take imaginary value (see Appendix A) and we can choose this particular $\nu = \nu_{0} =  i\bar\nu_{0}$ with $\bar\nu_{0} \in (0, \infty)$ since  $\lambda + \lambda^{-1} = 2 \cosh 2\pi \bar\nu_{0}$. This is actually the consequence of the matrix $w$ (\ref{matrixlw}) being  a general Lorentz transformation. It happens whenever the applied electric fluxes cannot be eliminated by a Lorentz transformation.  

We can now express the NSNS-amplitude (\ref{amplitudensns}) in terms of
the $\theta$-functions and the Dedekind $\eta$-function as (see, for example, \cite{polbookone} for their definitions) 
\bea\label{amp-nsns}
 \Gamma_{\rm NSNS} &=& \frac{ 2^{\left[\frac{p + 1}{2}\right]} \, V_{p + 1} \sqrt{\det (\eta + \hat F')\det(\eta + \hat F)}}{2\,(8 \pi^2 \alpha')^{\frac{1 + p}{2}}} \int_0^\infty \frac{d t} {t^{\frac{9 - p}{2}}} \frac{e^{- \frac{y^2}{2\pi\alpha' t}}}{\left[\eta (it)\right]^{12 - 3 \left[\frac{p + 1}{2}\right]}} \, \prod_{\alpha = 0}^{\left[\frac{p - 1}{2}\right]}  \frac{\sin \pi \nu_{\alpha}}{\theta_{1} (\nu_{\alpha} | it) } \nn
 &\,&\times \left[\left[\theta_{3} (0 | it) \right]^{4 - \left[\frac{p + 1}{2}\right]} \prod_{\alpha = 0}^{\left[\frac{p - 1}{2}\right]}  \theta_{3} (\nu_{\alpha} | it) - \left[\theta_{4} (0 | it)\right]^{4 - \left[\frac{p + 1}{2}\right]} \prod_{\alpha = 0}^{\left[\frac{p - 1}{2}\right]}  \theta_{4} (\nu_{\alpha} | it)\right].
  \eea 
Similarly for the RR-amplitude (\ref{amplituderr}), we have 
\bea\label{amp-rr}
\Gamma_{\rm RR}  &=& \frac{2^{\left[\frac{p + 1}{2}\right] - 4} \,V_{p + 1} \sqrt{\det(\eta + \hat F')\det(\eta + \hat F)}}{2\,(8\pi^2\alpha')^{\frac{1 + p}{2}}} {}_{\rm 0R}\langle B', \eta'| B, \eta\rangle_{\rm 0R}
\int_0^\infty \frac{dt}{t^{\frac{9 - p}{2}}} \, \frac{e^{- \frac{y^2}{2\pi\alpha' t}}}{\left[\eta (it)\right]^{12 - 3 \left[\frac{p + 1}{2}\right]}} \nn
&\,&\times \left[\theta_{2} (0 | it) \right]^{4 - \left[\frac{p + 1}{2}\right]} \prod_{\alpha = 0}^{\left[\frac{p - 1}{2}\right]} \frac{ \theta_{2} (\nu_{\alpha} | it)}{\theta_{1} (\nu_{\alpha} | it) } \, \prod_{\alpha = 0}^{\left[\frac{p - 1}{2}\right]}  \frac{\sin \pi \nu_{\alpha}}{\cos\pi \nu_{\alpha}}.
\eea
In Appendix B, we show that the zero-mode contribution (\ref{0mme-explicit}) can be written in terms of the $\nu_{\alpha}$ when $\nu_{\alpha} \in [0, 1)$ as
\be\label{0mme-nu}
{}_{\rm 0R}\langle B', \eta'| B, \eta\rangle_{\rm 0R} = - 2^{4}\,\delta_{\eta\eta', +}\, \prod_{\alpha =0}^{\left[\frac{p -1}{2}\right]} \cos\pi \nu_{\alpha}.
\ee 
We would like to stress that the above zero-mode expression plays a key role in writing the total amplitude in a nice form as given later in (\ref{t-amplitude-cylinder}).

With the above, we can have the total interaction amplitude for $0 \le p = p' \le 6$ as
\bea\label{t-amplitude}
\Gamma_{p, p} &=& \Gamma_{\rm NSNS} + \Gamma_{RR}\nn
&=&  \frac{ 2^{\left[\frac{p + 1}{2}\right]} \, V_{p + 1} \sqrt{\det (\eta + \hat F')\det(\eta + \hat F)}}{2\,(8 \pi^2 \alpha')^{\frac{1 + p}{2}}} \int_0^\infty \frac{d t} {t^{\frac{9 - p}{2}}} \frac{e^{- \frac{y^2}{2\pi\alpha' t}}}{\left[\eta (it)\right]^{12 - 3 \left[\frac{p + 1}{2}\right]}} \, \prod_{\alpha = 0}^{\left[\frac{p - 1}{2}\right]}  \frac{\sin \pi \nu_{\alpha}}{\theta_{1} (\nu_{\alpha} | it) } \nn
&\,&\times \left[\left[\theta_{3} (0 | it) \right]^{4 - \left[\frac{p + 1}{2}\right]} \prod_{\alpha = 0}^{\left[\frac{p - 1}{2}\right]}  \theta_{3} (\nu_{\alpha} | it) - \left[\theta_{4} (0 | it)\right]^{4 - \left[\frac{p + 1}{2}\right]} \prod_{\alpha = 0}^{\left[\frac{p - 1}{2}\right]}  \theta_{4} (\nu_{\alpha} | it) \right. \nn
&\,&\left.\qquad - \left[\theta_{2} (0 | it) \right]^{4 - \left[\frac{p + 1}{2}\right]} \prod_{\alpha = 0}^{\left[\frac{p - 1}{2}\right]}  \theta_{2} (\nu_{\alpha} | it)\right].
\eea
We would like to stress that in spite of its appearance, each term in the above square bracket is the product of four theta-functions of the same type.  For convenience, we list the three terms in the square bracket for each case in Table \ref{t2}. 
 \begin{table}
\begin{center}
 \begin{tabular}{|c|c|}   
\hline
 p & The $\theta$-terms in the square bracket in (\ref{t-amplitude}) and their simplification\\
 \hline
 0& $\theta^{4}_{3} (0 |it) - \theta^{4}_{4} (0 |it) - \theta^{4}_{2} (0 | it) =  2\, \theta^{4}_{1} (0 | it) = 0$\\
 \hline
 1  or 2 & $\theta^{3}_{3} (0 |it) \theta_{3} (\nu_{0} |it) - \theta^{3}_{4} (0 |it) \theta_{4} (\nu_{0} |it) - \theta^{3}_{2} (0 |it) \theta_{2} (\nu_{0} | it) = 2\, \theta^{4}_{1} \left(\left.\frac{\nu_{0}}{2}\right| it \right)$\\
 \hline
 3 or 4 &$\theta^{2}_{3} (0 |it) \theta_{3} (\nu_{0} |it) \theta_{3} (\nu_{1} |it)- \theta^{2}_{4} (0 |it) \theta_{4} (\nu_{0} |it) \theta_{4} (\nu_{1} |it)- \theta^{2}_{2} (0 |it) \theta_{2} (\nu_{0} | it)\theta_{2} (\nu_{1} | it)$\\
 & $= 2\, \theta^{2}_{1} \left(\left.\frac{\nu_{0} + \nu_{1}}{2}\right|it \right) \theta^{2}_{1} \left(\left.\frac{\nu_{0} - \nu_{1}}{2}\right|it \right)$\\
  \hline
  & $\theta_{3} (0 |it) \theta_{3} (\nu_{0} |it) \theta_{3} (\nu_{1} |it) \theta_{3} (\nu_{2} |it) - \theta_{4} (0 |it) \theta_{4} (\nu_{0} |it) \theta_{4} (\nu_{1} |it) \theta_{4} (\nu_{2} |it)$\\
 5 or 6 & $ - \theta_{2} (0 |it) \theta_{2} (\nu_{0} | it)\theta_{2} (\nu_{1} | it) \theta_{2} (\nu_{2} | it) $\\
 & $= 2\, \theta_{1} \left(\left.\frac{\nu_{0} + \nu_{1} + \nu_{2}}{2}\right| it \right) \theta_{1} \left(\left.\frac{\nu_{0} - \nu_{1} + \nu_{2}}{2}\right| it \right)
 \theta_{1} \left(\left.\frac{\nu_{0} + \nu_{1} - \nu_{2}}{2}\right| it \right)\theta_{1} \left(\left.\frac{\nu_{0} - \nu_{1} - \nu_{2}}{2}\right| it \right)$\\
 \hline
 \end{tabular}
\caption{The $\theta$-terms in the square bracket in (\ref{t-amplitude}) and their simplification.}\label{t2}
\end{center}
 \end{table}
In this Table, we also use the following identity for the $\theta$-functions from \cite{whittaker-watson} to simplify the formula
\bea\label{theta-identity}
&&2 \,\theta_{1} (w |\tau) \theta_{1} (x |\tau) \theta_{1} (y |\tau)  \theta_{1} (z |\tau) = \theta_{1} (w' |\tau) \theta_{1} (x' |\tau) \theta_{1} (y' |\tau)  \theta_{1} (z' |\tau) \nn
&&\qquad\qquad\qquad  \qquad + \theta_{2} (w' |\tau) \theta_{2} (x' |\tau) \theta_{2} (y' |\tau)  \theta_{2} (z' |\tau) - \theta_{3} (w' |\tau) \theta_{3} (x' |\tau) \theta_{3} (y' |\tau)  \theta_{3} (z' |\tau)\nn
&&  \qquad\qquad\qquad \qquad  + \theta_{4} (w' |\tau) \theta_{4} (x' |\tau) \theta_{4} (y' |\tau)  \theta_{4} (z' |\tau),
\eea    
where $w', x', y'$ and $z'$ are related to $w, x, y, z$ as 
\bea
&& 2 w' = - w + x + y + z, \quad 2 x' = w - x + y + z,\nn
&& 2 y' = w + x - y + z,\quad 2 z' = w + x + y - z.
\eea   
Note that $\theta_{1} (0 |\tau) = 0$. So we have computed the closed string cylinder amplitude for each of the $0 \le p = p' \le 6$ cases with the various $\theta$-function terms in the square bracket in (\ref{t-amplitude}) now given by the simplified term containing only the $\theta_{1}$-function listed in Table 2.  Carefully examining the integrand of the amplitude for each case, in particular the $\theta_{1}$-function factor, we observe that the amplitude can be expressed by a universal formula for each of $0 \le p = p' \le 6$ so long a proper choice or a limit is taken for the respective $\nu_{\alpha}$.  Concretely,  the general amplitude is 
 \bea\label{t-amplitude-cylinder}
\Gamma_{p,p} &=&\frac{ 2^3 \, V_{p + 1}\, \left[\det (\eta + \hat F')\det(\eta + \hat F)\right]^{\frac{1}{2}} }{(8 \pi^2 \alpha')^{\frac{p + 1}{2}}} \int_0^\infty \frac{d t} {t^{\frac{9 - p}{2}}} \frac{e^{- \frac{y^2}{2\pi\alpha' t}}}{\eta^{3} (it)} \nn 
&\,& \times \frac{ \theta_{1} \left(\left.\frac{\nu_{0} + \nu_{1} + \nu_{2}}{2}\right| it \right) \theta_{1} \left(\left.\frac{\nu_{0} - \nu_{1} + \nu_{2}}{2}\right| it \right)
 \theta_{1} \left(\left.\frac{\nu_{0} + \nu_{1} - \nu_{2}}{2}\right| it \right)\theta_{1} \left(\left.\frac{\nu_{0} - \nu_{1} - \nu_{2}}{2}\right| it \right)}{\theta_{1} (\nu_{0} | it)\theta_{1} (\nu_{1} | it)\theta_{1} (\nu_{2} | it)}  \prod_{\alpha =0}^{2} \sin \pi \nu_{\alpha} \nn
 &=&  \frac{ 2^2 \, V_{p+ 1} \left[\det (\eta + \hat F')\det(\eta + \hat F)\right]^{\frac{1}{2}}  \left[\sum_{\alpha = 0}^{2}\cos^{2}\pi \nu_{\alpha} - 2\prod_{\alpha =0}^{2} \cos\pi\nu_{\alpha} - 1\right]}{(8 \pi^2 \alpha')^{\frac{p + 1}{2}}}   \nn
 &\,&\times \int_0^\infty \frac{d t}{t^{\frac{9 - p}{2}}}  e^{- \frac{y^2}{2\pi\alpha' t}} \prod_{n=1}^{\infty} C_{n},
 \eea
 where 
 \be\label{C}
 C_{n} = \frac{\tilde C_{n} }{(1 - |z|^{2n})^{2} \prod_{\alpha = 0}^{2}
 \left[1 - 2 |z|^{2n} \cos2\pi \nu_{\alpha}  + |z|^{4n}\right]},
 \ee
 with 
 \bea
 \tilde C_{n} &=& \left[1 - 2 |z|^{2n} \cos\pi(\nu_{0} + \nu_{1} + \nu_{2}) + |z|^{4n}\right] \left[1 - 2 |z|^{2n} \cos\pi(\nu_{0} - \nu_{1} + \nu_{2}) + |z|^{4n}\right] \nn
 &&\times \left[1 - 2 |z|^{2n} \cos\pi(\nu_{0} + \nu_{1} - \nu_{2}) + |z|^{4n}\right]\left[1 - 2 |z|^{2n} \cos\pi(\nu_{0} - \nu_{1} - \nu_{2}) + |z|^{4n}\right] \nn
 &\,& \label{tildeC1}\\
 &\stackrel{\rm or}{=}& \left[1 - 2 |z|^{2n} e^{i\pi\nu_{0}} \cos\pi(\nu_{1} + \nu_{2}) + e^{2\pi i \nu_{0}} |z|^{4n}\right]\nn
 &\,& \times \left[1 - 2 |z|^{2n} e^{i\pi\nu_{0}} \cos\pi(\nu_{1} - \nu_{2}) + e^{2\pi i \nu_{0}} |z|^{4n}\right] \nn
 &\,&\times \left[1 - 2 |z|^{2n} e^{- i\pi\nu_{0}} \cos\pi(\nu_{1} + \nu_{2}) + e^{- 2\pi i \nu_{0}} |z|^{4n}\right]\nn
&\,&\times\left[1 - 2 |z|^{2n} e^{- i\pi\nu_{0}} \cos\pi(\nu_{1} - \nu_{2}) + e^{-2\pi i \nu_{0}} |z|^{4n}\right].\label{tildeC2}\eea
In the first line of equality in (\ref{t-amplitude-cylinder}), the following factor in the integrand 
\be\label{65-factor}
 \frac{ \theta_{1} \left(\left.\frac{\nu_{0} + \nu_{1} + \nu_{2}}{2}\right| it \right) \theta_{1} \left(\left.\frac{\nu_{0} - \nu_{1} + \nu_{2}}{2}\right| it \right)
 \theta_{1} \left(\left.\frac{\nu_{0} + \nu_{1} - \nu_{2}}{2}\right| it \right)\theta_{1} \left(\left.\frac{\nu_{0} - \nu_{1} - \nu_{2}}{2}\right| it \right)}{\theta_{1} (\nu_{0} | it)\theta_{1} (\nu_{1} | it)\theta_{1} (\nu_{2} | it)} \prod_{\alpha =0}^{2} \sin \pi \nu_{\alpha} 
 \ee 
 is actually for $p = p' = 6$ or $5$.  But it will reduce to the $p = p' = 4$ or $3$ case if we take the limit $\nu_{2} = 0$, to the  $p = p' = 2$ or $1$ if we take the limits $\nu_{2} = 0, \nu_{1} =0$ and to the $p = p' = 0$ case if we take $\nu_{2} = 0, \nu_{1} = 0, \nu_{0} = 0$.  We will explain this nice feature of the amplitude later on.
 
 This same amplitude can also be expressed in terms of the open string one-loop annulus one via the Jacobi transformation $t \to t' = 1/t$, from the first equality in (\ref{t-amplitude-cylinder}), as
\bea\label{t-amplitude-annulus}
\Gamma_{p,p} &=& - \frac{ 2^3 \, i \, V_{p + 1}\, \left[\det (\eta + \hat F')\det(\eta + \hat F)\right]^{\frac{1}{2}}\prod_{\alpha =0}^{2} \sin \pi \nu_{\alpha} }{(8 \pi^2 \alpha')^{\frac{p + 1}{2}}} \int_0^\infty \frac{d t'} {t'^{\frac{p - 3}{2}}} \frac{e^{- \frac{y^2 t'}{2\pi\alpha' }}}{\eta^{3} (it')} \nn 
&\,& \times \frac{ \theta_{1} \left(\left. \frac{\nu_{0} + \nu_{1} + \nu_{2}}{2} i t' \right| it' \right) \theta_{1} \left(\left. \frac{\nu_{0} - \nu_{1} + \nu_{2}}{2} it' \right| it' \right)
 \theta_{1} \left(\left.  \frac{\nu_{0} + \nu_{1} - \nu_{2}}{2} it' \right| it' \right)\theta_{1} \left(\left.  \frac{\nu_{0} - \nu_{1} - \nu_{2}}{2} it' \right| it' \right)}{\theta_{1} (\nu_{0} it' | it')\theta_{1} ( \nu_{1} it' | it')\theta_{1} ( \nu_{2} it' | it')}\nn
 &=&  \frac{ 2^2 \, V_{p+ 1} \left[\det (\eta + \hat F')\det(\eta + \hat F)\right]^{\frac{1}{2}} }{(8 \pi^2 \alpha')^{\frac{p + 1}{2}}}   \int_0^\infty \frac{d t}{t^{\frac{p - 3}{2}}}  e^{- \frac{y^2 t}{2\pi\alpha'}} \prod_{\alpha =0}^{2} \frac{\sin \pi \nu_{\alpha}}{\sinh \pi \nu_{\alpha} t} \nn
 &\,& \times \left[\sum_{\alpha = 0}^{2}\cosh^{2}\pi \nu_{\alpha} t- 2 \prod_{\alpha = 0}^{2} \cosh\pi \nu_{\alpha} t - 1\right] \prod_{n=1}^{\infty} Z_{n},
 \eea
 where  in obtaining the first equality in (\ref{t-amplitude-annulus}),  we have used the following relations for the $\theta_{1}$-function and the Dedekind $\eta$-function,
\be\label{jacobi}
\eta (\tau) = \frac{1}{(- i \tau)^{1/2}} \eta \left(- \frac{1}{\tau}\right), \quad \theta_1 (\nu | \tau) = i \frac{e^{- i \pi \nu^2/\tau}}{(- i \tau)^{1/2}} \theta_1 \left(\left.\frac{\nu}{\tau} \right| - \frac{1}{\tau}\right),
\ee 
in the second equality we have dropped the prime on $t$, and 
  \be\label{Z}
 Z_{n} = \frac{\tilde Z_{n} }{(1 - |z|^{2n})^{2} \prod_{\alpha = 0}^{2}
 \left[1 - 2 |z|^{2n} \cosh2\pi \nu_{\alpha} t  + |z|^{4n}\right]},
 \ee
 with
 \bea
 \tilde Z_{n} &=& \left[1 - 2 |z|^{2n} \cosh\pi(\nu_{0} + \nu_{1} + \nu_{2}) t + |z|^{4n}\right] \left[1 - 2 |z|^{2n} \cosh\pi(\nu_{0} - \nu_{1} + \nu_{2})t + |z|^{4n}\right] \nn
 &\times& \left[1 - 2 |z|^{2n} \cosh\pi(\nu_{0} + \nu_{1} - \nu_{2})t + |z|^{4n}\right]\left[1 - 2 |z|^{2n} \cosh\pi(\nu_{0} - \nu_{1} - \nu_{2})t + |z|^{4n}\right] \nn
 &\,& \label{tildeZ1}\\
 &\stackrel{\rm or}{=}& \left[1 - 2 |z|^{2n} e^{-\pi\nu_{0}t} \cosh\pi(\nu_{1} + \nu_{2})t + e^{-2\pi  \nu_{0} t} |z|^{4n}\right]\nn
 &\,&\times \left[1 - 2 |z|^{2n} e^{- \pi\nu_{0} t} \cosh\pi(\nu_{1} - \nu_{2})t + e^{- 2\pi  \nu_{0} t} |z|^{4n}\right] \nn
 &\,&\times \left[1 - 2 |z|^{2n} e^{\pi\nu_{0} t} \cosh\pi(\nu_{1} + \nu_{2})t + e^{ 2\pi  \nu_{0}t} |z|^{4n}\right]\nn
&\,&\times\left[1 - 2 |z|^{2n} e^{\pi\nu_{0}t} \cosh\pi(\nu_{1} - \nu_{2}) t + e^{2\pi  \nu_{0}t} |z|^{4n}\right].\label{tildeZ2}
\eea
In what follows, we will discuss each of the four cases: 1) $p = p' = 6$ or $5$, 2) $p = p' = 4$ or $3$, 3)$p = p' = 2$ or $1$, and 4) $p = p' =0$, separately.

\subsection{The $p = p' =6$ or $5$ case\label{subsubs3.1}}  
For the respective general worldvolume fluxes $\hat F$ and $\hat F'$, from the second equality of (\ref{t-amplitude-cylinder}), we have the interaction amplitude for $p = p' = 6$  or $5$ case as
\bea\label{tca6}
\Gamma_{p, p} &=&  \frac{ 2^2 \, V_{p + 1}\,  \left[\sum_{\alpha = 0}^{2 }\cos^{2}\pi \nu_{\alpha} - 2 \prod_{\alpha = 0}^{2} \cos\pi \nu_{\alpha}  - 1\right]  \sqrt{\det (\eta + \hat F')\det(\eta + \hat F)}}{(8 \pi^2 \alpha')^{\frac{p + 1}{2}}} \nn
 &\,& \times \int_0^\infty \frac{d t} {t^{\frac{9 - p}{2}}} e^{- \frac{y^2}{2\pi\alpha' t}} \prod_{n=1}^{\infty} C_{n},
 \eea
 where $C_{n}$ is given in (\ref{C}). For large brane separation $y$, the amplitude has its contribution mostly from the large $t$ integration for which $C_{n} \approx 1$. We have therefore 
 
 \bea\label{largeyp=65}
 \frac{\Gamma_{p, p}}{V_{p + 1}} &\approx&  \frac{ \left[\sum_{\alpha = 0}^{2 }\cos^{2}\pi \nu_{\alpha} - 2 \prod_{\alpha = 0}^{2} \cos\pi \nu_{\alpha}  - 1\right] \sqrt{\det (\eta + \hat F')\det(\eta + \hat F)}}{2^{-2}(8 \pi^2 \alpha')^{\frac{p + 1}{2}}} \nn
 &\,&\times \int_0^\infty \frac{d t} {t^{\frac{9 - p}{2}} }e^{- \frac{y^2}{2\pi\alpha' t}}\nn
 &=&  \frac{\left[\sum_{\alpha = 0}^{2 }\cos^{2}\pi \nu_{\alpha} - 2 \prod_{\alpha = 0}^{2} \cos\pi \nu_{\alpha}  - 1\right]  \sqrt{\det (\eta + \hat F')\det(\eta + \hat F)}}{2^{p - 1} \pi^{\frac{p + 1}{2}} (2 \pi \alpha')^{p - 3} y^{7 -p}}
 \int_0^\infty \frac{d t} {t^{\frac{9 -p }{2}}} e^{- \frac{1}{ t}}\nn
 &=&   \frac{\left[\sum_{\alpha = 0}^{2 }\cos^{2}\pi \nu_{\alpha} - 2 \prod_{\alpha = 0}^{2} \cos\pi \nu_{\alpha}  - 1\right]  \sqrt{\det (\eta + \hat F')\det(\eta + \hat F)}}{2^{p - 1} \pi^{\frac{p + 1}{2}}(2 \pi \alpha')^{p - 3} y^{7 -p}} \Gamma \left(\frac{7 - p}{2}\right),\nn
  \eea
where in the second equality, we have rescaled the integration variable $t$, and $\Gamma (7 - p/2)$ is the $\Gamma$-function with $\Gamma (1/2) = \sqrt{\pi}$ for $p = 6$ and $\Gamma (1) = 1$ for $p = 5$, respectively. 
 According to our conventions,   $\Gamma_{p, p} > 0$ gives an attractive interaction while $\Gamma_{p, p} < 0$ gives a repulsive one.  The sign of the factor ${\cal F} = \sum_{\alpha = 0}^{2 }\cos^{2}\pi \nu_{\alpha} - 2 \prod_{\alpha = 0}^{2} \cos\pi \nu_{\alpha}  - 1$ in the above determines that of the amplitude. When any of $\nu_{0}, \nu_{1}$ and $\nu_{2}$ vanishes, it is non-negative. For example, taking $\nu_{0} = 0$, we can write it as ${\cal F} = (\cos\nu_{2} - \cos\nu_{1})^{2} \ge 0$. When none of them vanishes, we have two cases to consider. 
 Case 1):  one of them is imaginary, say, $\nu_{0} = i \bar\nu_{0}$ with $\bar\nu_{0} \in (0, \infty)$, and $\nu_{1}, \nu_{2} \in (0, 1)$.  We have now
 \bea\label{oneimag}
 {\cal F} &=&  \sinh^{2} \pi \bar \nu_{0} + \cos^{2}\pi\nu_{1} + \cos^{2} \pi \nu_{2} - 2 \cosh \pi \bar\nu_{0} \cos\pi\nu_{1} \cos\pi\nu_{2}\nn
&=&  \sinh^{2}\pi\bar\nu_{0} +   \cos^{2}\pi\nu_{1} + \cos^{2} \pi \nu_{2} - 2 \left(1 + 2 \sinh^{2} \frac{\pi\bar\nu_{0}}{2}\right) \cos\pi\nu_{1} \cos\pi\nu_{2}\nn
&=&  4  \left(\cosh^{2} \frac{\pi\bar\nu_{0}}{2} - \cos\pi\nu_{1}\cos\pi\nu_{2}\right)\sinh^{2} \frac{\pi\bar\nu_{0}}{2} + (\cos\pi\nu_{1} - \cos\pi\nu_{2})^{2} > 0.
\eea
In other words, the large brane separation is always attractive.
Case 2): all $\nu_{0}, \nu_{1}, \nu_{2} \in (0, 1)$.  This can only be possible for $p = 6$ since it needs at least six worldvolume spatial directions.  For this case,  in order to determine the condition for the sign of ${\cal F}$, we rewrite it as
\be\label{all-real}
{\cal F} = 2^{2} \sin \pi \frac{\nu_{0} + \nu_{1} + \nu_{2}}{2} \sin\pi \frac{\nu_{0} - \nu_{1} + \nu_{2}}{2} \sin \pi \frac{\nu_{0} + \nu_{1} - \nu_{2}}{2} \sin\pi\frac{\nu_{0} - \nu_{1} - \nu_{2}}{2}.
\ee 
 Note that ${\cal F}$ is symmetric under the exchange of any two of $\nu$'s. So without loss of generality, we can assume $\nu_{0} \le \nu_{1} \le \nu_{2}$. Given this, we have three subcases to consider due to the allowed range for each of the $\nu$'s: a) $0 < \nu_{0} + \nu_{1} + \nu_{2} < 2$, b) $\nu_{0} + \nu_{1} + \nu_{2} = 2$ and c) $2 < \nu_{0} + \nu_{1} + \nu_{3} < 3$.  For subcase a), it is clear that the first two factors $\sin \pi (\nu_{0} + \nu_{1} + \nu_{2})/2 > 0$ and $\sin\pi (\nu_{0} - \nu_{1} + \nu_{2})/2 > 0$ while the last factor $\sin\pi(\nu_{0} - \nu_{1} - \nu_{2})/2 < 0$. So the sign of ${\cal F}$ is determined by that of third factor $\sin \pi (\nu_{0} + \nu_{1} - \nu_{2})/2$. If $\nu_{0} + \nu_{1} > \nu_{2}$, then ${\cal F} < 0$ while on the other hand if $\nu_{0} + \nu_{1} < \nu_{2}$, we have ${\cal F} > 0$.  ${\cal F} = 0$ when $\nu_{0} + \nu_{1} = \nu_{2}$. In other words, if the possible largest one of the three $\nu$'s is smaller than the sum of the remaining two, the long-range interaction is repulsive. Otherwise, it is attractive or vanishes.  For subcase b), ${\cal F} = 0$ and so we have the force vanishing, implying the underlying system to be supersymmetric. For subcase c),  we have both the first factor $\sin \pi (\nu_{0} + \nu_{1} + \nu_{2})/2  < 0$ and the last factor $\sin\pi(\nu_{0} - \nu_{1} - \nu_{2})/2 < 0$ while the second factor $\sin\pi (\nu_{0} - \nu_{1} + \nu_{2})/2 > 0$. For this subcase, we must have $\nu_{0} + \nu_{1} > \nu_{2}$ given each of the $\nu_{\alpha} \in (0, 1)$ and  we therefore have the third factor $\sin \pi (\nu_{0} + \nu_{1} - \nu_{2})/2 > 0$. Hence we have now ${\cal F} > 0$, therefore the interaction is again attractive.  In other words,  the long-range interaction can only be repulsive when the three $\nu$'s are non-vanishingly real and satisfy $\nu_{0} + \nu_{1} + \nu_{2} < 2$, and when the possible largest one of the three $\nu$'s  is smaller than the sum of the remaining two. The above discussion actually remains true for any brane separation as we will show this later on.  For $\nu_{0} + \nu_{1} + \nu_{2} < 2$, as  will be shown, when the possible largest one of the three $\nu$'s is smaller than the sum of the remaining two, the interaction is repulsive for any brane separation and there is no tachyonic instability.  However, when the largest one is larger than the sum of the remaining two,  the corresponding interaction remains attractive at least until the brane separation reaches the minimum determined by the so-called tachyonic shift and  after that there is a tachyonic instability to occur. This tachyonic instability occurs whenever the interaction is attractive and the brane separation reaches the minimum. When the possible largest one equals to the sum of the remaining two or $\nu_{0} + \nu_{1} + \nu_{2} = 2$, the interaction vanishes for any brane separation and this no-force implies the underlying system to be supersymmetric.  
 
We now come to explain the above.  It is well-known that when a D2 or D4 brane is placed parallel to a D6 at a separation transverse to the D6, the interaction between them is zero or attractive while by the same token, the interaction between the D6 and a D0 is repulsive. Each of these cases can be examined easily in the following section when we consider the case of $p - p' = 2, 4, 6$ with $p \le 6$.  For the present case under consideration, we have $p = p' = 6$ along with $\nu_{\alpha} \in (0, 1)$ for $\alpha = 0, 1, 2$.  In other words,  at least one of the two D6 carries a constant magnetic flux, for example, $\hat F$, with non-vanishing components $\hat F_{12}, \hat F_{34}$ and $\hat F_{56}$, which can give rise to the above three $\nu_{\alpha} \in (0, 1)$.  For such a flux,  it implies that the D6 has its delocalized D4, D2 and D0 within the D6, which can be easily understood from the following coupling on the D6 worldvolume as 
\be\label{d6brane-coupling}
T_{6} \int \left(C_{7 - 2n} \wedge \hat F^{n}_{2}\right)_{7},
\ee
 where $T_{6}$ is the D6 brane tension, $C_{p + 1}$ is the potential minimally coupled with Dp,  and $\hat F^{n}_{2} = \hat F_{2} \wedge \cdots \wedge \hat F_{2}$ stands for the wedge product of n $\hat F_{2}$.  So $n = 0$ gives the coupling of D6 with the R-R potential $C_{7}$, $n = 1$ gives the coupling of D4 with $C_{5}$, $n = 2$ gives the coupling of D2 with $C_{3}$ and $n = 3$ gives the coupling of D0 with $C_{1}$.  Given what we have for the non-vanishing components of $\hat F$, we can have $n = 0, 1, 2 , 3$.  So this gives an explanation of the delocalized D0, D2, D4 within D6 mentioned above. So when the possible largest one of the three $\nu$'s is smaller than the sum of the remaining two for the case $\nu_{0} + \nu_{1} + \nu_{2} < 2$,  the above cylinder amplitude shows that the repulsive interaction between the other D6 and the delocalized D0 on this D6 overtakes the attractive ones between the other D6 and the delocalized D2 or D4.  Otherwise, the attractive interaction overtakes the repulsive one.  The net interaction vanishes when the two equals. This also explains that when one of the three $\nu$'s is imaginary or vanishes, the net interaction is attractive since we must have one of the $\hat F_{12}, \hat F_{34}$ and $\hat F_{56}$ being zero which implies the absence of D0 branes on the D6.  
  
 For small brane separation, the small $t$ integration  in (\ref{tca6}) becomes important and has to be considered.  For small $t$, $C_{n}$ in (\ref{C}) can be large. In particular, $C_{n}$ blows up when $t \to 0$ due to the factor 
 $(1 - |z|^{2n})^{2}$ in its denominator. So overall we have a blowing up factor $\prod_{n = 1}^{\infty} (1 - |z|^{2n})^{-2}$ for $t \to 0$ in the integrand of the amplitude (\ref{tca6}).  Note also that the exponentially suppressing factor $e^{- y^{2}/(2\pi\alpha' t)}$ in the integrand becomes vanishingly small when $t \to 0$. So there is a competition between the two and one expects a potentially interesting physics 
to occur when $t \to 0$. This will become manifest when we transform the closed string cylinder amplitude to the corresponding open string one-loop annulus one and it is a potential open string tachyonic instability. For now, the nature of $\Gamma_{p, p}$ depends on that of the parameters $\nu_{0}, \nu_{1}$ and $\nu_{2}$.   

Following the previous discussion for large $y$, we have that the interaction is attractive whenever the three $\nu$'s are all real and the possible largest one of these is larger than the sum of the remaining two when $\nu_{0} + \nu_{1} + \nu_{2} < 2$ or is smaller than that when $2 < \nu_{0} + \nu_{1} + \nu_{2} < 3$ or one of them is imaginary which we will address later on.  We now take a close look for a general $y$. Note that the $C_{n}$ (\ref{C}) is still positive even for small $t$ and this can be easily checked. Each factor in the numerator $\tilde C_{n}$ (\ref{tildeC1}) is positive,  for example, the first factor $[1 - 2 |z|^{2n} \cos\pi(\nu_{0} + \nu_{1} + \nu_{2}) + |z|^{4n}] > 1 - 2 |z|^{2n} + |z|^{4n} = (1 - |z|^{2n})^{2} > 0$.  By the same token, each factor in the denominator is also positive.  In other words, the sign of $\Gamma_{p, p}$ is still determined by that of ${\cal F}$ given earlier.  For this,  the attractive 
interaction acting between the two D6 has a tendency to  move the two towards each other and to make the brane separation smaller.  Therefore the exponential factor $e^{- y^{2}/(2\pi\alpha' t)}$ will make its suppressing less important and one expects that the diverging effect from $C_{n}$ at small $t$ will become to dominate at a certain point. So we expect then a potential instability mentioned above to occur.    On the other hand, when the possible largest one of three $\nu$'s is less than the sum of the remaining two when $\nu_{0} + \nu_{1} + \nu_{2} < 2$, the interaction is repulsive and as such has a tendency to move the two D6 apart further.  So this makes the suppression of the exponential factor $e^{- y^{2}/(2\pi\alpha' t)}$ in the integration more important and disfavors the instability to occur.  So this appears to provide a correlation between the nature of interaction and the existence of potential tachyonic instability.  We will show later that this is indeed true.

For small $t$, there appears a new feature when one of three $\nu$'s takes an imaginary value.  For example, we choose $\nu_{0} = i \bar \nu_{0}$ with $\bar\nu_{0} \in (0, \infty)$.  Now $C_{n}$ can be negative. By the same token as given in the previous paragraph, each factor in $\tilde C_{n}$ (\ref{tildeC2}) continues to be positive, for example, the third factor $ \left[1 - 2 |z|^{2n} e^{\pi\bar \nu_{0}} \cos\pi(\nu_{1} + \nu_{2}) + e^{2\pi \bar \nu_{0}} |z|^{4n}\right] > 1 - 2 |z|^{2n} e^{\pi \bar\nu_{0}} + e^{2\pi\bar\nu_{0}} |z|^{4n} = (1 - e^{\pi\bar\nu_{0}} |z|^{2n})^{2} > 0$. However, the factor $[1 - 2 |z|^{2n} \cosh2\pi\bar\nu_{0} + |z|^{4n}] \approx 2 (1-  \cosh 2\pi\bar\nu_{0}) < 0 $, in the denominator of $C_{n}$, for small $t$. Since there are an infinite number of $C_{n}$ appearing as product in the integrand,  the sign of $\Gamma_{p, p}$ will then be indefinite. So for small $y$, the nature of the interaction becomes obscure for the case under consideration and this indicates that there should exist new physical process occurring in addition to the potential tachyonic instability mentioned above for small $t$.  This new physics is actually the decay of the underlying system via the so-called open string pair production under the action of applied electric fluxes which makes $\nu_{0}$ become imaginary.  All these will become 
manifest when the interaction is expressed in terms of the open string variable as the open string one-loop annulus amplitude (\ref{t-amplitude-annulus}) for $p = 6$ which we turn next.  Note that this consideration applies to both $p = p' = 6$ and $p = p' = 5$ cases. 
 
 The open string one-loop  annulus amplitude for $p = p' = 5, 6$, respectively, can be read from the second equality of (\ref{t-amplitude-annulus}) as 
 \bea\label{taa56}
\Gamma_{p, p} &=&  \frac{ 2^2 \, V_{p + 1} \left[\det (\eta + \hat F')\det(\eta + \hat F)\right]^{\frac{1}{2}} }{(8 \pi^2 \alpha')^{\frac{p + 1}{2}}}   \int_0^\infty \frac{d t}{t^{\frac{p - 3}{2}}}  e^{- \frac{y^2 t}{2\pi\alpha'}} \prod_{\alpha =0}^{2} \frac{\sin \pi \nu_{\alpha}}{\sinh \pi \nu_{\alpha} t} \nn
 &\,& \times \left[\sum_{\alpha = 0}^{2}\cosh^{2}\pi \nu_{\alpha} t- 2 \prod_{\alpha = 0}^{2} \cosh\pi \nu_{\alpha} t - 1\right] \prod_{n=1}^{\infty} Z_{n},
 \eea 
where $Z_{n}$ is given in (\ref{Z}).  Note that the closed string t-variable and the open string t-variable are inversely related to each other.  So small t in closed string case implies large t in open string one. So the potential open string tachyonic instability 
should show up for large t in the integrand of the above amplitude if it exists at all.  Let us examine this in detail.  

For large t, note that $n \ge 1$ and  $Z_{n} \approx 1$ from ({\ref{Z}) for either all $\nu$'s being real with $\nu_{\alpha} \in (0, 1)$ and $\nu_{0} + \nu_{1} + \nu_{2} < 2$ or one of them being imaginary and the rest being real but less than unity.   When all three $\nu_{\alpha}$ are real with $\nu_{\alpha} \in (0, 1)$ and $2 < \nu_{0} + \nu_{1} + \nu_{2} < 3$, we have instead,  for large t,  $Z_{1} \approx -  e^{\pi (\nu_{0} + \nu_{1} + \nu_{2} - 2) t}$ and $Z_{n} \approx 1$ for $n \ge 2$ again from (\ref{Z}).   When all three $\nu$'s are real (only valid for $p = p' = 6$),  we once again assume $\nu_{0} \le \nu_{1} \le \nu_{2}$ without loss of generality since the amplitude is symmetric under the exchange of any two of the three $\nu_{0}, \nu_{1}$ and $\nu_{2}$. From the discussion in the closed string variable, we know that when $\nu_{0} + \nu_{1} > \nu_{2}$ for $\nu_{0} + \nu_{1} + \nu_{2} < 2$, the interaction is repulsive and one expects no tachyonic instability. Let us check this explicitly here. From (\ref{taa56}), the terms in the square bracket can be expressed as 
\be\label{sb}
2^{2} \sinh \frac{\pi (\nu_{0} + \nu_{1} + \nu_{2}) t}{2}  \sinh \frac{\pi (\nu_{0} - \nu_{1} + \nu_{2}) t}{2}   \sinh \frac{\pi (\nu_{0} + \nu_{1} - \nu_{2}) t}{2}  \sinh \frac{\pi (\nu_{0} - \nu_{1} - \nu_{2}) t}{2}.
\ee 
So from this,  one sees for large $t$ and $\nu_{0} + \nu_{1} > \nu_{2}$ with $\nu_{0} + \nu_{1} + \nu_{2} < 2$ that the integrand in (\ref{taa56}) is negative and further it behaves like
\be
\sim - \frac{e^{- \frac{y^2 t}{2\pi\alpha'}} e^{\pi (\nu_{0} + \nu_{1} + \nu_{2}) t}}{ \sinh\pi\nu_{0} t \sinh\pi \nu_{1} t \sinh\pi\nu_{2}t} \stackrel{t \to \infty}{\longrightarrow}  - e^{- \frac{y^2 t}{2\pi\alpha'}},
\ee  
which shows no tachyonic shift and therefore no potential tachyonic instability.  This is consistent with our anticipation.  However, when $\nu_{0} + \nu_{1} < \nu_{2}$, we do expect to see the potential tachyonic instability. Now (\ref{sb}) gives $\sim e^{2\pi \nu_{2} t}$ and  the integrand is positive and behaves like
\be
\sim  \frac{e^{- \frac{y^2 t}{2\pi\alpha'}} e^{2\pi \nu_{2} t}}{ \sinh\pi\nu_{0} t \sinh\pi \nu_{1} \sinh\pi\nu_{2}t} \stackrel{t \to \infty}{\longrightarrow}  e^{- \frac{y^2 t}{2\pi\alpha'}} e^{\pi (\nu_{2} - \nu_{0} - \nu_{1})t},
\ee
where we have a so-called tachyonic shift $(\nu_{2} - \nu_{0} -\nu_{1})/2 > 0$ \cite{Ferrara:1993sq, Lu:2018nsc}.  The effective mass square for the open string is 
\be\label{msquare}
m^{2} = \frac{y^{2}}{(2\pi \alpha')^{2}} - \frac{\nu_{2} - \nu_{0} - \nu_{1}}{2 \alpha'},
\ee
which becomes tachyonic if $y < \pi \sqrt{2 (\nu_{2} - \nu_{0} - \nu_{1})\alpha'}$.  Let us now move to the case $2 < \nu_{0} + \nu_{1} + \nu_{2} < 3$. For this, we must have $\nu_{0} + \nu_{1} > \nu_{2}$.  Then (\ref{sb}) gives, for large t, $\sim - e^{\pi (\nu_{0} + \nu_{1} + \nu_{2}) t}$, a negative infinity,  and the integrand in (\ref{taa56}) is still positive and behaves as
\be
  - \frac{e^{- \frac{y^2 t}{2\pi\alpha'}} e^{\pi(\nu_{0} + \nu_{1} + \nu_{2}) t}}{ \sinh\pi\nu_{0} t \sinh\pi \nu_{1} \sinh\pi\nu_{2}t} Z_{1} \stackrel{t \to \infty}{\longrightarrow} e^{- \frac{y^2 t}{2\pi\alpha'}} e^{\pi(\nu_{0} + \nu_{1} + \nu_{2} - 2) t}, 
  \ee
  where we have used $Z_{1} \sim - e^{\pi (\nu_{0} + \nu_{1} + \nu_{2} - 2) t}$  and $Z_{n} \approx 1$ for $n \ge 2$ as given earlier. We have here the tachyonic shift as $(\nu_{0} + \nu_{1} + \nu_{2} - 2)/2$ and the effective mass for the open string is
\be\label{msquare-new}
m^{2} =  \frac{y^{2}}{(2\pi \alpha')^{2}} - \frac{\nu_{0} + \nu_{1} + \nu_{2} - 2}{2\alpha'},
\ee
which becomes tachyonic when $y < \pi \sqrt{2 (\nu_{0} + \nu_{1} + \nu_{2} - 2)\alpha'}$. Whenever this happens,  the integrand blows up for $t \to \infty$ and this reflects the onset of tachyonic instability. Then we will have a phase transition via the so-called tachyon condensation.  Once again, this is consistent with our expectation.  So this confirms our earlier assertion that there is indeed a correlation between the attractive nature of interaction and the existence of a tachyonic instability.   

The tachyonic shift and the appearance of tachyon mode can also be understood from the spectrum of the open string connecting the two D6 carrying magnetic fluxes which give rise to the $\nu_{0}, \nu_{1}$ and $\nu_{2}$ following \cite{Ferrara:1993sq, Bolognesi:2012gr}.  Let us use an explicit example for $p = p' = 6$ to demonstrate this. For this purpose, we choose the following magnetic flux for $\hat F$
\be\label{mf6}
\hat F = \left(\begin{array}{ccccccc}
0& &&&&&\\
&0 & - \hat g_{0}&&&&\\
&\hat g_{0} &0&&&&\\
&&&0& - \hat g_{1}&&\\
&&&\hat g_{1}&0&&\\
&&&&&0&-\hat g_{2}\\
&&&&&\hat g_{2}&0 
\end{array}\right),
\ee
and for $\hat F'$ we just replace each $g$'s in $\hat F$ with a prime on it. Following the prescription given earlier, we have 
\be\label{mf6eigenv}
\lambda_{\alpha} + \lambda^{-1}_{\alpha} = 2 \frac{(1 - \hat g_{\alpha}^{2})(1 - \hat g'^{2}_{\alpha}) + 4 \hat g_{\alpha} \hat g'_{\alpha}}{(1 + \hat g^{2}_{\alpha})(1 + \hat g'^{2}_{\alpha})},
\ee
where  $\alpha = 0, 1, 2$ and which gives, noting $\lambda_{\alpha} = e^{2\pi i \nu_{\alpha}}$ with $\nu_{\alpha} \in [0, 1)$,
\be\label{mf6nu}
\tan\pi \nu_{\alpha} =\frac{ \left|\hat g_{\alpha} - \hat g'_{\alpha}\right|}{1 + \hat g_{\alpha} \hat g'_{\alpha}}.
\ee
We have also now the amplitude  (\ref{taa56}) with 
\be
\left[\det (\eta + \hat F') \det( \eta + \hat F')\right]^{1/2} = \prod_{\alpha =0}^{2} \left[(1 + \hat g^{2}_{\alpha})(1 + \hat g'^{2}_{\alpha} )\right]^{1/2}.
\ee   
 Type I superstring in a single magnetic background, say, the magnetic field being in 56-directions, has been discussed in \cite{Ferrara:1993sq}. Here we have three magnetic fields, the first in 12-directions, the second in 34-directions and the third in 
 56-directions. The generalization of the discussion given there  to the present case in the R-sector is straightforward and the conclusion remains the same even if we exclude the contribution of $y^{2}/(2\pi\alpha')^{2}$ to the energy square. In other words, unlike the case in the NS-secor which we will turn next, there is no  possibility for the existence of tachyonic shift in the R-sector. The GSO-projected R-sector ground states (the eight fermions $8_{\rm F}$)  have  masses no less than $y/(2\pi \alpha')$.  In what follows, we focus here only on the generalization to the present NS-sector. As before, without loss of generality, we assume
 once again $\nu_{0} \le \nu_{1} \le \nu_{2}$.  The energy spectrum is now
 \be\label{energy-ns}
 \alpha' E^{2}_{\rm NS} =  \frac{y^{2}}{(2\pi)^{2}\alpha'} + \sum_{\alpha = 0}^{2}\left[(2 N_{\alpha} + 1) \frac{\nu_{\alpha}}{2} - \nu_{\alpha} S_{\alpha} \right] + L^{\rm free}_{\rm NS},  
\ee
where 
\bea\label{ns-sector}
N_{\alpha} &=& b^{+}_{\alpha, 0} b_{\alpha, 0}, \quad S_{\alpha} = \sum_{n = 1}^{\infty} \left(a^{+}_{\alpha, n} a_{\alpha, n} - b^{+}_{\alpha, n} b_{\alpha, n}\right) + \sum_{r = 1/2}^{\infty} \left(d^{+}_{\alpha, r} d_{\alpha, r} - \tilde d^{+}_{\alpha, r} \tilde d_{\alpha, r}\right),\nn
L^{\rm free}_{\rm NS} &=& \sum_{\alpha =0}^{2}\left[\sum_{n = 1}^{\infty} n (a^{+}_{\alpha, n} a_{\alpha, n} + b^{+}_{\alpha, n} b_{\alpha, n}) + \sum_{r = 1/2}^{\infty} r (d^{+}_{\alpha, r} d_{\alpha, r} + \tilde d^{+}_{\alpha, r} \tilde d_{\alpha, r})\right] - \frac{1}{2} + L^{\perp{\rm free} }_{\rm NS}.\quad
\eea
In the above,  $N_{\alpha}$ defines the corresponding Landau-level for $\alpha = 0, 1, 2$, respectively, $S_{\alpha}$ is the spin operator and $L^{\perp{\rm free}}_{\rm NS}$ is the part contributing to the zero-mode Virasoro generator from the 0, 7, 8, 9-directions. One can check when $\nu_{0} + \nu_{1} < \nu_{2}$ and $\nu_{0} + \nu_{1} + \nu_{2} < 2$, 
the possible lowest energy state is from the GSO-projected ground state\footnote{\label{fn8} Both the state $d^{+}_{0, 1/2}  |0\rangle_{\rm NS}$ and $d^{+}_{1, 1/2}  |0\rangle_{\rm NS}$ have their respective energy higher than that of  $d^{+}_{2, 1/2} |0\rangle_{\rm NS}$.} $d^{+}_{2, 1/2} |0\rangle_{\rm NS}$ and for this we have
\bea\label{grounds}
\alpha' E^{2}_{\rm NS} &=& \frac{y^{2}}{(2\pi)^{2} \alpha'} - \frac{\nu_{2} -  \nu_{0} - \nu_{1}}{2},\nn
 S_{2} &=& 1,  \quad S_{0} = S_{1} = 0, \quad N_{0} = N_{1} = N_{2} = 0, \quad  L^{\rm free}_{\rm NS} = 0.
\eea
Here the first equation is exactly the same as (\ref{msquare}).  In other words, when $\nu_{2} > \nu_{0} + \nu_{1}$ and $\nu_{0} + \nu_{1} + \nu_{2 } < 2$  we have a tachyonic shift and this gives a potential tachyonic instability. Otherwise we don't.  So the conclusion remains the same as before and we will not repeat it here. It is nice to see the same from a different approach. From either (\ref{msquare}) or (\ref{grounds}),  we see that the tachyonic shift would be just $\nu_{2} /2$, rather than the smaller one 
$(\nu_{2} - \nu_{0} - \nu_{1})/2$, in the absence of the other two fluxes.  In other words, in order to have the largest tachyonic shift, we need to choose to apply the largest magnetic one but no more. This largest tachyonic shift is also responsible for the largest open string pair production enhancement discussed in \cite{Lu:2018nsc, Jia:2018mlr}. We will also address this later when we discuss the open string pair production in the presence of electric fluxes.  

Similarly, when $ 2 < \nu_{0} + \nu_{1} + \nu_{2} < 3$ and $\nu_{0} + \nu_{1} > \nu_{2}$,  the lowest energy state is the GSO-projected one $ d^{+}_{0, 1/2} d^{+}_{1, 1/2} d^{+}_{2, 1/2}|0\rangle_{\rm NS}$ which gives $S_{0} = S_{1} = S_{2} = 1$ 
and $L^{\rm free}_{\rm NS} = 1$ and for this we have
\bea\label{grounds-new}
\alpha' E^{2}_{\rm NS} &=& \frac{y^{2}}{(2\pi)^{2} \alpha'} - \frac{\nu_{0} +  \nu_{1} +  \nu_{2} - 2}{2},\nn
S_{0} &=& S_{1} = S_{2} = 1, \quad N_{0} = N_{1} = N_{2} = 0,  \quad L^{\rm free}_{\rm NS} = 1.
\eea
The lowest energy is nothing but the effective mass given in (\ref{msquare-new}) with the same tachyonic shift $(\nu_{0} + \nu_{1} + \nu_{2} - 2)/2$. So the tachyonic instability can be discussed exactly the same and will not be repeated here. 

We now move to the case when one of three $\nu$'s is imaginary, say, $\nu_{0} = i \bar \nu_{0}$, and both $\nu_{1}$ and $\nu_{2}$ are real.  So this applies to both $p = p' = 5$ and $p = p' = 6$. Now the open string annulus amplitude (\ref{taa56}) becomes
 \bea\label{taa56-e}
\Gamma_{p, p} &=&  \frac{ 2^2 \, V_{
p + 1} \left[\det (\eta + \hat F')\det(\eta + \hat F)\right]^{\frac{1}{2}} }{(8 \pi^2 \alpha')^{\frac{p +1 }{2}}}   \int_0^\infty \frac{d t}{t^{\frac{p - 3}{2}}}  e^{- \frac{y^2 t}{2\pi\alpha'}} \frac{\sinh \pi \bar\nu_{0} \sin\pi\nu_{1} \sin\pi\nu_{2}}{\sin \pi\bar \nu_{0} t
\sinh\pi\nu_{1}t \sinh\pi\nu_{2}t} \nn
 &\,& \times \left[\left(\cosh\pi\nu_{1}t - \cosh\pi\nu_{2}t \right)^{2} + 4 \sin^{2} \frac{\pi\bar\nu_{0}t}{2} \left(\cosh\pi\nu_{1} t \cosh\pi\nu_{2} t - \cos^{2} \frac{\pi\bar\nu_{0} t}{2} \right) \right] \nn
 &\,& \times \prod_{n=1}^{\infty} Z_{n},
  \eea 
where $Z_{n}$, from (\ref{Z}), becomes
 \be\label{Z-e}
 Z_{n} = \frac{\tilde Z_{n} }{(1 - |z|^{2n})^{2} 
 \left[1 - 2 |z|^{2n} \cos 2\pi\bar \nu_{0} t  + |z|^{4n}\right]\prod_{\alpha =1}^{2} \left[1 - 2 |z|^{2n} \cosh 2\pi\nu_{\alpha} t  + |z|^{4n}\right] },
 \ee
with $\tilde Z_{n}$, from (\ref{tildeZ2}), as
\bea\label{tildeZ-e}
\tilde Z_{n} &=&  \left| \left[1 - 2 |z|^{2n} e^{-i \pi\bar\nu_{0}t} \cosh\pi(\nu_{1} + \nu_{2})t + e^{-2i \pi  \bar\nu_{0} t} |z|^{4n}\right]\right|^{2}\nn
 &\,&\times \left|\left[1 - 2 |z|^{2n} e^{- i\pi\bar\nu_{0} t} \cosh\pi(\nu_{1} - \nu_{2})t + e^{- 2i\pi  \bar\nu_{0} t} |z|^{4n}\right]\right|^{2} > 0.
\eea
Note that in the denominator of $Z_{n}$, the factor $1 - 2 |z|^{2n} \cos\pi\bar \nu_{0} t + |z|^{4n} > (1 - |z|^{2n})^{2} > 0$ and for $\alpha = 1, 2$, $1 - 2 |z|^{2n} \cosh\pi\nu_{\alpha} t + |z|^{4n} = (1 - |z|^{2n} e^{2\pi\nu_{\alpha} t})
 (1 - |z|^{2n} e^{- 2\pi \nu_{\alpha} t}) > 0$ due to $n \ge 1$ and $\nu_{\alpha} \in [0, 1)$. So we have $Z_{n} > 0$. Note also that every other factor in the integrand, except for the $\sin\pi\bar \nu_{0} t$ in the denominator, is also positive. 
  The interesting physics shows up precisely due to this factor $\sin\pi\bar\nu_{0} t$. It gives an infinite number of simple poles of the integrand at $t_{k} = k/\bar\nu_{0}$ with $k = 1, 2, \cdots$ along the positive t-axis.  This implies that the interaction amplitude  has an imaginary part, indicating the decay of the underlying system via the so-called open string pair production. By saying this, we first need to note that the integral has no singularity when we take  $t \to 0$. Secondly, we need to have $y > \pi \sqrt{2 |\nu_{1} - \nu_{2}|\alpha'}$ to avoid a potential tachyonic instability and to validate our amplitude computations since otherwise the integrand blows up for large $t$ as
\be\label{taa56-e-lt}
\sim e^{- \frac{y^2 t}{2\pi\alpha'}} e^{\pi |\nu_{1} - \nu_{2}| t} = e^{- 2 \pi t\left[\frac{y^{2}}{(2\pi)^{2}\alpha'} - \frac{|\nu_{1} - \nu_{2}|}{2}\right]},
\ee
and as such a phase transition, called tachyon condensation, occurs.  The decay rate of the underlying system per unit volume of Dp brane via the open string pair production can be computed, following \cite{Bachas:1992bh}, as the sum of the residues of the simple poles of the integrand in (\ref{taa56-e}) times $\pi$ and is given as
\bea\label{decay-rate}
{\cal W}_{p, p} &=& - \frac{2 \,{\rm Im} \Gamma}{V_{p + 1}}\nn
&=& \frac{ 2^3 \, \left[\det (\eta + \hat F')\det(\eta + \hat F) \right]^{\frac{1}{2}} \sinh \pi \bar\nu_{0} \sin\pi\nu_{1} \sin\pi\nu_{2}}{\bar \nu_{0} (8 \pi^2 \alpha')^{\frac{p + 1}{2}}} \sum_{k = 1}^{\infty} (-)^{k + 1} \left(\frac{\bar\nu_{0}}{k}\right)^{\frac{p - 3}{2}} e^{- \frac{k y^2 }{2\pi\bar\nu_{0} \alpha'}}\nn
&\,&\times \frac{\left(\cosh \frac{k\pi \nu_{1}}{\bar\nu_{0}} - (-)^{k} \cosh \frac{k \pi \nu_{2}}{\bar\nu_{0}} \right)^{2}}{\sinh\frac{k \pi\nu_{1}}{\bar\nu_{0}} \sinh\frac{k\pi\nu_{2}}{\bar\nu_{0}}}  Z_{k} (\bar\nu_{0}, \nu_{1},\nu_{2}),
\eea
where 
\be\label{Znk}
Z_{k} = \prod_{n = 1}^{\infty}\frac{[1 - 2 (-)^{k} |z_{k}|^{2n} \cosh\frac{k \pi(\nu_{1} + \nu_{2})}{\bar\nu_{0}} + |z_{k}|^{4n}]^{2}
[1 - 2 (-)^{k}|z_{k}|^{2n} \cosh\frac{k\pi(\nu_{1} - \nu_{2})}{\bar\nu_{0}} + |z_{k}|^{4n}]^{2} }{(1 - |z_{k}|^{2n})^{4} 
[1 - 2 |z_{k}|^{2n} \cosh \frac{2k\pi\nu_{1}}{\bar\nu_{0}}  + |z_{k}|^{4n}] [1 - 2 |z_{k}|^{2n} \cosh \frac{2k\pi\nu_{2}}{\bar\nu_{0}}  + |z_{k}|^{4n}]},
 \ee 
with $|z_{k}| = e^{-k\pi/\bar\nu_{0}}$.  Note that when $\bar\nu_{0} \to \infty$,  we have $Z_{k} \to \infty$ for $k =$ odd while $Z_{k} \to 1$ for $k = $ even due to $|z_{k}| \to 1$. For the rate, the odd-k terms, each is blowing up and positive,  are dominant over the almost vanishing negative even-terms, and so the rate blows up.  This gives another singularity. As we will see, this is due to the electric field reaching its critical value. The open strings break under the action of the critical field and their production cascades. 
 
According to  \cite{nikishov},  the rate (\ref{decay-rate}) should be more properly interpreted as the decay one of the underlying system rather than the pair production one.  The pair production rate is just the leading  $k = 1$ term in the above rate as
\bea\label{pp-rate}
{\cal W}^{(1)}_{p, p} &=& \frac{ 2^3 \, \left[\det (\eta + \hat F')\det(\eta + \hat F) \right]^{\frac{1}{2}} \sinh \pi \bar\nu_{0} \sin\pi\nu_{1} \sin\pi\nu_{2}}{\bar \nu_{0} (8 \pi^2 \alpha')^{\frac{p + 1}{2}}}  \bar\nu_{0}^{\frac{p - 3}{2}} e^{- \frac{ y^2 }{2\pi\bar\nu_{0} \alpha'}}\nn
&\,&\times \frac{\left(\cosh \frac{\pi \nu_{1}}{\bar\nu_{0}} + \cosh \frac{ \pi \nu_{2}}{\bar\nu_{0}} \right)^{2}}{\sinh\frac{ \pi\nu_{1}}{\bar\nu_{0}} \sinh\frac{\pi\nu_{2}}{\bar\nu_{0}}}  Z_{1} (\bar\nu_{0}, \nu_{1}, \nu_{2}).
 \eea
 So the pair production simply cascades when $\bar\nu_{0} \to \infty$. Since ${\cal W}^{(1)}$ is symmetric to $\nu_{1}$ and $\nu_{2}$, without loss of generality and for convenience, we assume $\nu_{1} \ge \nu_{2}$ for the following discussion.  Given $\nu_{1} \ge \nu_{2} \in [0, 1)$, one can check that $Z_{1}$, from (\ref{Znk}), increases as $\bar\nu_{0}$ increases. When $\bar\nu_{0} \gg \nu_{1}$,  the factor $\left(\cosh \frac{\pi \nu_{1}}{\bar\nu_{0}} + \cosh \frac{ \pi \nu_{2}}{\bar\nu_{0}} \right)^{2}/ \sinh\frac{ \pi\nu_{1}}{\bar\nu_{0}} \sinh\frac{\pi\nu_{2}}{\bar\nu_{0}} \sim 4 \bar\nu^{2}_{0}/(\pi^{2}\nu_{1}\nu_{2})$ increases also when we increase $\bar\nu_{0}$. Since $p \ge 5$,   all other factors have an overall increase  when we increase $\bar\nu_{0}$. This holds true at least for the most interesting cases with a large enhancement of the rate and  is also  expected since $\bar\nu_{0}$ is related to the applied electric fluxes and increases when any of them increases, which will be explicitly demonstrated in an example given later.  When $\bar\nu_{0} \sim \nu_{1} \ge \nu_{2}$, this same factor will not play important role for the rate. The rate is now controlled by the other factors and still increases with the $\bar \nu_{0}$.  If $\bar\nu_{0} \ll \nu_{2} \le \nu_{1}$,  this implies $\bar\nu_{0} \ll 1$. So $Z_{1} (\bar\nu_{0}, \nu_{1}, \nu_{2}) \approx 1$.  The pair production rate is
 \be\label{smallnu0}
 {\cal W}^{(1)}_{p, p} \approx \frac{ 8\pi \, \left[\det (\eta + \hat F')\det(\eta + \hat F) \right]^{\frac{1}{2}} \sin\pi\nu_{1} \sin\pi\nu_{2}}{(8 \pi^2 \alpha')^{\frac{p + 1}{2}}}  \bar\nu_{0}^{\frac{p - 3}{2}} e^{- \frac{ y^2 }{2\pi\bar\nu_{0} \alpha'}} e^{\frac{\pi (\nu_{1} - \nu_{2})}{\bar\nu_{0}}},
 \ee
where the factor $e^{\pi(\nu_{1} - \nu_{2})/\bar\nu_{0}} \gg 1$, a large enhancement of the rate in the presence of magnetic fluxes. If $y > \pi \sqrt{2 (\nu_{1} - \nu_{2})\alpha'}$ (for avoiding the  tachyonic instability), the rate still increases when we increase $\bar\nu_{0}$.  For the purpose of illustration, we consider the $p = p' = 5$ case and take the following simple flux for $\hat F$ as
\be\label{F56}
\hat F = \left(\begin{array}{ccccccc}
0& - \hat f&&&&&\\
\hat f &0 & &&&&\\
& &&0& - \hat g_{1}&&\\
&&&\hat g_{1}& 0&&\\
&&&&&0&- \hat g_{2}\\
&&&&&g_{2}&0\\ 
\end{array}\right),
\ee 
where $\hat f$ stands for the electric flux along $01$-directions while $\hat g_{1}, \hat g_{2}$ are the magnetic ones along 23- and 45-directions, respectively.  Similarly for $\hat F'$ but denoting every quantity with a prime.  We can then determine the eigenvalues as
\be\label{eigen5}
\lambda_{0} + \lambda^{-1}_{0} = 2 \frac{(1 + \hat f^{2})(1 + \hat f'^{2}) - 4 \hat f \hat f'}{(1 - \hat f^{2})(1 - \hat f'^{2})}, \quad \lambda_{a} + \lambda^{-1}_{a} 
= 2\frac{(1 - \hat g_{a}^{2})(1 - \hat g'^{2}_{a}) + 4 \hat g_{a} \hat g'_{a}}{(1 + \hat g^{2}_{a})(1 + \hat g'^{2}_{a})},
\ee
where $a = 1, 2$. By setting $\lambda_{\alpha} = e^{2i\pi \nu_{\alpha}}$ with $\alpha = (0, a)$, we have
\be\label{nu5}
\tanh\pi \bar\nu_{0} = \frac{|\hat f - \hat f'|}{1 - \hat f \hat f'}, \quad \tan\pi\nu_{a} =\frac{ \left|\hat g_{a} - \hat g'_{a}\right|}{1 + \hat g_{a}\hat g'_{a}},
\ee
where we have set $\nu_{0} = i \bar\nu_{0}$. Note that $|\hat f|, |\hat f'| < 1$ while $|\hat g_{a}|, |\hat g'_{a}| < \infty$. As explained earlier, we always have
$\bar\nu_{0} \in (0, \infty)$ and $\nu_{a} \in [0, 1)$ for the amplitude and the rate. It is clear that $\bar\nu_{0}$ increases when we increase  $\hat f$ or $\hat f'$ as mentioned earlier.  Now the factor $\det (\eta + \hat F')\det(\eta + \hat F)  = (1 - \hat f^{2})(1 - \hat f'^{2})\prod_{a =1}^{2}(1 + \hat g_{a}^{2})(1 + \hat g'^{2}_{a})$. The open string pair production rate can now be expressed as
\be\label{pprate56}
{\cal W}^{(1)}_{p, p}  =\frac{ 2^3 \, |\hat f - \hat f'| |\hat g_{1} - \hat g'_{1}| |\hat g_{2} - \hat g'_{2}| }{\bar \nu_{0} (8 \pi^2 \alpha')^{\frac{p + 1}{2}}}  \bar\nu_{0}^{\frac{p - 3}{2}} e^{- \frac{ y^2 }{2\pi\bar\nu_{0} \alpha'}}
 \frac{\left(\cosh \frac{\pi \nu_{1}}{\bar\nu_{0}} + \cosh \frac{ \pi \nu_{2}}{\bar\nu_{0}} \right)^{2}}{\sinh\frac{ \pi\nu_{1}}{\bar\nu_{0}} \sinh\frac{\pi\nu_{2}}{\bar\nu_{0}}}  Z_{1} (\bar\nu_{0}, \nu_{1}, \nu_{2}),
\ee
where $\bar\nu_{0}$ and $\nu_{a}$ with $a = 1, 2$ are given in (\ref{nu5}). This rate also applies to the  $p = p' = 6$ case. The earlier general discussion on how the pair production rate depends on the applied electric fluxes or $\bar \nu_{0}$ for fixed
$\nu_{a}$ with $a = 1, 2$ continues to hold and we will not  repeat it here. We focus now on how the rate depends on the applied magnetic fluxes or $\nu_{a}$ for fixed non-vanishing $\bar\nu_{0}$.  The rate for vanishing magnetic fluxes can be obtained from (\ref{pprate56}) as
\be\label{pprate56g=0}
{\cal W}^{(1)}_{p, p} (\hat g_{a} = \hat g'_{a} = 0) =\frac{ 2^5 \, |\hat f - \hat f'|  }{(8 \pi^2 \alpha')^{\frac{p + 1}{2}}}  \bar\nu_{0}^{\frac{p - 1}{2}} 
e^{- \frac{ y^2 }{2\pi\bar\nu_{0} \alpha'}}  Z_{1} (\bar\nu_{0}, \nu_{a} = 0).
 \ee
We have then 
\be\label{rate56-ratio}
\frac{{\cal W}^{(1)}_{p, p} (\hat g_{a}, \hat g'_{a} \neq 0)}{{\cal W}^{(1)}_{p, p} (\hat g_{a} = \hat g'_{a} = 0)}= \frac{ |\hat g_{1} - \hat g'_{1}| |\hat g_{2} - \hat g'_{2}|}{ 4 \bar\nu_{0}^{2}} \frac{\left(\cosh \frac{\pi \nu_{1}}{\bar\nu_{0}} + \cosh \frac{ \pi \nu_{2}}{\bar\nu_{0}} \right)^{2}}{\sinh\frac{ \pi\nu_{1}}{\bar\nu_{0}} \sinh\frac{\pi\nu_{2}} {\bar\nu_{0}}}  \frac{Z_{1} (\bar\nu_{0}, \nu_{a} \neq 0)}{Z_{1} (\bar\nu_{0}, \nu_{a} = 0)}.
\ee 
For non-vanishing $\nu_{a} \in (0, 1)$ and $|\hat g_{a} - \hat g'_{a}| \sim {\cal O} (1)$,  if  $\bar\nu_{0}$ is not too small, the presence of magnetic fluxes will not give 
a significant enhancement of the rate as can be seen from the above. However, if instead $\nu_{a}/\bar\nu_{0} \gg 1$ and $|\hat g_{a} - \hat g'_{a}| \ge 1$ or $\nu_{a}/\bar\nu_{0} \ll 1$ but all $\hat g_{a}, \hat g'_{a}$ are large, with $a = 1, 2$, the rate has a significant enhancement.  Let us consider the latter case for which all $\hat g_{a}$ and $\hat g'_{a}$ are large. Now $\nu_{a}$ are small. So from (\ref{nu5}), we have $\hat g_{a} \hat g'_{a} \pi \nu_{a} = |\hat g_{a} - \hat g'_{a}|$. The ratio of (\ref{rate56-ratio}) becomes 
\be  
\frac{{\cal W}^{(1)}_{p, p} (\hat g_{a}, \hat g'_{a} \neq 0)}{{\cal W}^{(1)}_{p, p} (\hat g_{a} = \hat g'_{a} = 0)} \approx \hat g_{1} \hat g'_{1} \hat g_{2} \hat g'_{2} \gg 1,
\ee
much enhanced.  In the above, we have used $Z_{1} (\bar\nu_{0}, \nu_{a} \ll 1) \approx Z_{1} (\bar\nu_{0}, \nu_{a} = 0)$.

Unless we consider relevant physics in string scale, the fluxes $\hat f, \hat f', \hat g_{a}$ and $\hat g'_{a}$ are in general small in terms of string scale. In other words, 
$|\hat f| \ll 1, |\hat f'| \ll 1, |\hat g_{a}| \ll 1$ and $|\hat g'_{a}| \ll 1$. We then have $\pi \bar\nu_{0} = |\hat f - \hat f'| \ll 1, \pi \nu_{a} = |\hat g_{a} - \hat g'_{a}|\ll 1$. The 
rate (\ref{pprate56}) becomes 
\be\label{weak-rate56}
{\cal W}^{(1)}_{p, p}  =\frac{ 8 \,\pi^{3} \nu_{1}\nu_{2} }{ (8 \pi^2 \alpha')^{\frac{p + 1}{2}}}  \bar\nu_{0}^{\frac{p - 3}{2}} e^{- \frac{ y^2 }{2\pi\bar\nu_{0} \alpha'}}
 \frac{\left(\cosh \frac{\pi \nu_{1}}{\bar\nu_{0}} + \cosh \frac{ \pi \nu_{2}}{\bar\nu_{0}} \right)^{2}}{\sinh\frac{ \pi\nu_{1}}{\bar\nu_{0}} \sinh\frac{\pi\nu_{2}}{\bar\nu_{0}}},
\ee
where $Z_{1} (\bar\nu_{0}, \nu_{1}, \nu_{2}) \approx 1$. In the weak field limit, we showed in \cite{Jia:2018mlr} that adding magnetic fluxes $\hat g_{2}$ and $\hat g'_{2}$, assuming $\nu_{2} \le \nu_{1}$, in general diminishes rather than enhances the rate.  This can also be understood via the tachyonic shift discussed earlier. So for the purpose of enhancing the rate via adding magnetic fluxes, we merely need to add only the possible largest fluxes $\hat g_{1}$ and $\hat g'_{1}$.  In other words,  for given $\bar\nu_{0} \ll 1$ and the largest possible $\nu_{1}$, the corresponding largest possible rate is 
\be\label{largest-rate}
{\cal W}^{(1)}_{p,p}  =\frac{ 8 \,\pi^{2}\bar\nu_{0} \nu_{1} }{ (8 \pi^2 \alpha')^{\frac{p + 1}{2}}} \bar\nu_{0}^{\frac{p - 3}{2}} e^{- \frac{ y^2 }{2\pi\bar\nu_{0} \alpha'}}
 \frac{\left(\cosh \frac{\pi \nu_{1}}{\bar\nu_{0}} + 1 \right)^{2}}{\sinh\frac{ \pi\nu_{1}}{\bar\nu_{0}} }.
\ee
This rate formula is actually valid for $p \ge 3$\cite{Lu:2018suj}. For given $\bar\nu_{0} \ll 1$ and $\nu_{1} \ll 1$, it is clear that the smallest $p = 3$ case gives the largest rate \cite{Jia:2018mlr, Lu:2018nsc}.  The enhancement due to the added magnetic fluxes is
\be\label{p3enhancement}
\frac{{\cal W}^{(1)}_{p, p} (\bar\nu_{0}, \nu_{1} \neq 0)}{{\cal W}^{(1)}_{p, p} (\bar\nu_{0}, \nu_{1} = 0)} = \frac{\pi \nu_{1}}{\bar\nu_{0}} \frac{\left[1 + \cosh\frac{\pi\nu_{1}}{\bar\nu_{0}}\right]^{2}}{4 \sinh\frac{\pi\nu_{1}}{\bar\nu_{0}}},
\ee
which is always greater than unity for $\pi\nu_{1}/\bar\nu_{0} > 0$.   One can check this numerically.  In particular, when $\nu_{1}/\bar\nu_{0} \gg 1$, this ratio becomes 
\be 
\frac{{\cal W}^{(1)}_{p, p} (\bar\nu_{0}, \nu_{1} \neq 0)}{{\cal W}^{(1)}_{p, p} (\bar\nu_{0}, \nu_{1} = 0)} = \frac{\pi\nu_{1}}{8 \bar\nu_{0}} e^{\frac{\pi\nu_{1}}{\bar\nu_{0}}} \gg 1.
\ee 

\subsection{The $p = p' < 5$ cases \label{subs3.2}}

 Given the discussion for $p = p' = 5, 6$ cases in the previous subsection, the relevant discussion for $p = p' < 5$ is straightforward. We will spell out this in detail in this subsection.  To be concrete, let us explain the rationale behind the integral representation structure of the cylinder amplitude (\ref{t-amplitude-cylinder}) for $p = p' < 5$.  
  
     
 In using the closed string boundary state representation of D-brane to compute the cylinder interaction amplitude between two parallel placed D-branes of the same or different dimensionality at a separation,  we note that the worldvolume dimensionality of the respective D-brane  is encoded in its $M$-matrix (\ref{mmatrix}), the bosonic zero-mode (\ref{bzm}) in the bosonic sector and the fermionic zero-mode (\ref{rrzm}) in the R-R sector in the matter part.  The rest are independent of this dimensionality. Let us first examine carefully the $M$-matrix 
(\ref{mmatrix}) which we rewrite here for convenience, 
\be\label{mmatrix-new}
M = ([(\eta -
\hat{F})(\eta + \hat{F})^{-1}]_{\alpha\beta},  -
\delta_{ij}),
\ee     
where $\alpha, \beta$ are along the brane directions while $i, j$ are along the directions transverse to the brane. For example, let us first consider the D6 brane. In other words, $\alpha, \beta = 0, 1, \cdots 6$ and $i, j = 7, 8, 9$. For any other Dp with even $p < 6$, we denote their $\alpha', \beta' = 0, 1, \cdots p$ along its brane directions and $i', j' = p + 1, \cdots
9 - p$ as directions transverse to this brane. Its corresponding $M_{p}$-matrix with a general worldvolume flux $(\hat F_{p})_{\alpha'\beta'}$ can be taken as a special case of the D6 brane, namely  $M_{6}$, in the following sense.  For the D$p$ with even $p < 6$, we have
\be\label{mmatrixp}
  M_{p} = ([(\eta_{p} -
\hat{F}_{p})(\eta_{p} + \hat{F}_{p})^{-1}]_{\alpha'\beta'},  -
\delta_{i'j'}).
\ee 
We now extend $\alpha', \beta' =0,  1, \cdots, p$ to $\alpha, \beta = 0, 1, \cdots 6$ and $\hat F_{p}$ to $\hat F_{6}$ taking the following special form,
\be\label{flux6}
 (\hat F_{6})_{\alpha\beta} = \left(\begin{array}{cccccc}
 (\hat F_{p})_{\alpha'\beta'} &&&&&\\
 &0& \hat g_{1}&&&\\
 &- \hat g_{1}&0&&&\\
  &&&\ddots&&\\
  &&&&0&  \hat g_{\frac{6 - p}{2}}\\
  &&&&- \hat g_{\frac{6 - p}{2}}&0\end{array}\right)_{7\times 7}.
  \ee  
 With this special choice, the $M_{6}$ turns out to give just $M_{p}$ when we  take the special magnetic fluxes $\hat g_{k} \to \infty$ with $ k = 1, \cdots, (6 - p)/2$. Let us see this in detail. With the special flux (\ref{flux6}), we have
 \be
 (M_{6})_{\alpha\beta} = \left(\begin{array}{cccccc}
 (M_{p})_{\alpha'\beta'} &&&&&\\
 &\frac{1 - \hat g^{2}_{1}}{1 + \hat g^{2}_{1}}& \frac{2\, \hat g_{1}}{1 + \hat g^{2}_{1}}&&&\\
 &\frac{2\,\hat g_{1}}{1 + \hat g^{2}_{1}}&\frac{1 - \hat g^{2}_{1}}{1 + \hat g^{2}_{1}}&&&\\
 &&&\ddots&&\\
 &&&&\frac{1 - \hat g^{2}_{\frac{6 - p}{2}}}{1 + \hat g^{2}_{\frac{6 - p}{2}}}&\frac{2\, \hat g_{\frac{6 - p}{2}}}{1 + \hat g^{2}_{\frac{6 - p}{2}}}\\
 &&&&\frac{2\, \hat g_{\frac{6 - p}{2}}}{1 + \hat g^{2}_{\frac{6 - p}{2}}}&\frac{1 - \hat g^{2}_{\frac{6 - p}{2}}}{1 + \hat g^{2}_{\frac{6 - p}{2}}}\\
 \end{array}\right),
 \ee 
 which becomes $(M_{6})_{\alpha\beta} = ((M_{p})_{\alpha'\beta'}, - \delta_{k'l'})$ with $k', l' = p + 1, \cdots 6$ when we take $g_{k} \to \infty$ with $k = 1, \cdots (6 - p)/2$.   So we have $M_{6} = ((M_{p})_{\alpha'\beta'}, - \delta_{i'j'}) = M_{p}$ for the above special choice of the flux $\hat F_{6}$ (\ref{flux6})  when we take $\hat g_{k} \to \infty$ with $k = 1, \cdots (6 - p)/2$.  In other words, $M_{p}$ is just a special case of $M_{6}$ when the worldvolume flux of D6 takes a special choice as indicated above.  This same discussion applies to the odd $p < 5$ from $p = 5$.  

This same applies to the R-R zero-mode contribution (\ref{0mme-nu}) to the amplitude. We discuss this in great detail in Appendix B and refer there for detail. These two considerations  explain the following part of the integrand in the amplitude (\ref{t-amplitude-cylinder}),
\be\label{6-to-p}
  \frac{ \theta_{1} \left(\left.\frac{\nu_{0} + \nu_{1} + \nu_{2}}{2}\right| it \right) \theta_{1} \left(\left.\frac{\nu_{0} - \nu_{1} + \nu_{2}}{2}\right| it \right)
 \theta_{1} \left(\left.\frac{\nu_{0} + \nu_{1} - \nu_{2}}{2}\right| it \right)\theta_{1} \left(\left.\frac{\nu_{0} - \nu_{1} - \nu_{2}}{2}\right| it \right)}{\theta_{1} (\nu_{0} | it)\theta_{1} (\nu_{1} | it)\theta_{1} (\nu_{2} | it) \eta^{3} (it)} \prod_{\alpha =0}^{2} \sin \pi \nu_{\alpha},
 \ee
which is valid in general for $p = p' = 5$ or $6$ but will reduce to the corresponding expected one for $p = p' < 5$ once the respective special flux such as (\ref{flux6}) is chosen and the corresponding limit is taken.
 However, the story for the bosonic zero-mode (\ref{bzm}) is different. Except for the overall factor $ [\det (\eta_{p} + \hat F_{p})]^{1/2}$,  the other part of the zero mode has nothing to do with the applied worldvolume flux and therefore this same trick as used for the $M$-matrix and the RR zero-mode does not apply here.  This zero-mode contribution to the amplitude gives essentially the other part of the integrand as
\be\label{bzm-contribution}
 \sqrt{\det(\eta_{p'} + \hat F'_{p'})\det(\eta_{p} + \hat F_{p})} \,V_{\rm NN} \left(2 \pi^{2} \alpha' t\right)^{- \frac{\rm DD}{2}} e^{- \frac{y^{2}}{2\pi\alpha' t}},
\ee 
where $V_{\rm NN} = V_{p' + 1}$ denotes the volume of the Dp$'$ worldvolume following the conventions given in footnote (\ref{fn1}) and DD denotes the DD-directions. Here DD
$= 9 - p$ with our conventions.  It is obvious that the $t^{- (9 -p)/2}$-factor in the amplitude (\ref{t-amplitude-cylinder}) for $p = p' < 5$ cannot be obtained from $p = p' = 5$ or $6$ even we choose the respective special fluxes and take the corresponding limits.

We therefore give an explanation to the cylinder amplitude (\ref{t-amplitude-cylinder}) for the case of $p = p' < 5$. Given the extended flux (\ref{flux6}) for $\hat F_{6}$ (or $\hat F_{5}$) and similarly for $\hat F'_{6}$ (or $\hat F'_{5}$) but with now $\hat g'_{k}$ with a prime,  we have now at least 
\be
\tan\pi \nu_{2} = \left\{\begin{array}{ll}
\frac{\left|\hat g_{\frac{6 - p}{2}} - \hat g'_{\frac{6 - p}{2}}\right|}{1 + \hat g_{\frac{6 -p}{2}} \hat g'_{\frac{6 - p}{2}}}&{\rm for\,\, even\,\,} p < 6,\\
&\\
 \frac{\left|\hat g_{\frac{5 - p}{2}} - \hat g'_{\frac{5 - p}{2}}\right|}{1 + \hat g_{\frac{5 - p}{2}} \hat g'_{\frac{5 - p}{2}}} & {\rm for \,\, odd\,\,}\, p < 5,
 \end{array}\right.
\ee
which gives $\nu_{2} \to 0$ when we take the limits $\hat g_{(6 - p)/2}  \to \infty$ and  $\hat g'_{(6 - p)/2} \to \infty$ (or $\hat g_{(5 - p)/2} \to 
\infty$ and  $\hat g'_{(5 - p)/2} \to \infty $). So with a vanishing $\nu_{2} = 0$,  we have $\theta_{1} (\nu_{2}| it) \to 2 \,\eta^{3} (it)\, \sin\pi \nu_{2}$ and the closed string tree-level cylinder amplitude (\ref{t-amplitude-cylinder}) now becomes

 \bea\label{t-amplitude-cylinder-p<5}
\Gamma_{p, p} &=&\frac{ 2^2 \, V_{p + 1}\, \left[\det (\eta_{p} + \hat F'_{p})\det(\eta_{p} + \hat F_{p})\right]^{\frac{1}{2}} }{(8 \pi^2 \alpha')^{\frac{p + 1}{2}}} \int_0^\infty \frac{d t} {t^{\frac{9 - p}{2}}} \frac{e^{- \frac{y^2}{2\pi\alpha' t}}}{\eta^{6} (it)}  \frac{ \theta^{2}_{1} \left(\left.\frac{\nu_{0} + \nu_{1}}{2}\right| it \right) \theta^{2}_{1} \left(\left.\frac{\nu_{0} - \nu_{1}}{2}\right| it \right)}{\theta_{1} (\nu_{0} | it)\theta_{1} (\nu_{1} | it)}\nn
&\,&\times \prod_{\alpha =0}^{1} \sin \pi \nu_{\alpha}\nn
 &=&  \frac{ 2^{2} \, V_{p+ 1} \left[\det (\eta_{p} + \hat F'_{p})\det(\eta_{p} + \hat F_{p})\right]^{\frac{1}{2}}  \left(\cos\pi\nu_{0} - \cos\pi\nu_{1}\right)^{2}}{(8 \pi^2 \alpha')^{\frac{p + 1}{2}}}   \int_0^\infty \frac{d t}{t^{\frac{9 - p}{2}}}  e^{- \frac{y^2}{2\pi\alpha' t}}  \prod_{n=1}^{\infty} C_{n},\nn
 \eea     
 where $C_{n}$ from (\ref{C}) becomes
  \be\label{C-p<5}
 C_{n} = \frac{\tilde C_{n} }{(1 - |z|^{2n})^{4} \prod_{\alpha = 0}^{1}
 \left[1 - 2 |z|^{2n} \cos2\pi \nu_{\alpha}  + |z|^{4n}\right]},
 \ee
 with $\tilde C_{n}$ from (\ref{tildeC1}) or (\ref{tildeC2}) as
 \bea
 \tilde C_{n} &=& \left[1 - 2 |z|^{2n} \cos\pi(\nu_{0} + \nu_{1}) + |z|^{4n}\right]^{2} \left[1 - 2 |z|^{2n} \cos\pi(\nu_{0} - \nu_{1}) + |z|^{4n}\right]^{2}  \label{tildeC1-p<5}\\
 &\stackrel{\rm or}{=}& \left[1 - 2 |z|^{2n} e^{i\pi\nu_{0}} \cos\pi\nu_{1} + e^{2\pi i \nu_{0}} |z|^{4n}\right]^{2} \left[1 - 2 |z|^{2n} e^{- i\pi\nu_{0}} \cos\pi\nu_{1} + e^{- 2\pi i \nu_{0}} |z|^{4n}\right]^{2}.\label{tildeC2-p<5}\nn
 \eea
The corresponding open string one-loop annulus amplitude, from (\ref{t-amplitude-annulus}), is now
\bea\label{t-amplitude-annulus-p<5}
\Gamma_{p, p} &=& - \frac{ 2^2  \, V_{p + 1}\, \left[\det (\eta_{p} + \hat F'_{p})\det(\eta_{p} + \hat F_{p})\right]^{\frac{1}{2}} \prod_{\alpha =0}^{1} \sin \pi \nu_{\alpha}}{(8 \pi^2 \alpha')^{\frac{p + 1}{2}}} \int_0^\infty \frac{d t} {t^{\frac{p - 1}{2}}} \frac{e^{- \frac{y^2 t}{2\pi\alpha' }}}{\eta^{6} (it)}  \nn
&\,& \times \frac{ \theta^{2}_{1} \left(\left. \frac{\nu_{0} + \nu_{1}}{2} i t \right| it \right) \theta^2_{1} \left(\left. \frac{\nu_{0} - \nu_{1}}{2} it \right| it \right)}{\theta_{1} (\nu_{0} it | it)\theta_{1} ( \nu_{1} it | it)}\nn
 &=&  \frac{ 2^2 \, V_{p+ 1} \left[\det (\eta_{p} + \hat F'_{p})\det(\eta_{p} + \hat F_{p})\right]^{\frac{1}{2}} \prod_{\alpha =0}^{1} \sin \pi \nu_{\alpha}}{(8 \pi^2 \alpha')^{\frac{p + 1}{2}}}   \int_0^\infty \frac{d t}{t^{\frac{p - 1}{2}}}  e^{- \frac{y^2 t}{2\pi\alpha'}} \nn
 &\,& \times \frac{\left(\cosh\pi\nu_{0} t - \cosh\pi\nu_{1} t \right)^{2}}{\sinh\pi\nu_{0} t \sinh\pi\nu_{1}t} \prod_{n=1}^{\infty} Z_{n},
 \eea 
 where $Z_{n}$, from (\ref{Z}), becomes  
\be\label{Z-p<5}
 Z_{n} = \frac{\tilde Z_{n} }{(1 - |z|^{2n})^{4} \prod_{\alpha = 0}^{1}
 \left[1 - 2 |z|^{2n} \cosh2\pi \nu_{\alpha} t  + |z|^{4n}\right]},
 \ee
 with $\tilde Z_{n}$, from (\ref{tildeZ1}) or (\ref{tildeZ2}), as
 \bea
 \tilde Z_{n} &=& \left[1 - 2 |z|^{2n} \cosh\pi(\nu_{0} + \nu_{1}) t + |z|^{4n}\right]^{2} \left[1 - 2 |z|^{2n} \cosh\pi(\nu_{0} - \nu_{1})t + |z|^{4n}\right]^{2}\label{tildeZ1-p<5}\\
 &\stackrel{\rm or}{=}& \left[1 - 2 |z|^{2n} e^{-\pi\nu_{0}t} \cosh\pi\nu_{1}t + e^{-2\pi  \nu_{0} t} |z|^{4n}\right]^{2} \left[1 - 2 |z|^{2n} e^{\pi\nu_{0} t} \cosh\pi\nu_{1} t + e^{ 2\pi  \nu_{0}t} |z|^{4n}\right]^{2}.\label{tildeZ2-p<5}\nn
\eea
As before, the large $y$ interaction can be obtained from (\ref{t-amplitude-cylinder-p<5}) with the large $t$-integration as
 \be\label{largeyp<5}
 \frac{\Gamma_{p, p}}{V_{p + 1}} \approx   \frac{\left(\cos\pi\nu_{0} - \cos\pi\nu_{1}\right)^{2}  \sqrt{\det (\eta_{p} + \hat F'_{p})\det(\eta_{p} + \hat F_{p})}}{2^{p - 1} \pi^{\frac{p + 1}{2}}(2 \pi \alpha')^{p - 3} y^{7 -p}} \Gamma \left(\frac{7 - p}{2}\right),\nn
  \ee
which is always non-negative, therefore implying an attractive interaction in general.    This is consistent with what has been discussed for the $p = p' = 5$ or $6$ case given in the previous subsection.   In other words, whenever $p = p' < 6$, the long-range interaction is always non-repulsive. The possible long-range repulsive interaction for $p = p'$ can occur only for  $p = p' = 6$ with all three $\nu_{0}, \nu_{1}, \nu_{2}$ real and non-vanishing.  As discussed in the previous subsection, it actually happens when the possible largest of the three $\nu_{\alpha}$'s is smaller than the sum of the remaining two when $\nu_{0} + \nu_{1} + \nu_{2} < 2$. When $\nu_{0}$ is imaginary, given as $\nu_{0} = i \bar\nu_{0}$ with $\bar\nu_{0} \in (0, \infty)$, once again the integrand of the open string one-loop amplitude (\ref{t-amplitude-annulus-p<5}) has an infinite number of simples poles at $t_{k} = k/\bar\nu_{0}$ with $k = 1, 2, \cdots $ along the positive $t$-axis, indicating the decay of the underlying system via the open string pair production. The decay rate and the corresponding open string pair production rate can be computed as before, respectively, as
\bea\label{decayrate-p<5}
{\cal W}_{p, p} &=& \frac{ 2^3 \, \left[\det (\eta_{p} + \hat F'_{p})\det(\eta_{p} + \hat F_{p}) \right]^{\frac{1}{2}} \sinh \pi \bar\nu_{0} \sin\pi\nu_{1}}{\bar \nu_{0} (8 \pi^2 \alpha')^{\frac{p + 1}{2}}} \sum_{k = 1}^{\infty} (-)^{k + 1} \left(\frac{\bar\nu_{0}}{k}\right)^{\frac{p - 1}{2}} e^{- \frac{k y^2 }{2\pi\bar\nu_{0} \alpha'}}\nn
&\,&\times \frac{\left(\cosh \frac{k\pi \nu_{1}}{\bar\nu_{0}} - (-)^{k} \right)^{2}}{\sinh\frac{k \pi\nu_{1}}{\bar\nu_{0}}}  Z_{k} (\bar\nu_{0}, \nu_{1}),
\eea
where 
\be\label{Zk-p<5}
Z_{k} (\bar\nu_{0}, \nu_{1}) = \prod_{n = 1}^{\infty}\frac{\left[1 - 2 (-)^{k} |z_{k}|^{2n} \cosh\frac{k \pi\nu_{1}}{\bar\nu_{0}} + |z_{k}|^{4n}\right]^{4} }{(1 - |z_{k}|^{2n})^{6} 
\left[1 - 2 |z_{k}|^{2n} \cosh \frac{2k\pi\nu_{1}}{\bar\nu_{0}}  + |z_{k}|^{4n}\right] },
 \ee 
with $|z_{k}| = e^{-k\pi/\bar\nu_{0}}$, and
\bea\label{pprate-p<5}
{\cal W}^{(1)}_{p, p} &=& \frac{ 2^3 \, \left[\det (\eta_{p} + \hat F'_{p})\det(\eta_{p} + \hat F_{p}) \right]^{\frac{1}{2}} \sinh \pi \bar\nu_{0} \sin\pi\nu_{1}}{\bar \nu_{0} (8 \pi^2 \alpha')^{\frac{p + 1}{2}}}  \bar\nu_{0}^{\frac{p - 1}{2}} e^{- \frac{ y^2 }{2\pi\bar\nu_{0} \alpha'}}\nn
&\,&\times \frac{\left(\cosh \frac{\pi \nu_{1}}{\bar\nu_{0}} + 1 \right)^{2}}{\sinh\frac{ \pi\nu_{1}}{\bar\nu_{0}}}  Z_{1} (\bar\nu_{0}, \nu_{1}).
 \eea
 The above two rates can be obtained from (\ref{decay-rate}) and (\ref{pp-rate}), respectively,  by taking the limit $\nu_{2} \to 0$. 
  We now discuss the cases for $p = p' < 5$  one by one in what follows.\\ 
     
\noindent
{\bf The $p = p' = 3$ or $4$ case:}  The $p = p' = 3$ can be obtained from the $p = p' = 5$ while $p = p' = 4$ can be obtained from the $p = p' = 6$ in the sense described in the present subsection given above. In either case, the worldvolume flux can be extended the following way,
\be\label{flux34}
\hat F_{\alpha\beta} = \left(\begin{array}{ccc}
\hat F_{\alpha'\beta'}&&\\
&0& \hat g_{2}\\
&-\hat g_{2} &0\\
\end{array}\right),\qquad \hat F'_{\alpha\beta} = \left(\begin{array}{ccc}
\hat F'_{\alpha'\beta'}&&\\
&0& \hat g'_{2}\\
&-\hat g'_{2} &0\\
\end{array}\right).
\ee  
We then have here 
\be\label{nu2}
\tan\pi\nu_{2} = \frac{\left|\hat g_{2} - \hat g'_{2}\right|}{1 + \hat g_{2} \hat g'_{2}},
\ee
which gives $\nu_{2} \to 0$ when we take $\hat g_{2} \to \infty$ and $\hat g'_{2} \to \infty$.  The general closed string tree-level cylinder amplitude is just given by 
(\ref{t-amplitude-cylinder-p<5}) while the corresponding open string one-loop annulus one is given by (\ref{t-amplitude-annulus-p<5}).  The respective physics of these amplitudes
such as the nature of the interaction, the relevant instabilities and the potential open string pair production and its enhancement can be similarly discussed in general following what we have done for the respective $p = p' = 5$ and $p = p' = 6$ cases. So we will not repeat the same discussion here.  For example, one typical interesting case is the $p = p' = 3$ one for the following choice of fluxes,
\be\label{flux3}
\hat F_{\alpha'\beta'} = \left(\begin{array}{cccc}
0& \hat f & &\\
- \hat f &0& &\\
& &0& \hat g_{1}\\
&&-\hat g_{1} &0
\end{array}\right)_{4\times 4},\qquad \hat F'_{\alpha'\beta'} = \left(\begin{array}{cccc}
0& \hat f' & &\\
- \hat f' &0& &\\
& &0& \hat g'_{1}\\
&&-\hat g'_{1} &0
\end{array}\right)_{4\times 4}.
\ee  
With the above fluxes, we have
\be\label{nu0nu1-3}
\tanh\pi\bar\nu_{0} = \frac{|\hat f - \hat f'|}{1 - \hat f \hat f'}, \qquad \tan \pi \nu_{1} =  \frac{\left|\hat g_{1} - \hat g'_{1}\right|}{1 + \hat g_{1} \hat g'_{1}}.
\ee
The closed string cylinder amplitude, the open string annulus amplitude, the decay rate and open string pair production rate of this system can be directly read from (\ref{t-amplitude-cylinder-p<5}), (\ref{t-amplitude-annulus-p<5}), (\ref{decayrate-p<5}) and (\ref{pprate-p<5}), respectively, for the present case. Their explicit expressions will not be written down here.
Their analysis, in particular the open string pair production and its enhancement along with their potential applications, has been discussed in great detail in \cite{Lu:2017tnm,Lu:2018nsc,
Jia:2018mlr}.  Again we will not repeat it here and refer there for detail.  For the $p = p' = 3$ case,  the discussion with the most general worldvolume fluxes is given explicitly in the  paper by one of the present authors\cite{wu} and the basic conclusion remains the same. For example, the interaction amplitude can be given in terms of six Lorentz invariants constructed from the fluxes. \\ 

\noindent
{\bf The $p = p' = 1$ or $2$ case:} The $p = p' = 1$ can be obtained from the $p = p' = 5$ while $p = p' = 2$ can be obtained from the $p = p' = 6$ again in the sense described in the present subsection given earlier. In either case, the worldvolume flux can be extended the following way as
\be\label{flux12}
\hat F_{\alpha\beta} = \left(\begin{array}{ccccc}
\hat F_{\alpha'\beta'}&&&&\\
&0& \hat g_{1}&&\\
&-\hat g_{1} &0&&\\
&&&0&\hat g_{2}\\
&&&- \hat g_{2}&0\\
\end{array}\right),\qquad \hat F'_{\alpha\beta} = \left(\begin{array}{ccccc}
\hat F'_{\alpha'\beta'}&&&&\\
&0& \hat g'_{1}&&\\
&-\hat g'_{1} &0&&\\
&&&0&\hat g'_{2}\\
&&&- \hat g'_{2}&0\\
\end{array}\right).
\ee  
We then have
\be\label{nu1nu2}
 \tan\pi \nu_{1} = \frac{\left|\hat g_{1} - \hat g'_{1}\right|}{1 + \hat g_{1} \hat g'_{1}}, \quad \tan\pi\nu_{2} = \frac{\left|\hat g_{2} - \hat g'_{2}\right|}{1 + \hat g_{2} \hat g'_{2}},
 \ee
  where both $\nu_{1} \to 0$ and $\nu_{2} \to 0$ when we take $\hat g_{k} \to \infty, \hat g'_{k} \to \infty$ with $k = 1, 2$.  The closed string cylinder amplitude can be obtained 
  from (\ref{t-amplitude-cylinder-p<5}) along with (\ref{C-p<5}) and (\ref{tildeC1-p<5}) or (\ref{tildeC2-p<5}) by taking $\nu_{1} \to 0$ limit. The open string one-loop annulus amplitude can be obtained from (\ref{t-amplitude-annulus-p<5}) along with (\ref{Z-p<5}) and (\ref{tildeZ1-p<5}) or (\ref{tildeZ2-p<5}) also by taking $\nu_{1} \to 0$ limit.  Since either of these is straightforward, we will not 
  rewrite the corresponding amplitude here.  The nature of interaction, the potential instabilities as well as the open string pair production can also be similarly discussed and will not be present here.  However, we would like to stress for the present case that we don't have the same enhancement of the open string pair production as discussed in the $p = p ' = 5$ or $6$ as well as in the previous work \cite{Lu:2009au,Lu:2017tnm,Lu:2018suj,Lu:2018nsc }, which requires $p = p ' \ge 3$ so that the needed magnetic flux can be added.  There can be some mild enhancement of open string pair production for the system of $p = p' = 2$ case, which also occurs for $2 \le p = p' \le 6$, as discussed in \cite{Lu:2018suj} by one of the present authors, when the added fluxes satisfy certain conditions.  We refer this to \cite{ Lu:2018suj} for detail.  \\
 
 \noindent 
 {\bf The $p = p' = 0$ case:}  This is a trivial one and can be obtained from the $p = p' = 6$, similarly as above, by setting $\nu_{0} \to 0$ and $\nu_{1} \to 0$ in the amplitude (\ref{t-amplitude-cylinder-p<5}) or (\ref{t-amplitude-annulus-p<5}). As expected, we simply have here $\Gamma_{0, 0}  = 0$.  For this system, there are no fluxes which can be added to the worldvolume and therefore this system remains still as a 1/2 BPS one. The $\Gamma_{0, 0} = 0$ is just the usual ``no-force'' condition.    
     
 In summary, in this section, we compute the closed string cylinder as well as the corresponding open string one-loop annulus amplitude for the system of two Dp branes, placed parallel at separation, with $0 \le p = p' \le 6$, carrying the most general respective constant worldvolume fluxes. We observe that the $\theta_{1}$-function factor (\ref{6-to-p}) in the integrand of the closed string cylinder amplitude for $p = p' = 6$ or $5$ works also for the corresponding factor for $p = p' < 5$ cases so long the respective limits $\nu_{2} \to 0$ or $\nu_{2} \to 0, \nu_{1} \to 0$ or $\nu_{2}, \nu_{1}, \nu_{0} \to 0$ are taken.   This gives us an opportunity to express the respective closed string cylinder or the open string annulus amplitude as a universal form as given either in (\ref{t-amplitude-cylinder}) or in (\ref{t-amplitude-annulus}) for each of the $0 \le p = p' \le 6$ cases so long the aforementioned limits are taken.  We provide a physical explanation to this universal feature based on the properties of the matrix $M$ (\ref{mmatrix}) and the various zero-modes in the matter sector of the closed string boundary state representation of Dp-brane.  This nice feature, as we will see, provides us a trick to compute the closed string cylinder amplitude (as well as the open string annulus amplitude) for the system of Dp and Dp$'$ with $p - p' = 2, 4, 6$ in the following section from the one computed in this section for the system with $p = p'$.  This will greatly simplify the computations for $p \neq p'$ cases given in the following section.   Using the computed amplitudes in this section,  we give also a general discussion on the properties of the amplitudes
 such as the nature of the interaction, the onset of potential tachyonic instability which is associated with the added worldvolume magnetic fluxes and the open string pair production when an electric flux is added.  In particular, we find that the interaction can be repulsive only for $p = p' = 6$ and when the added fluxes are all magnetic with the possible largest one  of three $\nu_{\alpha}$'s  smaller than the sum of  the remaining two when $\nu_{0} + \nu_{1} + \nu_{2} < 2$. Otherwise, it is attractive or vanishes.  The later case occurs either for the largest one equaling the sum of the remaining two or  $\nu_{0} + \nu_{1} + \nu_{2} = 2$. Whenever this happens, it implies that the underlying system is supersymmetric.  We also find that the nature of the interaction is correlated with the existence of a potential tachyonic instability.  When the interaction is repulsive, there is no tachyonic instability. While the interaction is attractive, there is one. We give also a detail discussion on the open string pair production enhancement following the line given in \cite{Lu:2009au,Lu:2017tnm,Lu:2018suj,Lu:2018nsc}.    
     
\section{Amplitude and its properties: the $p \neq p'$ case\label{section4}}     
 In this section, we move to compute the closed string tree-level cylinder amplitude between one Dp and the other Dp$'$, placed parallel at a separation transverse to the Dp, with $p - p' = \kappa = 2, 4, 6$ and $p \le 6$ and with each brane carrying a general constant worldvolume flux.  Here without loss of generality, we assume $p > p'$.  The discussion given in the previous section for computing the cylinder amplitude for $p = p'$ case makes it easier to carry out the computations in the present section. Once the cylinder amplitude is obtained, we can again use a Jacobi transformation to obtain the corresponding open string one-loop annulus amplitude.     
    
  The  trick used in the  subsection \ref{subs3.2} helps us here in obtaining the amplitude for $p \neq p'$ from that of $p = p'$ if we extend the general flux $\hat F'_{p'}$ on the Dp$'$ to the $\hat F'_{p}$ on a Dp in a similar fashion as we did in extending a Dp brane flux for $p < 5$ to the one on D5 or D6 there.  In other words, we first have the following extension of the flux $\hat F'_{p'}$ on the Dp$'$ as,  
  \be\label{extendedfluxp'}
 \hat F'_{\alpha\beta} = \left(\begin{array}{cccccc}
 (\hat F'_{p'})_{\alpha'\beta'} &&&&&\\
 &0& \hat g'_{1}&&&\\
 &- \hat g'_{1}&0&&&\\
  &&&\ddots&&\\
  &&&&0&  \hat g'_{\frac{\kappa}{2}}\\
  &&&&- \hat g'_{\frac{\kappa}{2}}&0\end{array}\right)_{(p + 1)\times (p + 1)},
  \ee       
 where $\alpha, \beta =0, 1, \cdots, p$ and $\alpha', \beta' = 0, 1, \cdots  p'$.  Here $\kappa = 2, 4, 6$. As before, at the end of computations, we need to send $\hat g'_{k} \to
 \infty$ with $k = 1, \cdots, \kappa/2$.  As discussed in the subsection \ref{subs3.2}, this extension will not change anything about the corresponding matrix $M'_{p'}$ given in (\ref{mmatrix}) for the present case and the RR zero-mode contribution in the matter part to the amplitude so long the above limit is taken at the end of computations. Moreover, unlike the extension given there, we have here a bonus for the extension of the bosonic zero-mode contribution to the amplitude so long things are taken care of properly.  Let us explain this in detail. If one computes the bosonic zero-mode contribution in the matter part to the amplitude for the present case, as already given in (\ref{bzm-contribution}),  it is 
\be\label{bzm-contribution-pp'}
 \sqrt{\det(\eta_{p'} + \hat F'_{p'})\det(\eta_{p} + \hat F_{p})} \,V_{\rm NN} \left(2 \pi^{2} \alpha' t\right)^{- \frac{\rm DD}{2}} e^{- \frac{y^{2}}{2\pi\alpha' t}}.
\ee  
In the present context, we have $V_{\rm NN} = V_{p' + 1}$ and ${\rm DD} = 9 - p$. If we use the trick mentioned above, we have then the following
\be\label{bzm-extended}
 \sqrt{\det(\eta_{p} + \hat F'_{p})\det(\eta_{p} + \hat F_{p})} \,V_{\rm p + 1} \left(2 \pi^{2} \alpha' t\right)^{- \frac{9 - p}{2}} e^{- \frac{y^{2}}{2\pi\alpha' t}}.
 \ee     
  Comparing the two, the nice thing here is that the t-dependent part is the same and their difference occurs only in the t-independent part.   From (\ref{extendedfluxp'}), we have
  $\det (\eta_{p} + \hat F'_{p}) = (1 + \hat g'^{2}_{1}) \cdots (1 + \hat g'^{2}_{\kappa/2}) \det(\eta_{p'} + \hat F'_{p'}) = \hat g'^{2}_{1} \cdots \hat g'^{2}_{\kappa/2} \det(\eta_{p'} + \hat F'_{p'})$  when we take $\hat g'_{k} \to \infty$ with $k = 1, \cdots \kappa/2$. So we have 
  \be\label{dbi-factor}
  \sqrt{\det(\eta_{p} + \hat F'_{p})\det(\eta_{p} + \hat F_{p})} = \hat g'_{1} \cdots \hat g'_{\kappa/2}      
 \sqrt{\det(\eta_{p'} + \hat F'_{p'})\det(\eta_{p} + \hat F_{p})}.
 \ee
  Note also that $V_{p + 1} = V_{p' + 1} V_{\kappa}$.  For a Dp brane with the flux (\ref{extendedfluxp'}), following the discussion of (\ref{d6brane-coupling}), we have the following coupling among others,
 \be\label{pcoupling}
 T_{p} \int (C_{p' + 1} \wedge \hat F'\wedge \hat F'\cdots \wedge \hat F')_{p + 1},
 \ee  
 where the number of $\hat F'$'s is $(p -p')/2$ and the $C_{p ' +1}$ is the $(p' + 1)$-form RR potential which can couple with Dp$'$ brane.  It is clear that when we take all $\hat g'_{k} \to \infty$ with $k = 1, \cdots, \kappa/2$, the only dominant coupling is the following one
 \be\label{coupling-limit}
 T_{p} V_{\kappa} \hat g'_{1} \cdots \hat g'_{\kappa/2} \int C_{p' + 1} ,
 \ee     
where we have now $p - p' = \kappa$ and the coefficient in front of the coupling denotes the quantized charge $N$ of Dp$'$ brane in terms of its tension.  In other words, we have 
\be\label{ndp'}
N T_{p'} = T_{p} V_{\kappa} \hat g'_{1} \cdots \hat g'_{\kappa/2},
\ee     
which gives 
\be\label{volume}
V_{\kappa} = \frac{N}{\hat g'_{1} \cdots \hat g'_{\kappa/2}} \frac{T_{p'}}{T_{p}}.
\ee     
With the above considerations, now (\ref{bzm-extended}) becomes
\bea\label{bzm-extended-new}
&&\sqrt{\det(\eta_{p} + \hat F'_{p})\det(\eta_{p} + \hat F_{p})} \,V_{\rm p + 1} \left(2 \pi^{2} \alpha' t\right)^{- \frac{9 - p}{2}} e^{- \frac{y^{2}}{2\pi\alpha' t}}\nn
&&=  \hat g'_{1} \cdots \hat g'_{\kappa/2}   \sqrt{\det(\eta_{p'} + \hat F'_{p'})\det(\eta_{p} + \hat F_{p})} V_{p' + 1} \frac{N}{\hat g'_{1} \cdots \hat g'_{\kappa/2}} \frac{T_{p'}}{T_{p}}  \left(2 \pi^{2} \alpha' t\right)^{- \frac{9 - p}{2}} e^{- \frac{y^{2}}{2\pi\alpha' t}}\nn
&&= N \frac{c_{p'}}{c_{p}} \sqrt{\det(\eta_{p'} + \hat F'_{p'})\det(\eta_{p} + \hat F_{p})} V_{p' + 1} \left(2 \pi^{2} \alpha' t\right)^{- \frac{9 - p}{2}} e^{- \frac{y^{2}}{2\pi\alpha' t}},
\eea    
where we have used the relation $T_{p'}/T_{p} = c_{p'}/c_{p}$ with the normalization $c_{p} = \sqrt{\pi} (2\pi \sqrt{\alpha'})^{3 - p}$ for the boundary state given right after (\ref{mgbs}).  This factor $c_{p'}/c_{p}$ is the one just needed to convert the factor $c^{2}_{p}$, which is used to compute the cylinder amplitude when $p = p'$ (for example, see (\ref{amplitude})),  to $c^{2}_{p} \times c_{p'}/c_{p} = c_{p} c_{p'}$, the correct one for the present amplitude. The large integer $N$ here implies what has been computed using the trick described is actually between one single Dp and N Dp$'$ branes (not a single Dp$'$).  To obtain the wanted amplitude with a single Dp$'$, we need to divide so obtained amplitude by $N$.   Given what has been discussed, the closed string tree-level cylinder amplitude for $p - p' = \kappa \neq 0$ can be obtained from (\ref{t-amplitude-cylinder}) as
\bea\nonumber
\begin{aligned}
\Gamma_{p,p'} &= \frac{\Gamma_{p, p}}{N}\\
 &=\frac{ 2^3 \, V_{p + 1}\, \left[\det (\eta_{p} + \hat F'_{p})\det(\eta_{p} + \hat F_{p})\right]^{\frac{1}{2}}\prod_{\alpha =0}^{2} \sin \pi \nu_{\alpha} }{N (8 \pi^2 \alpha')^{\frac{p + 1}{2}}} \int_0^\infty \frac{d t} {t^{\frac{9 - p}{2}}} \frac{e^{- \frac{y^2}{2\pi\alpha' t}}}{\eta^{3} (it)} \\ 
&\, \times \frac{ \theta_{1} \left(\left.\frac{\nu_{0} + \nu_{1} + \nu_{2}}{2}\right| it \right) \theta_{1} \left(\left.\frac{\nu_{0} - \nu_{1} + \nu_{2}}{2}\right| it \right)
 \theta_{1} \left(\left.\frac{\nu_{0} + \nu_{1} - \nu_{2}}{2}\right| it \right)\theta_{1} \left(\left.\frac{\nu_{0} - \nu_{1} - \nu_{2}}{2}\right| it \right)}{\theta_{1} (\nu_{0} | it)\theta_{1} (\nu_{1} | it)\theta_{1} (\nu_{2} | it)}\\
 &= \frac{c_{p'}}{c_{p}} \frac{ 2^3 \, V_{p' + 1}\, \left[\det (\eta_{p'} + \hat F'_{p'})\det(\eta_{p} + \hat F_{p})\right]^{\frac{1}{2}}\prod_{\alpha =0}^{2} \sin \pi \nu_{\alpha} }{ (8 \pi^2 \alpha')^{\frac{p + 1}{2}}} \int_0^\infty \frac{d t} {t^{\frac{9 - p}{2}}} \frac{e^{- \frac{y^2}{2\pi\alpha' t}}}{\eta^{3} (it)} \\ 
&\, \times \frac{ \theta_{1} \left(\left.\frac{\nu_{0} + \nu_{1} + \nu_{2}}{2}\right| it \right) \theta_{1} \left(\left.\frac{\nu_{0} - \nu_{1} + \nu_{2}}{2}\right| it \right)
 \theta_{1} \left(\left.\frac{\nu_{0} + \nu_{1} - \nu_{2}}{2}\right| it \right)\theta_{1} \left(\left.\frac{\nu_{0} - \nu_{1} - \nu_{2}}{2}\right| it \right)}{\theta_{1} (\nu_{0} | it)\theta_{1} (\nu_{1} | it)\theta_{1} (\nu_{2} | it)}\\ 
 &=  \frac{ 2^3 \, V_{p' + 1}\, \left[\det (\eta_{p'} + \hat F'_{p'})\det(\eta_{p} + \hat F_{p})\right]^{\frac{1}{2}}\prod_{\alpha =0}^{2} \sin \pi \nu_{\alpha} }{ 2^{\frac{\kappa}{2}} (8 \pi^2 \alpha')^{\frac{p' + 1}{2}}} \int_0^\infty \frac{d t} {t^{\frac{9 - p}{2}}} \frac{e^{- \frac{y^2}{2\pi\alpha' t}}}{\eta^{3} (it)} \\ 
&\, \times \frac{ \theta_{1} \left(\left.\frac{\nu_{0} + \nu_{1} + \nu_{2}}{2}\right| it \right) \theta_{1} \left(\left.\frac{\nu_{0} - \nu_{1} + \nu_{2}}{2}\right| it \right)
 \theta_{1} \left(\left.\frac{\nu_{0} + \nu_{1} - \nu_{2}}{2}\right| it \right)\theta_{1} \left(\left.\frac{\nu_{0} - \nu_{1} - \nu_{2}}{2}\right| it \right)}{\theta_{1} (\nu_{0} | it)\theta_{1} (\nu_{1} | it)\theta_{1} (\nu_{2} | it)}\\
 \end{aligned}
 \eea
\bea\label{t-amplitude-cylinder-pp'} 
&=&  \frac{ 2^2 \, V_{p'+ 1} \left[\det (\eta_{p'} + \hat F'_{p'})\det(\eta_{p} + \hat F_{p})\right]^{\frac{1}{2}}  \left[\sum_{\alpha = 0}^{2}\cos^{2}\pi \nu_{\alpha} - 2\prod_{\alpha =0}^{2} \cos\pi\nu_{\alpha} - 1\right]}{ 2^{\frac{\kappa}{2}}(8 \pi^2 \alpha')^{\frac{p' + 1}{2}}} \nn
 &\,& \times  \int_0^\infty \frac{d t}{t^{\frac{9 - p}{2}}}  e^{- \frac{y^2}{2\pi\alpha' t}}  \prod_{n=1}^{\infty} C_{n},
 \eea
 where in the first equality the $\Gamma_{p, p}$ is the cylinder amplitude (\ref{t-amplitude-cylinder}) for the extended flux $\hat F'_{p}$ given in (\ref{extendedfluxp'}),  in the second equality we have used $V_{p + 1} = V_{p' + 1} V_{\kappa}$ and (\ref{bzm-extended-new}),  in the third equality we have used the explicit expression for $c_{p} = \sqrt{\pi}
 (2\pi\sqrt{\alpha'})^{3 - p}$, and $C_{n}$ continues to be given by (\ref{C}) and the extension trick discussed in section \ref{section3} for $\nu_{\alpha}$'s still applies here. It is clear that the basic structure of the above cylinder amplitude is the same as that for the $p = p'$ case discussed in the previous section. So we expect the same properties of the amplitude as discussed there such as the nature of the interaction and the potential instabilities. So we will not repeat this discussion here.  Moreover, we also expect some special features to arise here which will be discussed later in this section.  
     
Once we have the  closed string tree-level cylinder amplitude (\ref{t-amplitude-cylinder-pp'}), the corresponding open string one-loop annulus amplitude can be obtained from the next to the last equality of this amplitude above by a Jacobi transformation following the standard prescription given earlier in the previous section.  This open string one-loop annulus amplitude is then
\bea\label{t-amplitude-annulus-pp'}
\Gamma_{p,p'} &=& - \frac{ 2^3 \, i \, V_{p' + 1}\, \left[\det (\eta_{p'} + \hat F'_{p'})\det(\eta_{p} + \hat F_{p})\right]^{\frac{1}{2}}\prod_{\alpha =0}^{2} \sin \pi \nu_{\alpha} }{2^{\frac{\kappa}{2}}(8 \pi^2 \alpha')^{\frac{p' + 1}{2}}} \int_0^\infty \frac{d t} {t^{\frac{p - 3}{2}}} \frac{e^{- \frac{y^2 t}{2\pi\alpha' }}}{\eta^{3} (it)} \nn 
&\,& \times \frac{ \theta_{1} \left(\left. \frac{\nu_{0} + \nu_{1} + \nu_{2}}{2} i t \right| it \right) \theta_{1} \left(\left. \frac{\nu_{0} - \nu_{1} + \nu_{2}}{2} it \right| it \right)
 \theta_{1} \left(\left.  \frac{\nu_{0} + \nu_{1} - \nu_{2}}{2} it \right| it \right)\theta_{1} \left(\left.  \frac{\nu_{0} - \nu_{1} - \nu_{2}}{2} it \right| it \right)}{\theta_{1} (\nu_{0} it | it)\theta_{1} ( \nu_{1} it | it)\theta_{1} ( \nu_{2} it | it)}\nn
 &=&  \frac{ 2^2 \, V_{p'+ 1} \left[\det (\eta_{p'} + \hat F'_{p'})\det(\eta_{p} + \hat F_{p})\right]^{\frac{1}{2}} }{2^{\frac{\kappa}{2}} (8 \pi^2 \alpha')^{\frac{p' + 1}{2}}}   \int_0^\infty \frac{d t}{t^{\frac{p - 3}{2}}}  e^{- \frac{y^2 t}{2\pi\alpha'}} \prod_{\alpha =0}^{2} \frac{\sin \pi \nu_{\alpha}}{\sinh \pi \nu_{\alpha} t} \nn
 &\,& \times \left[\sum_{\alpha = 0}^{2}\cosh^{2}\pi \nu_{\alpha} t- 2 \prod_{\alpha = 0}^{2} \cosh\pi \nu_{\alpha} t - 1\right] \prod_{n=1}^{\infty} Z_{n},
 \eea     
where $Z_{n}$ is still given by (\ref{Z}).  The use of this open string one-loop annulus amplitude is for analyzing the small $y$ behavior such as the onset of tachyonic instability and that when $\nu_{0}$ is imaginary, the underlying system will decay via the so-called open string pair production.  So we will give here a general discussion of both. 

When all three $\nu_{0}, \nu_{1}$ and $\nu_{2}$ are real and non-vanishing, once again without loss of generality and for convenience, we assume $\nu_{0} \le \nu_{1} \le \nu_{2}$.  Let us first focus on the case $\nu_{0} + \nu_{1} + \nu_{2} < 2$.  If $\nu_{0} + \nu_{1} > \nu_{2}$, the interaction (\ref{t-amplitude-cylinder-pp'}) is repulsive and there is no potential tachyonic instability which can be checked from the integrand of (\ref{t-amplitude-annulus-pp'}) for large $t$. If $\nu_{0} + \nu_{1} = \nu_{2}$, the amplitude vanishes and this indicates the preservation of certain supersymmetry.  On the other hand, if $\nu_{0} + \nu_{1} < \nu_{2}$, the interaction is attractive and for large $t$, it behaves as
\be\label{tachyonic-pp'}
\sim e^{- \frac{y^2 t}{2\pi\alpha'}} e^{\pi (\nu_{2} - \nu_{1} - \nu_{0}) t} = e^{- 2\pi t\left(\frac{y^{2}}{(2\pi)^{2} \alpha'} - \frac{\nu_{2} - \nu_{1} - \nu_{0}}{2}\right)},
\ee
which blows up if $y < \pi \sqrt{2 (\nu_{2} - \nu_{1} - \nu_{0})\alpha'}$, indicating the onset of tachyonic instability.  For the case $\nu_{0} + \nu_{1} + \nu_{2} = 2$, the interaction again vanishes and the underlying system preserves certain supersymmetry.  For $2 < \nu_{0} + \nu_{1} + \nu_{2} < 3$,  we can only have $\nu_{0} + \nu_{1} > \nu_{2}$ and the interaction is also attractive.  For large t, the amplitude behaves as
\be\label{tachyonic-pp'-new}
\sim  e^{- \frac{y^2 t}{2\pi\alpha'}} e^{\pi (\nu_{2} + \nu_{1} + \nu_{0} - 2) t} = e^{- 2\pi t\left(\frac{y^{2}}{(2\pi)^{2} \alpha'} - \frac{\nu_{2} + \nu_{1} + \nu_{0} - 2}{2}\right)},
\ee
which blows up if $y < \pi \sqrt{2 (\nu_{0} + \nu_{1} + \nu_{2} - 2)\alpha'}$, indicating again the tachyonic instability.
So everything here is consistent with what has been discussed in the previous section for $p = p'$ case. 

 If $\nu_{0} = i \bar\nu_{0}$, i.e., imaginary, with $\bar\nu_{0} \in (0, \infty)$,  so the factor $\sin\pi\nu_{0} /\sinh\pi\nu_{0} t$ in the integrand of (\ref{t-amplitude-annulus-pp'}) becomes $\sinh\pi\bar\nu_{0}/\sin\pi\bar\nu_{0} t$, indicating the appearance of an infinite number of simple poles of the integrand at $t_{k} = k/\bar\nu_{0}$ with 
$k = 1, 2, \cdots$. So this implies that the amplitude has an imaginary part,  indicating the decay of the underlying system via the so-called open string pair production.  The decay rate per unit volume of Dp$'$ brane worldvolume can be computed as before as
\bea\label{decayrate-pp'}
{\cal W}_{p, p'} &=& -  \frac{2\, {\rm Im} \Gamma}{V_{p' + 1}}\nn
&=& \frac{ 2^{3- \frac{\kappa}{2}} \, \left[\det (\eta_{p'} + \hat F'_{p'})\det(\eta_{p} + \hat F_{p})\right]^{\frac{1}{2}} \sinh\pi\bar\nu_{0} \sin\pi\nu_{1} \sin\pi\nu_{2}} {\bar\nu_{0} (8 \pi^2 \alpha')^{\frac{p' + 1}{2}}}  \sum_{k = 1}^{\infty} (-)^{k + 1} \left(\frac{\bar\nu_{0}}{k}\right)^{\frac{p - 3}{2}} \nn
&\,&\times \frac{\left(\cosh\frac{k \pi \nu_{1}}{\bar\nu_{0}} - (-)^{k} \cosh\frac{k\pi\nu_{2}}{\bar\nu_{0}}\right)^{2}}{\sinh\frac{k\pi\nu_{1}}{\bar\nu_{0}}\sinh\frac{k\pi\nu_{2}}{\bar\nu_{0}}} e^{- \frac{k y^2 }{2\pi\alpha' \bar\nu_{0}}}\, Z_{k} (\bar\nu_{0}, \nu_{1}, \nu_{2}),
\eea
where
\be\label{Zk}
Z_{k}  = \prod_{n = 1}^{\infty} \frac{\left[ \left(1 + |z_{k}|^{4n } - 2 (-)^{k} |z_{k}|^{2n} \cosh\frac{k \pi \nu_{1}}{\bar\nu_{0}}\cosh\frac{k\pi \nu_{2}}{\bar\nu_{0}}\right)^{2} - 
4  |z_{k}|^{4n} \sinh^{2}\frac{k\pi\nu_{1}}{\bar\nu_{0}} \sinh^{2}\frac{k \pi \nu_{2}}{\bar\nu_{0}}\right]^{2}} { (1 - |z_{k}|^{2n})^{2}\left(1 - 2 |z_{k}|^{2n} \cosh\frac{2 \pi k \nu_{1}}{\bar\nu_{0}}  
+ |z_{k}|^{4n}\right)\left(1 - 2 |z_{k}|^{2n} \cosh\frac{2 \pi k \nu_{2}}{\bar\nu_{0}}  + |z_{k}|^{4n}\right)},
 \ee
 with $|z_{k}| = e^{- k \pi/\bar\nu_{0}}$.  As before, the open string pair production rate is given by the $k = 1$ term of the above as
 \bea\label{pprate-pp'}
 {\cal W}^{(1)}_{p, p'} &=&  \frac{ 2^{3- \frac{\kappa}{2}} \, \left[\det (\eta_{p'} + \hat F'_{p'})\det(\eta_{p} + \hat F_{p})\right]^{\frac{1}{2}} \sinh\pi\bar\nu_{0} \sin\pi\nu_{1} \sin\pi\nu_{2}} { (8 \pi^2 \alpha')^{\frac{p' + 1}{2}}} \bar\nu_{0}^{\frac{p - 5}{2}} \, e^{- \frac{y^2 }{2\pi\alpha' \bar\nu_{0}}}\nn
 &\,&\times \frac{\left(\cosh\frac{ \pi \nu_{1}}{\bar\nu_{0}} + \cosh\frac{\pi\nu_{2}}{\bar\nu_{0}}\right)^{2}}{\sinh\frac{\pi\nu_{1}}{\bar\nu_{0}}\sinh\frac{\pi\nu_{2}}{\bar\nu_{0}}} \, Z_{1} (\bar\nu_{0}, \nu_{1}, \nu_{2}).
 \eea
 In the above, we assume $p \ge 5$.    The $p = 3, 4$ amplitude or rate can be obtained by sending $\nu_{2} \to 0$ and the $p = 1, 2$ amplitude or rate can further be obtained by sending $\nu_{1}, \nu_{2} \to 0$.  This becomes clear when we discuss the $p - p' = 2$, $p - p' = 4$ and $p - p' = 6$ one by one in the following. 
 
 Before we move to that, let us check one thing mentioned in the previous section.  It is that the interaction between a Dp and a Dp$'$ for $6 \ge p > p'$, placed parallel at a separation and without any brane flux present, is attractive when $ p - p' = 2$, vanishes when $p - p' = 4$ and is repulsive when $p  - p' = 6$ ($p = 6, p' = 0$).   Let us check each of them explicitly here. For $p = 6$ and $p' = 0$, we have here $\nu_{2} = \nu_{1} = \nu_{0} = 1/2$ and so we have, for example,  the cylinder amplitude from the last equality of (\ref{t-amplitude-cylinder-pp'}) as
 \be\label{t-amplitude-cylinder-60-special}
\Gamma_{6, 0} = - \frac{V_{1}}{ 2 (8 \pi^2 \alpha')^{\frac{1}{2}}}   \int_0^\infty \frac{d t}{t^{\frac{3}{2}}}  e^{- \frac{y^2}{2\pi\alpha' t}}  \prod_{n=1}^{\infty} \frac{(1 + |z|^{4n})^{4}}{(1 - |z|^{4n})^{2} (1 + |z|^{2n})^{4}},
 \ee
 where we have used (\ref{C}) for $C_{n}$.  It is indeed repulsive since $\Gamma_{6, 0} < 0$ for any $y$.  For $p - p' = 2$, we have one of three $\nu_{0},\nu_{1}, \nu_{2}$ being half and the remaining two being zero while for $p - p' = 4$, we have two of them being half and the remaining one being zero.  For the former case, the cylinder amplitude (\ref{t-amplitude-cylinder-pp'}) is now,
 \be\label{t-amplitude-cylinder-kappa2-special}
 \Gamma_{\kappa = 2} =    \frac{ 2 \, V_{p'+ 1}}{(8 \pi^2 \alpha')^{\frac{p' + 1}{2}}}  \int_0^\infty \frac{d t}{t^{\frac{9 - p}{2}}}  e^{- \frac{y^2}{2\pi\alpha' t}}  \prod_{n=1}^{\infty} \frac{(1 + |z|^{4n})^{4}}{(1 - |z|^{2n})^{4} (1 - |z|^{4n})^{2}},
 \ee
 where we have used (\ref{C}) for $C_{n}$.  It is indeed attractive since $\Gamma_{\kappa = 2} > 0$ for any $y$.  For the latter case, the amplitude $\Gamma_{\kappa = 4}$ simply vanishes due to the constant factor $[\sum_{\alpha = 0}^{2}\cos^{2}\pi \nu_{\alpha} - 2\prod_{\alpha =0}^{2} \cos\pi\nu_{\alpha} - 1] $ in the amplitude (\ref{t-amplitude-cylinder-pp'}) being zero for the present case. All are as expected. We now discuss each separate case mentioned earlier.

\subsection {The $p - p' = 2$ case\label{subsection4.1}}
 In this subsection, we will focus on the $p - p' = 2$ case, specifically.  We will discuss here each of $p = 6$ or $5$; $p = 4$ or $3$ and  $p = 2$, separately.  Let us begin with $p = 6$ or $5$ case.\\

 \noindent
 {\bf The $p = 6$ or $5$ case:}   For either of these two cases, we can extend the flux $\hat F'_{p'}$ to $\hat F'_{p}$, prescribed in (\ref{extendedfluxp'}), as  
\be\label{extendedfluxp'43}
(\hat F'_{p})_{\alpha\beta} = \left(\begin{array}{ccc}
(\hat F'_{p'})_{\alpha'\beta'} &&\\
&0& \hat g'\\
&- \hat g'&0\end{array}\right),
\ee
where we will take $\hat g' \to \infty$ at the end of computations.  For illustration purpose, we consider the flux $\hat F_{p}$ on the Dp brane the following form
\be\label{fluxp56}
\hat F_{\alpha\beta}  = \left(\begin{array}{ccc}
(\hat F_{p'})_{\alpha'\beta'} &&\\
&0& \hat g\\
&- \hat g&0\end{array}\right),
\ee
where $g$ is finite.  We can then determine $\nu_{2}$ as
\be\label{nu2}
\tan\pi\nu_{2} = \frac{\left|\hat g' - \hat g\right|}{1 + \hat g' \hat g},
\ee
which gives $\tan\pi\nu_{2} = 1/\hat g$ when we take $\hat g' \to \infty$ limit.  For a given fixed $\hat g$, the discussion goes exactly the same as we did for the $p = p' = 6$ case given in the previous section. For this reason, we will not repeat it here.  We here focus on vanishingly small $\hat g$ (in practice  $\hat g \ll 1$) for which we have $\nu_{2} \to 1/2$.  
 From (\ref{t-amplitude-cylinder-pp'}), we have the closed string cylinder amplitude for the present case as
\be\label{t-amplitude-cylinder-65-2}
\Gamma_{p,p'}  =  \frac{ 2 \, V_{p'+ 1} \left[\det (\eta_{p'} + \hat F'_{p'})\det(\eta_{p} + \hat F_{p})\right]^{\frac{1}{2}}  \left(\cos^{2}\pi\nu_{0} - \sin^{2}\pi\nu_{1}\right)}{(8 \pi^2 \alpha')^{\frac{p' + 1}{2}}}   \int_0^\infty \frac{d t}{t^{\frac{9 - p}{2}}}  e^{- \frac{y^2}{2\pi\alpha' t}}  \prod_{n=1}^{\infty} C_{n},
 \ee
 where $C_{n}$, from (\ref{C}), is now
  \be\label{C65-2}
 C_{n} = \frac{\tilde C_{n} }{(1 - |z|^{4n})^{2} \prod_{\alpha = 0}^{1}
 \left[1 - 2 |z|^{2n} \cos2\pi \nu_{\alpha}  + |z|^{4n}\right]},
 \ee
 with 
 \bea
 \tilde C_{n} &=& \left[(1+ |z|^{4n})^{2} - 4 |z|^{4n} \sin^{2}\pi(\nu_{0} + \nu_{1}) \right] \left[ (1 + |z|^{4n})^{2} -  4 |z|^{4n} \sin^{2}\pi(\nu_{0} - \nu_{1})\right]\nn
 &\,& \label{tildeC1-65-2}\\
 &\stackrel{\rm or}{=}& \left[ (1 + e^{2\pi i \nu_{0}} |z|^{4n})^{2} - 4 |z|^{4n} e^{2 i\pi\nu_{0}} \sin^{2}\pi \nu_{1} \right] \nn
 &\,&\times \left[(1 +  e^{- 2\pi i \nu_{0}} |z|^{4n})^{2} - 4 |z|^{4n} e^{- 2 i\pi\nu_{0}} \sin^{2}\pi \nu_{1}\right]. \label{tildeC2-65-2}
\eea 
When $p = 6$, we have two choices: 1) $\nu_{0}, \nu_{1} \in [0, 1)$ and 2) $\nu_{0} = i \bar\nu_{0}$ with $\bar \nu_{0} \in (0, \infty)$ and $\nu_{1} \in [0, 1)$. For the first case,
the interaction is attractive if $\cos^{2}\pi\nu_{0} > \sin^{2}\pi\nu_{1}$, vanishes if $\cos^{2}\pi \nu_{0} = \sin^{2}\pi\nu_{1}$, and is repulsive if $\cos^{2}\pi\nu_{0} < \sin^{2}\pi\nu_{1}$. 
Let us discuss each of these in detail. For the attractive interaction,  when  $\nu_{0} < 1/2$, we have, from $\cos^{2}\pi \nu_{0} > \sin^{2}\pi \nu_{1}$, $\cos\pi\nu_{0} > \sin \pi \nu_{1}$ which gives $\nu_{0} + \nu_{1} < 1/2 = \nu_{2}$.  In other words, the largest $\nu_{2}$ is larger than the sum of the remaining two and this is consistent with what we have achieved in the previous section since now $\nu_{0} + \nu_{1}  + \nu_{2} < 1 < 2$.  While for $\nu_{0} > 1/2$ (Note that $\nu_{0}$ cannot be $1/2$),  we have then $- \cos\pi \nu_{0} > \sin\pi\nu_{1}$ which gives $\nu_{0} > \nu_{1} + 1/2 = \nu_{1} + \nu_{2}$ (we still have $\nu_{0} + \nu_{1} + \nu_{2} < 2 \nu_{0} < 2$) and again the largest $\nu_{0}$ is larger than the sum of the remaining two. 
 The interaction vanishes when $\nu_{0} + \nu_{1} = 1/2 = \nu_{2}$ if $\nu_{0} \le 1/2$ or $\nu_{0} = \nu_{1} + 1/2 = \nu_{1} + \nu_{2}$ if $\nu_{0} > 1/2$. The interaction is repulsive when $\nu_{0} + \nu_{1} > 1/2 = \nu_{2}$ if $\nu_{0} \le 1/2$ and $\nu_{1} \le 1/2$ or $\nu_{1} < \nu_{0} + 1/2 = \nu_{0} + \nu_{2}$ if $\nu_{0} \le 1/2$ and $\nu_{1} > 1/2$  or $\nu_{0} < \nu_{1} + 1/2 = \nu_{1} + \nu_{2}$ if $\nu_{0} > 1/2$ and $\nu_{1} \le 1/2$ or $1 < \nu_{0} + \nu_{1} < 3/2$ if $\nu_{0} > 1/2$ and $\nu_{1} > 1/2$ for which if $\nu_{0} > \nu_{1}$ we will
 have $\nu_{0} < \nu_{1} + 1/2 = \nu_{1} + \nu_{2}$ or if $\nu_{1} > \nu_{0}$ we will have $\nu_{1} < \nu_{0} + \nu_{2}$. Note that for the repulsive interaction, we have $\nu_{0} + \nu_{1} + \nu_{2} < 2$.    Everything here is consistent with what has been discussed in the $ p = p' = 6$ case in the previous section if we take the present $\nu_{2} = 1/2$.  So here is just a special case with $\nu_{2} = 1/2$ of the general discussion of the $p = p' = 6$ in the previous section. 

For small $y$,  the small $t$ integration becomes important for the amplitude.  This gives also a potential singularity of this amplitude since we have two potential sources for this. 
One is from the $t^{- (9 - p)/2}$ factor in the integrand and the other comes from infinite product of $C_{n}$, each of which has a factor $(1 - |z|^{4n})^{2} \sim t^{2}$ for small $t$ in the denominator of $C_{n}$. Both of them blow up for small $t$. The discussion will again be the special case of what has been discussed in the previous section for the $p = p'$ cases and will not be repeated here.

For the second case, the sign of the integrand becomes again indefinite for small $t$ and the discussion goes the same as we did for the $p = p'$ case and will not be repeated here. 

The underlying physics for either of these two cases will become clear if we examine it from the corresponding open string one-loop annulus amplitude.  This annulus amplitude can be read from (\ref{t-amplitude-annulus-pp'}) for the present case as 
\bea\label{t-amplitude-annulus-65-2}
\Gamma_{p, p'} &=&  \frac{ 2 \, V_{p'+ 1} \left[\det (\eta_{p'} + \hat F'_{p'})\det(\eta_{p} + \hat F_{p})\right]^{\frac{1}{2}}}{(8 \pi^2 \alpha')^{\frac{p' + 1}{2}}}   \int_0^\infty \frac{d t}{t^{\frac{p - 3}{2}}}  e^{- \frac{y^2 t}{2\pi\alpha'}} \frac{\sinh \pi \bar\nu_{0} \sin\pi\nu_{1}}{\sin \pi\bar \nu_{0} t
\sinh\pi\nu_{1}t \sinh\frac{\pi t}{2}} \nn
 & \times&  \left[\left(\cosh\pi\nu_{1}t - \cosh\frac{\pi t}{2} \right)^{2} + 4 \sin^{2} \frac{\pi\bar\nu_{0}t}{2} \left(\cosh\pi\nu_{1} t \cosh\frac{\pi t}{2} - \cos^{2} \frac{\pi\bar\nu_{0} t}{2} \right) \right] 
 \prod_{n=1}^{\infty} Z_{n},\nn
 \eea     
 where $Z_{n}$ can be read from (\ref{Z}) as
  \be\label{Z-65-2}
 Z_{n} = \frac{\left[1 - 2 |z|^{2n} \cos2 \pi \bar\nu_{0} t + |z|^{4n}\right]^{-1} \tilde Z_{n}}{(1 - |z|^{2n})^{2} \left[1 - 2 |z|^{2n} \cosh \pi t + |z|^{4n}\right] 
 \left[1 - 2 |z|^{2n} \cosh2\pi \nu_{1} t  + |z|^{4n}\right]},
 \ee
 with
 \bea
 &&\tilde Z_{n} = \left[\left(1 - 2 |z|^{2n} \cos\pi\bar\nu_{0} t \cosh\pi\left(\nu_{1} + \frac{1}{2}\right)t + |z|^{4n}\right)^{2} \right.\nn
 &&\left. +  4 |z|^{4n}\sin^{2}\pi\bar\nu_{0}t \sinh^{2}\pi \left(\nu_{1} + \frac{1}{2}\right) t\right] \left[\left(1 - 2 |z|^{2n} \cos\pi\bar\nu_{0} t \cosh\pi\left(\nu_{1} - \frac{1}{2}\right)t + |z|^{4n}\right)^{2} \right.\nn
 &&\left. +  4 |z|^{4n} \sin^{2} \pi\bar\nu_{0} t \sinh^{2} \pi\left(\nu_{1} - \frac{1}{2}\right) t\right] > 0.\label{tildeZ2-65-2}
\eea  
For large $t$, $Z_{n} \approx 1$ and the integrand is
\be
\sim e^{- \frac{y^2 t}{2\pi\alpha'}}  e^{\pi \left|\frac{1}{2} - \nu_{1}\right|t} = e^{- 2\pi t \left[\frac{y^{2}}{(2\pi)^{2}\alpha'} - \frac{|1/2 - \nu_{1}|}{2}\right]},
\ee
which blows up when $y < \pi \sqrt{|1 - 2 \nu_{1}|\alpha'}$, indicating the onset of tachyonic instability mentioned above.  Once again, the factor $\sin\pi \bar\nu_{0} t$ in the denominator of the integrand of the amplitude gives an infinite number of simple poles along the positive t-axis (note that the integrand is regular as $t \to 0$) at $t_{k} = k/\bar\nu_{0}$ with $k = 1, 2, \cdots$.  This implies that the amplitude has an imaginary part, indicating the decay of the underlying system via the so-called open string pair production. The decay rate per unit $p'$-brane volume  can be computed as before to give
\bea\label{decay-rate-65-2}
{\cal W}_{p, p'} &=& - \frac{2 {\rm Im} \Gamma}{V_{p' + 1}}\nn
&=&  \frac{ 2^{2} \,\left[\det (\eta_{p'} + \hat F'_{p'})\det(\eta_{p} + \hat F_{p}) \right]^{\frac{1}{2}} \sinh\pi \bar\nu_{0} \sin\pi\nu_{1}}{\bar\nu_{0} (8 \pi^2 \alpha')^{\frac{p' + 1}{2}}} \sum_{k = 1}^{\infty} (- )^{k + 1}
\left(\frac{\bar\nu_{0}}{k}\right)^{\frac{p - 3}{2}} e^{- \frac{k y^{2}}{2\pi\alpha' \bar\nu_{0}}} \nn
&\,&\times \frac{\left(\cosh\frac{k \pi \nu_{1}}{\bar\nu_{0}} - (-)^{k} \cosh\frac{k \pi}{2\bar\nu_{0}}\right)^{2}}{\sinh\frac{k\pi\nu_{1}}{\bar\nu_{0}}\sinh\frac{k\pi}{2\bar\nu_{0}}} Z_{k} (\bar\nu_{0}, \nu_{1}),
\eea 
where 
\bea\label{Zk-65-2}
Z_{k} (\bar\nu_{0}, \nu_{1}) &=& \prod_{n = 1}^{\infty}\frac{\left[1 - 2 (-)^{k}|z_{k}|^{2n} \cosh \frac{k \pi}{\bar\nu_{0}}\left(\nu_{1} + \frac{1}{2}\right) + |z_{k}|^{4n}\right]^{2} 
}{(1 - |z_{k}|^{2n})^{4} \left[1 - 2 |z_{k}|^{2n} \cosh\frac{ k\pi }{\bar\nu_{0}} + |z_{k}|^{4n}\right] 
 }\nn
 &\,&\quad \times \frac{\left[1 - 2 (-)^{k}|z_{k}|^{2n} \cosh \frac{k \pi}{\bar\nu_{0}}\left(\nu_{1} - \frac{1}{2}\right) + |z_{k}|^{4n}\right]^{2}}{\left[1 - 2 |z_{k}|^{2n} \cosh\frac{2 k \pi \nu_{1}}{\bar\nu_{0}}  + |z_{k}|^{4n}\right]},
  \eea
with $|z_{k}| = e^{- k \pi/\bar\nu_{0}}$. The open string pair production rate is the $k = 1$ term of the above and it is
\bea\label{pprate-65-2}
{\cal W}^{(1)}_{p, p'} &=& \frac{ 2^{2} \,\left[\det (\eta_{p'} + \hat F'_{p'})\det(\eta_{p} + \hat F_{p}) \right]^{\frac{1}{2}} \sinh\pi \bar\nu_{0} \sin\pi\nu_{1}}{ (8 \pi^2 \alpha')^{\frac{p' + 1}{2}}} {\bar\nu_{0}}^{\frac{p - 5}{2}} e^{- \frac{ y^{2}}{2\pi\alpha' \bar\nu_{0}}} \nn
&\,&\times \frac{\left(\cosh\frac{ \pi \nu_{1}}{\bar\nu_{0}} +  \cosh\frac{ \pi}{2\bar\nu_{0}}\right)^{2}}{\sinh\frac{\pi\nu_{1}}{\bar\nu_{0}}\sinh\frac{\pi}{2\bar\nu_{0}}} Z_{1} (\bar\nu_{0}, \nu_{1}).
\eea
One can check easily that the above decay rate or the open string pair production rate is just the special case of  (\ref{decayrate-pp'}) or (\ref{pprate-pp'}) for $\nu_{2} = 1/2$ and $\kappa = 2$, respectively.  There is an interesting enhancement of the pair production rate even in the absence of magnetic flux for which we have $\nu_{1} = 0$ for small $\bar\nu_{0}$. This rate can be obtained from the above by taking $\nu_{1} \to 0$ limit as
\be\label{pprate-65-2-0nu1}
{\cal W}^{(1)}_{p, p'} =  \frac{ 2^{2} \,\left[\det (\eta_{p'} + \hat F'_{p'})\det(\eta_{p} + \hat F_{p}) \right]^{\frac{1}{2}} \sinh\pi \bar\nu_{0}}{ (8 \pi^2 \alpha')^{\frac{p' + 1}{2}}} {\bar\nu_{0}}^{\frac{p - 3}{2}} e^{- \frac{ y^{2}}{2\pi\alpha' \bar\nu_{0}}}  \frac{\left(1 +  \cosh\frac{ \pi}{2\bar\nu_{0}}\right)^{2}}{\sinh\frac{\pi}{2\bar\nu_{0}}} Z_{1} (\bar\nu_{0}, 0), 
\ee
where 
\be\label{Z1}
Z_{1} (\bar\nu_{0}, 0) = \prod_{n = 1}^{\infty} \frac{\left[1 + 2 |z_{1}|^{2n} \cosh \frac{\pi}{2\bar\nu_{0}}  + |z_{1}|^{4n}\right]^{4} 
}{(1 - |z_{1}|^{2n})^{6} \left[1 - 2 |z_{1}|^{2n} \cosh\frac{ \pi }{\bar\nu_{0}} + |z_{1}|^{4n}\right]}.
\ee   
We would like to remark here that given the form of flux $\hat F_{p}$ (\ref{fluxp56}), the above decay or pair production rate is also valid for $p \ge 3$ when we take $\nu_{1} = 0$. For illustration, let us have a consideration of the following special choice of fluxes $\hat F_{p}$ and $\hat F'_{p'}$ for $p = 5$ as
\be\label{specificfluxp=5}
\hat F_{5} = \left(\begin{array}{cccccc}
0 & \hat f &0&&&\\
- \hat f & 0&0&\cdots&&\\
0&0&0& &&\\
&\vdots&&\ddots&&\\
  &&&&&\\
  &&&&&\end{array}\right)_{6\times 6},\quad \hat F'_{3} = \left(\begin{array}{cccc}
  0&\hat f'&0&0\\
  -\hat f'&0&0&0\\
  0&0&0&0\\
  0&0&0&0\end{array}\right)_{4\times 4},
  \ee 
  where there is no magnetic flux present.  This gives the $\hat g = 0$ in (\ref{fluxp56}) and so we have $\nu_{2} = 1/2$.  With this special choice of fluxes, we have $\nu_{1} = 0$ and
  \be\label{nu0p=5}
  \tanh \pi\bar\nu_{0} = \frac{|\hat f - \hat f'|}{1 - \hat f\hat f'}, 
  \ee
  The pair production rate (\ref{pprate-65-2-0nu1}) becomes 
 \be\label{pprate-5-2}
 {\cal W}^{(1)}_{p, p'} = \frac{ 2^{2} \,|\hat f - \hat f'|}{ (8 \pi^2 \alpha')^{\frac{p' + 1}{2}}} {\bar\nu_{0}}^{\frac{p - 3}{2}} e^{- \frac{ y^{2}}{2\pi\alpha' \bar\nu_{0}}} 
 \frac{\left( 1 +  \cosh\frac{ \pi}{2\bar\nu_{0}}\right)^{2}}{\sinh\frac{\pi}{2\bar\nu_{0}}} Z_{1} (\bar\nu_{0}, 0),
\ee
 where $p = 5$ and $p' = 3$.  As pointed out above, this rate is also valid for $p = 3$ and $p' = 1$. For small $\bar \nu_{0}$, $Z_{1} (\bar\nu_{0}, 0) \approx 1$ and we have a large enhancement factor of $e^{\pi/(2\bar\nu_{0})} \gg 1$ which is not seen in the $p = p'$ case. This large enhancement was also considered by one of the present authors in \cite{Lu:2009pe} and it is essentially due to the Dp$'$ brane acting effectively as a stringy magnetic flux. 
 \\ 
 
 \noindent 
 {\bf The $p = 4$ or $3$ case:}  For this case, we need to set $\nu_{2} = 0$ from the outset.  Now the role of $\nu_{2}$ in the above $p = 6$ or $5$ case is replaced by that of $\nu_{1}$
 in the present one.  By the same token, we extend the flux $\hat F'_{p'}$ to $\hat F'_{p}$ the following way,
 \be\label{extendedfluxp'21}
(\hat F'_{p})_{\alpha\beta} = \left(\begin{array}{ccc}
(\hat F'_{p'})_{\alpha'\beta'} &&\\
&0& \hat g'\\
&- \hat g'&0\end{array}\right),
\ee
where we will take $\hat g' \to \infty$ at the end of computations.  For illustration purpose, we consider the flux $\hat F_{p}$ on the Dp brane the following form
\be\label{fluxp43}
\hat F_{\alpha\beta}  = \left(\begin{array}{ccc}
(\hat F_{p'})_{\alpha'\beta'} &&\\
&0& \hat g\\
&- \hat g&0\end{array}\right),
\ee
where $\hat g$ is finite.  We can then determine $\nu_{1}$ as
\be\label{nu1}
\tan\pi\nu_{1} = \frac{\left|\hat g' - \hat g\right|}{1 + \hat g' \hat g},
\ee
which gives $\tan\pi\nu_{1} = 1/\hat g$ when we take $\hat g' \to \infty$.   For a general $\hat g$, the present discussion is not different from its correspondence in the $p = p'$ case in the previous section and we will not repeat it here.  We here also focus on the  small or vanishing $\hat g$ for which we have $\nu_{1} \to 1/2$.  The closed string cylinder amplitude can be read from the last equality in (\ref{t-amplitude-cylinder-pp'}) as
  \be\label{t-amplitude-cylinder-43}
\Gamma_{p,p'}   =  \frac{ 2 \, V_{p'+ 1} \left[\det (\eta_{p'} + \hat F'_{p'})\det(\eta_{p} + \hat F_{p})\right]^{\frac{1}{2}} \cos^{2} \pi \nu_{0} }{ (8 \pi^2 \alpha')^{\frac{p' + 1}{2}}} 
  \int_0^\infty \frac{d t}{t^{\frac{9 - p}{2}}}  e^{- \frac{y^2}{2\pi\alpha' t}}  \prod_{n=1}^{\infty} C_{n},
 \ee
  where $C_{n}$ can be read from (\ref{C}) as
  \be\label{C-43-2}
  C_{n} = \frac{\left(1 + |z|^{2n}\right)^{2} \left(1 + 2 |z|^{4n} \cos2\pi\nu_{0} + |z|^{8n}\right)^{2}}{\left(1 - |z|^{4n}\right)^{4} \left(1 - 2 |z|^{2n} \cos2\pi\nu_{0} + |z|^{4n} \right)}.
  \ee
  It is clear that this interaction can only be attractive which is consistent with what we have achieved in the previous section.  This interaction  vanishes if $\nu_{0} = 1/2$ which can only be true for $p = 4$. This vanishing interaction can easily 
  be understood as follows. The $\nu_{0} =  1/2$ can be understood either from that the D4, assuming along 1, 2, 3, 4-directions, carries an infinite large magnetic flux and the D2, assuming along 1, 2-directions, carries no flux or from that the D4 carries no flux but the D2 carries such a magnetic flux. In the former case, 
  the contribution to the interaction from the D4 is actually dominated by the infinitely large magnetic flux which gives an infinite many of D2 branes within the D4 but along the 3, 4-directions. In other words, these infinite many D2 are along directions orthogonal to the above D2. If we perform T-dualities along 1, 2-directions, the D2 will become D0 while the infinite many D2 will become D4 along 1, 2, 3, 4-directions.  Since there is no interaction between this D0 and the infinite many D4 and the corresponding system preserves 1/4 SUSY, this explains the result since T-duality will not change this property.    For the latter, by the same token, the D2 behaves effectively as infinitely many D0 branes. So now the interaction is between one D4 and infinitely many D0-branes, placed parallel at a separation, which vanishes since there does not exist any interaction between D-branes whose dimensionality differs by 4.  Given what has been said, the two cases are not much different, both of which preserve 1/4 spacetime supersymmetries.
  
 Again the small $y$ physics can be best described in terms of the corresponding open string one-loop annulus amplitude (\ref{t-amplitude-annulus-pp'}). For the present case, it is
\bea\label{t-amplitude-annulus-43-2}
\Gamma_{p,p'} &=&  \frac{ 2\, V_{p'+ 1} \left[\det (\eta_{p'} + \hat F'_{p'})\det(\eta_{p} + \hat F_{p})\right]^{\frac{1}{2}} \sin\pi\nu_{0}}{ (8 \pi^2 \alpha')^{\frac{p' + 1}{2}}}   \int_0^\infty \frac{d t}{t^{\frac{p - 1}{2}}}  e^{- \frac{y^2 t}{2\pi\alpha'}} \nn
&\,&\times \frac{\left(\cosh\pi\nu_{0} t - \cosh\frac{\pi t}{2}\right)^{2}}{\sinh\pi\nu_{0} t \sinh\frac{\pi t}{2}} \prod_{n=1}^{\infty} Z_{n},
 \eea    
 where $Z_{n}$ can be read from (\ref{Z}) as
 \be\label{Z-43-2}
 Z_{n} = \frac{\left[ \left(1 + |z|^{4n} - 2 |z|^{2n} \cosh \pi\nu_{0} t \cosh \frac{\pi t}{2}\right)^{2} - 4 |z|^{4n} \sinh^{2} \pi\nu_{0} t \sinh^{2}\frac{\pi t}{2}\right]^{2}}{(1 - |z|^{2n})^{4} (1 - 2 |z|^{2n} \cosh\pi t + |z|^{4n})(1 - 2 |z|^{2n} \cosh2\pi\nu_{0} t + |z|^{4n})}.
 \ee 
  When $\nu_{0} = 1/2$, once again the amplitude vanishes and we have explained this in the cylinder amplitude. For $p = 4$, we have in general  $\nu_{0} \in [0, 1)$.  For large $t$, we have $Z_{n} \approx 1$ and the integrand behaves 
  \be\label{tachyon-43-2}
  \sim e^{- \frac{y^2 t}{2\pi\alpha'}}  e^{\pi\frac{|1 - 2\nu_{0}|}{2} t} = e^{- 2\pi t \left[\frac{y^{2}}{(2\pi)^{2} \alpha'} - \frac{|1 - 2 \nu_{0}|}{4}\right]},
  \ee 
  which blows up when $y < \pi \sqrt{|1 - 2 \nu_{0}|\alpha'}$, indicating again the onset of tachyonic instability.  We now consider an imaginary $\nu_{0} = i \bar\nu_{0}$ with $\bar\nu_{0} \in (0, \infty)$.
  We can use the following specific fluxes $\hat F_{p}$ and $\hat F'_{p'}$ to give a representative discussion,
  \be\label{specificfluxp43}
 ( \hat F_{p})_{01} = - (\hat F_{p})_{10} = \hat f,  \qquad (\hat F'_{p'})_{01} = - (\hat F'_{p'})_{10} = \hat f', 
  \ee
 where the rest components of both $\hat F_{p}$ and $\hat F'_{p'}$ are zero and we have also taken the $\hat g = 0$ given in (\ref{fluxp43}).  With this choice, we have
 \be\label{specificnu0-43}
 \tanh\pi \bar\nu_{0} = \frac{|\hat f - \hat f'|}{1 - \hat f' \hat f}.
 \ee
 We have now the amplitude (\ref{t-amplitude-annulus-43-2}) as
\be\label{t-amplitude-annulus-43-2-nu0-i}
\Gamma_{p,p'} =  \frac{ 2\, V_{p'+ 1} |\hat f - \hat f'|}{ (8 \pi^2 \alpha')^{\frac{p' + 1}{2}}}   \int_0^\infty \frac{d t}{t^{\frac{p - 1}{2}}}  e^{- \frac{y^2 t}{2\pi\alpha'}}  \frac{\left(\cos\pi\bar\nu_{0} t - \cosh\frac{\pi t}{2}\right)^{2}}{\sin\pi\bar\nu_{0} t \sinh\frac{\pi t}{2}} \prod_{n=1}^{\infty} Z_{n},
\ee    
where $Z_{n}$ continues to be given by (\ref{Z-43-2}) but with $\nu_{0} = i \bar\nu_{0}$.  This amplitude has now a tachyonic instability when $y < \pi \sqrt{\alpha'}$. In addition, the $\sin\pi\bar\nu_{0} t$ factor in the denominator of the integrand of the above amplitude gives again an infinite number of simple poles at $t_{k} = k/\bar\nu_{0}$ with $k = 1, 2, \cdots$ and therefore the amplitude has an imaginary part, indicating the decay of the underlying system via the so-called open string pair production.  The decay rate per unit volume of Dp$'$ brane can be computed to give
\bea\label{decayrate-43-2}
{\cal W}_{p, p'} &=& - \frac{2\, {\rm Im \Gamma}}{V_{p' + 1}}\nn
&=& \frac{4|\hat f - \hat f'|}{\bar\nu_{0}  (8 \pi^2 \alpha')^{\frac{p' + 1}{2}}} \sum_{k = 1}^{\infty} (-)^{k + 1} \left(\frac{\bar\nu_{0}}{k}\right)^{\frac{p - 1}{2}}e^{- \frac{k y^2 }{2\pi\alpha' \bar\nu_{0}}} \frac{\left[ \cosh\frac{k \pi }{2 \bar\nu_{0}} - (-)^{k}\right]^{2}}{ \sinh\frac{k \pi }{2\bar\nu_{0}}} Z_{k}(\bar\nu_{0}, \nu_{1} =1/2),\nn
\eea  
where 
\be\label{Zk-43-2}
Z_{k} (\bar\nu_{0}, \nu_{1} = 1/2) = \prod_{n = 1}^{\infty}\frac{\left[1 - 2 (-)^{k} |z_{k}|^{2n} \cosh\frac{k \pi}{2\bar\nu_{0}} + |z_{k}|^{4n}\right]^{4}}{(1 - |z_{k}|^{2n})^{6} (1 - 2 |z_{k}|^{2n} \cosh\frac{k \pi}{\bar\nu_{0}} + |z_{k}|^{4n})},
\ee  
with $|z_{k}| = e^{- k\pi/\bar\nu_{0}}$.  The open string pair production rate is given by the first $k = 1$ term of the above and it is
\be\label{pprate-43-2}
{\cal W}^{(1)}_{p, p'} =  \frac{4|\hat f - \hat f'|}{ (8 \pi^2 \alpha')^{\frac{p' + 1}{2}}}  \bar\nu_{0}^{\frac{p - 3}{2}} e^{- \frac{ y^2 }{2\pi\alpha' \bar\nu_{0}}} \frac{\left[ \cosh\frac{ \pi }{2 \bar\nu_{0}} + 1\right]^{2}}{ \sinh\frac{ \pi }{2\bar\nu_{0}}} Z_{1} (\bar\nu_{0}, \nu_{1} = 1/2),
 \ee
 where 
 \be\label{Z1-43-2}
  Z_{1} (\bar\nu_{0}, \nu_{1} = 1/2) = \prod_{n = 1}^{\infty}\frac{\left[1 + 2 |z_{1}|^{2n} \cosh\frac{\pi}{2\bar\nu_{0}} + |z_{1}|^{4n}\right]^{4}}{(1 - |z_{1}|^{2n})^{6} (1 - 2 |z_{1}|^{2n} \cosh\frac{ \pi}{\bar\nu_{0}} + |z_{1}|^{4n})}.
\ee  
This pair production rate is, as expected, the same as that given in (\ref{pprate-5-2}).  The same discussion applies here, too. Note also that the decay rate (\ref{decayrate-43-2}) and the pair production rate (\ref{pprate-43-2}) are just special cases of (\ref{decayrate-pp'}) and (\ref{pprate-pp'}) when we take $\nu_{2} = 0$ and $\nu_{1} = 1/2$, as expected.  \\

\noindent
{\bf The $p = 2$ case:}  This is the last case we will discuss in this subsection.  The D0 brane cannot carry any worldvolume flux.  However, by the same token as before, we can have the following extension  as
\be\label{extendedfluxp'2}
{\hat F}'_{2} = \left(\begin{array}{ccc}
0&0&0\\
0&0&\hat g'\\
0&-\hat g'&0\end{array}\right),
\ee
where we will set $\hat g' \to \infty$ at the end of computations.  For this case, we will consider the most general D2 worldvolume flux as an example.  This flux can be expressed as
\be\label{fluxp2}
\hat F_{2} =  \left(\begin{array}{ccc}
0&\hat f_{1}& \hat f_{2}\\
- \hat f_{1}&0&\hat g\\
- \hat f_{2}&-\hat g&0\end{array}\right).
\ee
Using (\ref{matrixlw}), we have\footnote{We here take this simple case as a direct check of the trick used. For the general flux (\ref{fluxp2}) on D2, we can compute its M-matrix using (\ref{mmatrix}) as $M = (M_{\alpha}\,^{\beta},  - \delta_{i}\,^{j})$ with $\alpha, \beta = 0, 1, 2$ and $i, j = 3, 4, \cdots 9$, where
\be\label{mmd2}
M_{\alpha}\,^{\beta} = (1 + \hat g^{2} - \hat f_{1}^{2} - \hat f_{2}^{2})^{-1}\left(\begin{array}{ccc}
1 + \hat g^{2} + \hat f_{1}^{2} + \hat f_{2}^{2} & - 2 (\hat f_{1} + \hat f_{2} \hat g) & 2 (\hat f_{1} \hat g - \hat f_{2})\\
2 (\hat f_{2} \hat g - \hat f_{1})& 1 - \hat g^{2} + \hat f_{1}^{2} - \hat f_{2}^{2} & 2 (\hat f_{1} \hat f_{2} - \hat g)\\
- 2(\hat f_{2} + \hat f_{1} \hat g) & 2 (\hat f_{1} \hat f_{2} + \hat g) & 1 - \hat g^{2} - \hat f_{1}^{2} + \hat f_{2}^{2}\end{array}\right),
\ee
while for D0 brane, we have $M' = (1, - \delta_{i'}\,^{j'})$ with $i', j' = 1, 2, \cdots 9$. So we have 
\be\label{Wpp'}
W = M M'^{T} = \left(\begin{array}{cc} 
w_{\alpha}\,^{\beta}&0\\
0& \mathbb{I}_{7\times 7}\end{array}\right),
\ee
where
\be\label{w20}
w_{\alpha}\,^{\beta} = (1 + \hat g^{2} - \hat f_{1}^{2} - \hat f_{2}^{2})^{-1}\left(\begin{array}{ccc}
1 + \hat g^{2} + \hat f_{1}^{2} + \hat f_{2}^{2} &  2 (\hat f_{1} + \hat f_{2} \hat g) & - 2 (\hat f_{1} \hat g - \hat f_{2})\\
2 (\hat f_{2} \hat g - \hat f_{1})& - (1 - \hat g^{2} + \hat f_{1}^{2} - \hat f_{2}^{2}) & - 2 (\hat f_{1} \hat f_{2} - \hat g)\\
- 2(\hat f_{2} + \hat f_{1} \hat g) & - 2 (\hat f_{1} \hat f_{2} + \hat g) &- (1 - \hat g^{2} - \hat f_{1}^{2} + \hat f_{2}^{2})\end{array}\right),
\ee
which is nothing but the limit of $w_{\alpha}\,^{\beta}$ given in (\ref{lmwp2-2}) when we set $\hat g' \to \infty$.  This confirms that the trick used in obtaining the corresponding eigenvalues of $w$ works indeed.  
 }

\be\label{lmwp2-2} 
 w_{\alpha}\,^{\beta} = \left(\begin{array}{ccc}
\frac{1 + \hat g^{2} + \hat f_{1}^{2} + \hat f^{2}_{2}}{1 + \hat g^{2} - \hat f_{1}^{2} - \hat f^{2}_{2}} & -  \frac{2(\hat f_{1} + \hat f_{2} \hat g)(1 - \hat g'^{2}) - 4 (\hat f_{2} - 
\hat f_{1} \hat g)\hat g'}{(1 + \hat g^{2} - \hat f_{1}^{2} - \hat f^{2}_{2})(1 + \hat g'^{2})} & - \frac{4 (\hat f_{1} + \hat f_{2}\hat g) \hat g' + 2 (\hat f_{2} - \hat f_{1} \hat g)(1 - \hat g'^{2})}{(1 + \hat g^{2} - \hat f_{1}^{2} - \hat f^{2}_{2})(1 + \hat g'^{2})}\\
- \frac{2 (\hat f_{1} - \hat f_{2}\hat g)}{1 + \hat g^{2} - \hat f_{1}^{2} - \hat f^{2}_{2}}& \frac{(1 - \hat g^{2} + \hat f_{1}^{2} - \hat f^{2}_{2})(1 -\hat g'^{2}) + 4 (\hat g - \hat f_{1}\hat f_{2})\hat g'}{(1 + \hat g^{2} - \hat f_{1}^{2} - \hat f^{2}_{2})(1 + \hat g'^{2})}&  \frac{ 2 (1 - \hat g^{2} + \hat f_{1}^{2} - \hat f^{2}_{2})\hat g' - 2 (\hat g - \hat f_{1}\hat f_{2})(1 - \hat g'^{2})}{(1 + \hat g^{2} - \hat f_{1}^{2} - \hat f^{2}_{2})(1 + \hat g'^{2})}\\
- \frac{2 (\hat f_{2} + \hat f_{1} \hat g)}{1 + \hat g^{2} - \hat f_{1}^{2} - \hat f^{2}_{2}}& \frac{2 (\hat g + \hat f_{1}\hat f_{2})(1 - \hat g'^{2}) - 2 (1 - \hat g^{2} - \hat f_{1}^{2} + \hat f^{2}_{2})\hat g'}{(1 + \hat g^{2} - \hat f_{1}^{2} - \hat f^{2}_{2})(1 + \hat g'^{2})}&\frac{(1 - \hat g^{2} - \hat f_{1}^{2} + \hat f^{2}_{2})(1 - \hat g'^{2}) + 4 (\hat g + \hat f_{1}\hat f_{2})\hat g'}{(1 + \hat g^{2} - \hat f_{1}^{2} - \hat f^{2}_{2})(1 + \hat g'^{2})}\end{array}\right).
\ee
One can check explicitly that the above $w$ has one eigenvalue unity and the other two $\lambda_{0}$ and $\lambda^{-1}_{0}$ satisfy
\be\label{eigenvp2-2}
\lambda_{0} + \lambda^{-1}_{0} = \frac{2 (\hat g^{2} + \hat f_{1}^{2} + \hat f_{2}^{2} - 1)}{1 + \hat g^{2} - \hat f_{1}^{2} - \hat f_{2}^{2}},
\ee
where we have taken $\hat g' \to \infty$. Setting $\lambda_{0} = e^{2\pi i\nu_{0}}$, we have
\be\label{nu0p2-2}
\tan\pi\nu_{0} = \frac{\sqrt{1 - \hat f_{1}^{2} - \hat f_{2}^{2}}}{\hat g}.
\ee
We have two cases to consider: 1) $\hat f_{1}^{2} + \hat f_{2}^{2} < 1$, 2) $ 1 < \hat f_{1}^{2} + \hat f_{2}^{2} < 1 + \hat g^{2}$.  For the first case, $\nu_{0} \in [0, 1)$. If $\hat g$ is finite, the discussion goes the same as the pure magnetic case of $p = p' = 2$ discussed in the previous section and we will not repeat it here and refer there for detail. If there is no magnetic flux, i.e., $\hat g  = 0$, we have then $\nu_{0} = 1/2$. The closed string cylinder amplitude can be obtained from (\ref{t-amplitude-cylinder-pp'}) with $\nu_{2} = \nu_{1} = 0$ and $\nu_{0} = 1/2$ as
\be\label{t-amplitude-cylinder-20}
\Gamma_{2, 0} = \frac{ 2 \, V_{1} \sqrt{1  - \hat f^{2}_{1} - \hat f_{2}^{2}} }{(8 \pi^2 \alpha')^{\frac{1}{2}}}   \int_0^\infty \frac{d t}{t^{\frac{7}{2}}}  e^{- \frac{y^2}{2\pi\alpha' t}}  \prod_{n=1}^{\infty} \frac{(1 + |z|^{4n})^{4}}{(1 - |z|^{2n})^{4} (1 - |z|^{4n})^{2}},
 \ee
where we also use (\ref{C}) for $C_{n}$. The integrand of this amplitude has a potential divergence but has no sign ambiguity for small $t$, indicating a potential tachyonic instability but no open string pair production, even though there exist applied electric fluxes.  To see both of these clearly, we need the corresponding open string one-loop annulus amplitude which can be read from (\ref{t-amplitude-annulus-pp'}) as
\bea\label{t-amplitude-20}
\Gamma_{2, 0} &=& \frac{ 2 \, V_{1} \sqrt{1 - \hat f^{2}_{1} - \hat f_{2}^{2}} }{(8 \pi^2 \alpha')^{\frac{1}{2}}}   \int_0^\infty \frac{d t}{t^{\frac{3}{2}}}  e^{- \frac{y^2 t}{2\pi\alpha'}} \frac{\left(\cosh\frac{\pi t}{2} - 1\right)^{2}}{\sinh \frac{\pi  t}{2}} \nn
&\,&\times  \prod_{n=1}^{\infty} \frac{[1 - 2 |z|^{2n} \cosh\frac{\pi t}{2} + |z|^{4n}]^{4}}{(1 - |z|^{2n})^{6} [1 - 2 |z|^{2n} \cosh\pi t + |z|^{4n}]},
 \eea    
where we have used (\ref{Z}) for $Z_{n}$.  For large $t$, the integrand of the above behaves like
\be\label{tachyonic-20}
\sim e^{- \frac{y^2 t}{2\pi\alpha'}} e^{\frac{\pi t}{2}} = e^{- 2\pi t\left[\frac{y^{2}}{(2\pi)^{2}\alpha'} - \frac{1}{4}\right]},
\ee
which blows up when $y < \pi \sqrt{\alpha'}$, indicating the onset of tachyonic instability.  The integrand is regular at $t \to 0$ and has no simple poles and so as anticipated there is no open string pair production even though there are applied electric fluxes on the D2 brane.  The explanation for this is similar to that a single D-brane carrying a constant electric flux 
cannot give rise to open string pair productions.  Here the story is that the two ends of virtual open string and virtual anti open string attracted on the D2 can be pulled away while the other two ends on the D0 cannot. So the electric fields applied can only stretch the virtual open string and the virtual anti open string to certain extend but cannot separate them even if we take $1 - \hat f_{1}^{2} - \hat f_{2}^{2} = \epsilon \to 0^{+}$. With $\hat g = 0$, from (\ref{eigenvp2-2}), we always have $\nu_{0} = 1/2$ and it holds true even in  the limit $1 - \hat f_{1}^{2} - \hat f_{2}^{2} = \epsilon \to 0^{+}$. Due to the tachyonic instability when $y < \pi \sqrt{\alpha'}$, we need to have $y > \pi \sqrt{\alpha'}$ to validate the amplitude computations.  Once this holds, the effective tension on the virtual open strings is less than the critical one even if we take $1  - \hat f_{1}^{2} - \hat f_{2}^{2} \to 0^{+}$. So this limiting tension cannot break the open strings and therefore there is no open string pair production. 

We now move to the second case for which we cannot set $\hat g$ vanish.   So we have $ \hat g^{2} < \hat g^{2} + \hat f^{2}_{1} + \hat f^{2}_{2} -  1 < 2 \hat g^{2}$ and $0 < \hat g^{2} + 1 - \hat f^{2}_{1} - \hat f^{2}_{2} < \hat g^{2}$. From (\ref{eigenvp2-2}), this must imply that $\nu_{0}$ is imaginary, i.e., $\nu_{0} = i \bar \nu_{0}$ with $\bar \nu_{0} \in (0, \infty)$. This can also be seen directly from (\ref{nu0p2-2}) and it is now
\be\label{barnu0p2-2}
\tanh\pi\bar\nu_{0} = \frac{\sqrt{\hat f_{1}^{2} + \hat f_{2}^{2} - 1}}{|\hat g|}.
\ee
The present closed string cylinder amplitude can be read from (\ref{t-amplitude-cylinder-pp'}) with $\nu_{2} = \nu_{1} = 0$ and $\nu_{0} = i \bar\nu_{0}$ as
\bea\label{t-amplitude-cylinder-20-barnu0}
 \Gamma_{2, 0} &=& \frac{ 2 \, V_{1} \,\sqrt{1 + \hat g^{2} - \hat f^{2}_{1} - \hat f^{2}_{2}}\, \left(\cosh\pi\bar\nu_{0} - 1\right)^{2}}{(8 \pi^2 \alpha')^{\frac{1}{2}}} 
  \int_0^\infty \frac{d t}{t^{\frac{7}{2}}}  e^{- \frac{y^2}{2\pi\alpha' t}} \nn
  &\,&\times \prod_{n=1}^{\infty} \frac{\left[1 - 2 |z|^{2n} \cosh\pi\bar\nu_{0} + |z|^{4n}\right]^{4}}{(1 - |z|^{2n})^{6} (1 - 2 |z|^{2n} \cosh2\pi\bar\nu_{0} + |z|^{4n})},
 \eea
where we have used (\ref{C}) for $C_{n}$.  As before, the large separation interaction is obviously attractive but the integrand for small $t$ has an ambiguity of its sign  in addition to
a potential singularity.  The sign ambiguity implies a decay of the underlying system via the open string pair production while the potential singularity implies a potential tachyonic instability.  To check both of these explicitly, we need to examine the corresponding open string one-loop annulus amplitude which can be read from (\ref{t-amplitude-annulus-pp'}) as
\bea\label{t-amplitude-annulus-20-bar}
\Gamma_{2, 0} &=& \frac{2\, V_{1}\,\sqrt{ \hat f^{2}_{1} + \hat f^{2}_{2} - 1}}{ (8 \pi^2 \alpha')^{\frac{1}{2}}}   \int_0^\infty \frac{d t}{t^{\frac{3}{2}}}  e^{- \frac{y^2 t}{2\pi\alpha'}} \frac{(1 - \cos\pi\bar\nu_{0}t)^{2}}{\sin\pi\bar\nu_{0}t}\nn
&\,&\times \prod_{n=1}^{\infty} \frac{\left[1 - 2 |z|^{2n} \cos\pi\bar\nu_{0} t + |z|^{4n}\right]^{4}}{(1 - |z|^{2n})^{6} \left[1 - 2 |z|^{2n} \cos 2\pi \bar\nu_{0} t + |z|^{4n}\right]},
 \eea     
 where we have used (\ref{Z}) for $Z_{n}$.  For large $t$, the integrand of this annulus amplitude does not have a blowing up behavior and therefore there is no potential tachyonic singularity.  However, the integrand does have an infinite number of simple poles at  $t_{k} = (2k - 1)/\bar\nu_{0}$ with $k = 1, 2, \cdots$, indicating the decay of the system via the open 
 string pair production.  The decay rate and the open string pair production rate can be read from (\ref{decayrate-pp'}) and (\ref{pprate-pp'}), respectively, with $\nu_{2} = \nu_{1} = 0$, as
 \be\label{decayrate-20}
 {\cal W}_{p, p'} =\frac{ 16 \,\bar\nu_{0} \sqrt{\hat f^{2}_{1} + \hat f^{2}_{2} - 1}} { (8 \pi^2 \alpha')^{\frac{1}{2}}}  \sum_{k = 1}^{\infty} \left(\frac{\bar\nu_{0}}{2k - 1}\right)^{\frac{p - 3}{2}} 
 e^{- \frac{(2k -1) y^2 }{2\pi\alpha' \bar\nu_{0}}}\,\prod_{n =1}^{\infty} \frac{\left(1 + |z_{2k - 1}|^{2n}\right)^{8}}{\left(1 - |z_{2k - 1}|^{2n}\right)^{8}},
\ee
where we have used (\ref{Zk}) for $Z_{k}$ and $|z_{k}| = e^{- k \pi /\bar\nu_{0}}$, and 
\be\label{pprate-20}
{\cal W}^{(1)}_{p. p'} = \frac{ 16 \, \sqrt{\hat f^{2}_{1} + \hat f^{2}_{2} - 1}} { (8 \pi^2 \alpha')^{\frac{1}{2}}}  \bar\nu_{0}^{\frac{p - 1}{2}} 
 e^{- \frac{ y^2 }{2\pi\alpha' \bar\nu_{0}}}\,\prod_{n =1}^{\infty} \frac{\left(1 + |z_{1}|^{2n}\right)^{8}}{\left(1 - |z_{1}|^{2n}\right)^{8}}.
\ee
Both of the rates blow up when $\bar \nu_{0} \to \infty$ for which $\hat f^{2}_{1} + \hat f^{2}_{2} -  1 \to \hat g^{2}$, the critical limit. 

In the above, we have an interesting thing happening.  Note that the above discussion for $\hat f^{2}_{1} + \hat f^{2}_{2} < 1$ holds true also for $\hat g \neq 0$.  For given $\hat g \neq 0$, there is a potential open string tachyonic instability but no open string pair production if $\hat f^{2}_{1} + \hat f^{2}_{2} < 1$ while there is open string pair production but no open string tachyonic instability if $\hat f^{2}_{1} + \hat f^{2}_{2} > 1$\footnote{\label{fn10}Note that in both cases we need to have $\hat f^{2}_{1} + \hat f^{2}_{2} < 1 + \hat g^{2}$ and for the former case, it satisfies trivially.}.  For the former,  the electric fluxes representing the respective delocalized fundamental strings (see footnote (\ref{fn4})) have no interaction with the D0 brane \cite{Ouyang:2014bha}. So their presence just gives certain modifications of the pure magnetic case of the underlying system but not its characteristic behavior, as discussed above.  So a potential tachyonic instability is expected when the brane separation reaches the distance determined by the tachyonic shift. For the latter, we have to admit that we don't have a better explanation of it except for the following observation. Before that, we would also like to point out that when $\hat f^{2}_{1} + \hat f^{2}_{2} = 1$, the above amplitudes and rates computed all vanish. 

For this, let us examine the matrix $w$ given in (\ref{lmwp2-2}) while keeping $\hat g'$ large.  Note that the ${\rm tr} w$ is a D2 worldvolume Lorentz invariant and 
the eigenvalue equation (\ref{eigenvp2-2}) is now replaced by
\be\label{eigenvp2-2-new}
\lambda_{0}  + \lambda^{-1}_{0}  = {\rm tr} w - 1 = 2 \frac{(1 + \hat g \hat g')^{2} - (\hat g - \hat g')^{2} + (1 + \hat g'^{2})(\hat f^{2}_{1} + \hat f^{2}_{2})}{(1 + \hat g'^{2})(1 + \hat g^{2} - \hat f^{2}_{1} - \hat f^{2}_{2})},
\ee 
which gives (\ref{eigenvp2-2}) if we send $\hat g' \to \infty$.  For the present purpose, we keep  $\hat g'$ large and take the limit $\hat g' \to \infty$ only at the end of the discussion. 
If we set $\lambda_{0} = e^{2\pi i \nu_{0}}$, we have from the above,
\bea\label{nu0p2-2-new}
\cos\pi\nu_{0} &=& \frac{1 + \hat g \hat g'}{\sqrt{(1 + \hat g'^{2}) (1 + \hat g^{2} - \hat f^{2}_{1} - \hat f^{2}_{2})}}, \nn
 \sin\pi\nu_{0} &=& \frac{\sqrt{(\hat g - \hat g')^{2} - (1 + \hat g'^{2})(\hat f^{2}_{1} + \hat f^{2}_{2})}}{\sqrt{(1 + \hat g'^{2}) (1 + \hat g^{2} - \hat f^{2}_{1} - \hat f^{2}_{2})}}.
\eea
Note that we have a few Lorentz invariants of D2 brane worldvolume: $\hat F_{\alpha\beta} \hat F^{\alpha\beta} = 2[\hat g^{2} - (\hat f_{1}^{2} + \hat f_{2}^{2})],   \hat F'_{\alpha\beta} \hat F'^{\alpha\beta} = 2 \hat g'^{2},  \hat F_{\alpha\beta} \hat  F'^{\alpha\beta} = 2 \hat g \hat g'$.  Therefore the numerator on the right side of $\sin \pi \nu_{0}$ in (\ref{nu0p2-2-new}) is also Lorentz invariant. In other words, the $\nu_{0}$ is a Lorentz invariant.  Let us now examine this numerator  which can be rewritten as $[\hat g'^{2} (1 - \hat f^{2}_{1} - \hat f^{2}_{2}) + \hat g^{2} - 2 \hat g' \hat g - (\hat f^{2}_{1} + \hat f^{2}_{2})]^{1/2}$. Due to the $\hat g'$-factor in the denominator, we need to have a $\hat g'$-factor in the numerator to give a non-vanishing $\nu_{0}$ and the numerator becomes $[(1 - \hat f_{1}^{2} - \hat f_{2}^{2}) \hat g'^{2}]^{1/2}$ when we take $\hat g' \to \infty$.  Note  that $(1 - \hat f_{1}^{2} - \hat f_{2}^{2}) \hat g'^{2}$ is also a Lorentz invariant of D2 worldvolume since it is related to $(\epsilon_{\alpha\beta\gamma} \epsilon^{\alpha\beta\delta} + \hat F_{\alpha_{1}\beta_{1}}\epsilon^{\alpha_{1}\beta_{1}\delta} \hat F^{\alpha_{2}\beta_{2}}\,_{\gamma}) \hat F'^{\gamma\tau}\hat F_{\tau\delta} = 4 \hat g'^{2} (1 - \hat f_{1}^{2} - \hat f_{2}^{2})$. So it is clear now that the sign of $1 - \hat f^{2}_{1} - \hat f^{2}_{2}$ determines the nature of $\nu_{0}$, real or imaginary!  When $\hat f^{2}_{1} + \hat f^{2}_{2} < 1$, $\nu_{0}$ is real and the underlying system with $\hat g' \to 
\infty$ resembles a pure magnetic case.  When $\hat f_{1}^{2} + \hat f_{2}^{2} =  1$, $\nu_{0}$ vanishes. While $\hat f_{1}^{2} + \hat f_{2}^{2} > 1$,  the $\nu_{0}$ is imaginary and the underlying system with $\hat g' \to \infty$ has a long-range attractive interaction but at small brane separation the amplitude has a sign ambiguity, indicating a decay via the open string pair production as described above. 
 
\subsection{The $p - p' = 4$ case\label{subsection4.2}}
For $p \le 6$, we have only three cases to consider in this subsection: 1) $p = 6, p' = 2$; 2) $p = 5, p' = 1$ and 3) $p = 4, p' = 0$. The extension of $\hat F'_{p'}$ on the Dp$'$ brane
 to $\hat F'_{p}$ given in (\ref{extendedfluxp'}) in the present context takes the following form
\be\label{extendedfluxp'-4}
(\hat F'_{p})_{\alpha\beta} = \left(\begin{array}{ccccc}
(\hat F'_{p'})_{\alpha'\beta'} &&&&\\
&0&\hat g'_{1}&&\\
&-\hat g'_{1}&0&&\\
&&&0&\hat g'_{2}\\
&&&-\hat g'_{2}&0
\end{array}\right),
\ee
where we need to take both $\hat g'_{1} \to \infty$ and $\hat g'_{2} \to \infty$ at the end of relevant computations. \\
 
 \noindent
 {\bf The $p = 6$ case:}  For a general flux $\hat F_{6}$ on D6,  even with the above extension (\ref{extendedfluxp'-4}) for $\hat F'_{2}$ on D2,  the characteristic behavior of the closed string cylinder amplitude or the corresponding open string one-loop annulus amplitude is similar to that for the $p = p' = 6$ discussed in the previous section. We here specify the $\hat F_{6}$ to the following form along with the extension of $\hat F'_{2}$ as,
\be\label{flux6-4}
\hat F_{\alpha\beta} =  \left(\begin{array}{ccccc}
(\hat F_{2})_{\alpha'\beta'} &&&&\\
&0&\hat g_{1}&&\\
&-\hat g_{1}&0&&\\
&&&0&\hat g_{2}\\
&&&-\hat g_{2}&0
\end{array}\right), \quad \hat F'_{\alpha\beta} =  \left(\begin{array}{ccccc}
(\hat F'_{2})_{\alpha'\beta'} &&&&\\
&0&\hat g'_{1}&&\\
&-\hat g'_{1}&0&&\\
&&&0&\hat g'_{2}\\
&&&-\hat g'_{2}&0
\end{array}\right).
\ee
We have therefore 
\be\label{nu1nu2-62-4}
\tan\pi\nu_{1} = \frac{1}{\hat g_{1}}, \qquad \tan\pi\nu_{2} = \frac{1}{\hat g_{2}}, 
 \ee
 where we have taken $\hat g'_{1} \to \infty$ and $\hat g'_{2} \to \infty$.  For general $\hat g_{1}$ and $\hat g_{2}$, the discussion continues to be the same as that of the $p = p' = 6$ case. We further specify to the case of both $\hat g_{1} = 0$ and $\hat g_{2} = 0$ for which $\nu_{1} = \nu_{2} = 1/2$. Now the closed string cylinder amplitude can be read from (\ref{t-amplitude-cylinder-pp'}) with $\nu_{1} = \nu_{2} = 1/2$ as
\bea\label{t-amplitude-cylinder-62-4}
\Gamma_{6,2} &=& - \frac{V_{3} \left[\det (\eta_{2} + \hat F'_{2})\det(\eta_{2} + \hat F_{2})\right]^{\frac{1}{2}}\, \sin^{2} \pi\nu_{0}  }{(8 \pi^2 \alpha')^{\frac{3}{2}}}  \int_0^\infty \frac{d t}{t^{\frac{3}{2}}}  e^{- \frac{y^2}{2\pi\alpha' t}}\nn
 &\,& \times   \prod_{n=1}^{\infty} \frac{\left[(1 + |z|^{4n})^{2} - 4 |z|^{4n} \cos^{2}\pi\nu_{0}\right]^{2}}{(1 - |z|^{2n})^{2} (1 + |z|^{2n})^{4} (1 - 2 |z|^{2n} \cos2\pi\nu_{0} + |z|^{4n})},
 \eea
where we have used (\ref{C}) for $C_{n}$.  The amplitude vanishes for $\nu_{0} = 0$ when $\hat F_{2} = \hat F'_{2} = 0$. This is consistent with the fact that there is no interaction between a Dp and a Dp$'$, placed parallel at a separation,  with $p - p' = 4$. When $\nu_{0} \in (0, 1)$,   this amplitude is negative since all factors in the integrand are positive.  So the interaction is repulsive.  This is also consistent with the conclusion reached for $p = p' = 6$ in the previous section since we have here $\nu_{0} + \nu_{1} > \nu_{2}$ if $\nu_{0} \le 1/2$ or $\nu_{1} + \nu_{2} > \nu_{0}$ if $\nu_{0} > 1/2$, i.e., when the possible largest one among the three $\nu_{0}, \nu_{1}, \nu_{2}$ is less than the sum of the remaining two.   We therefore don't expect to have a potential open string tachyonic instability which is obvious from the above amplitude.

When $\nu_{0}$ is imaginary, i.e., $\nu_{0} = i \bar\nu_{0}$ with $\bar\nu_{0} \in (0, \infty)$, the large separation interaction becomes attractive. For small $y$, the small $t$ becomes important in the integration. But for small $t$,  the factor $(1 - 2 |z|^{2n} \cosh2\pi\bar \nu_{0} + |z|^{4n})$ in the denominator of the infinite product in the integrand can be negative and this gives the ambiguity about the sign of the integrand.  As before, we expect the decay of the underlying system via the so-called open string pair production.  

Either of the above will become manifest if we look from the corresponding open string one-loop annulus amplitude which can be read from (\ref{t-amplitude-annulus-pp'}) in the present context as  
\bea\label{t-amplitude-annulus-62-4}
 \Gamma_{6, 2} &=&  \frac{4 V_{3} \left[\det (\eta_{2} + \hat F'_{2})\det(\eta_{2} + \hat F_{2})\right]^{\frac{1}{2}} \sin\pi\nu_{0}}{ (8 \pi^2 \alpha')^{\frac{3}{2}}}   \int_0^\infty \frac{d t}{t^{\frac{3}{2}}}  e^{- \frac{y^2 t}{2\pi\alpha'}}  \nn
 &\,& \times  \frac{\sinh^{2}\frac{\pi\nu_{0} t}{2} \left(\cosh^{2}\frac{\pi \nu_{0} t}{2} -   \cosh^{2} \frac{\pi t}{2}\right)}{\sinh \pi \nu_{0} t   \sinh^{2} \frac{\pi t}{2}} \prod_{n=1}^{\infty} Z_{n},
 \eea      
 where $Z_{n}$ can be read from (\ref{Z}) as
 \bea\label{Z-62-4}
 Z_{n} &=& \frac{\left[1 - 2 |z|^{2n} \cosh\pi\nu_{0} t  + |z|^{4n}\right]^{2}}{(1 - |z|^{2n})^{2} (1 - 2 |z|^{2n} \cosh\pi t + |z|^{4n})^{2} }\nn
 &\,& \times \frac{ \left[\left(1 + |z|^{4n} - 2 |z|^{2n} \cosh\pi\nu_{0} t \cosh\pi t\right)^{2} - 4 |z|^{4n} \sinh^{2}\pi\nu_{0} t \sinh^{2}\pi t \right]}{(1 - 2 |z|^{2n} \cosh2 \pi \nu_{0} t + |z|^{4n})}.  \eea
Since $\nu_{0}\in [0, 1)$,  the interaction is always repulsive and so we don't expect a tachyonic instability. For large $t$, $Z_{n} \approx 1$ and the integrand behaves like
\be\label{tachyon-62-4}
\sim   e^{- \frac{y^2 t}{2\pi\alpha'}},
\ee  
 which indicates no tachyonic instability as expected.   When $\nu_{0} = i \bar\nu_{0}$ with $\bar\nu_{0} \in (0, \infty)$,  the integrand has an infinite number of simple poles 
 occurring at $t_{k} = k/\bar\nu_{0}$ with $k = 1, 2, \cdots$, indicating the decay of the underlying system via the so-called open string pair production. The decay rate and the open string pair production rate are given by the (\ref{decayrate-pp'}) and (\ref{pprate-pp'}), respectively, with $\nu_{1} = \nu_{2} = 1/2$, and are not given here explicitly.  In particular, we would like to point out that there is no open string enhancement here even for small $\bar\nu_{0}$ since it is in general given by $e^{\pi |\nu_{1} - \nu_{2}|/\bar\nu_{0}}$ which is unity here.  However, for general non-vanishing  fluxes $\hat g_{1}$ and $\hat g_{2}$, this enhancement can still be significant.\\
 
 \noindent 
 {\bf The $p = 5$ case:}  Here $p' = 1$.  The extension of a general flux $\hat F'_{1}$ on D1 to $\hat F'_{5}$, following (\ref{extendedfluxp'}), is  
\be\label{extendedflux-1-4}
 (\hat F'_{5})_{\alpha\beta} = \left(\begin{array}{cccccc}
0 & \hat f' &&&&\\
- \hat f'&0& &&&\\
&&0& \hat g'_{1}&\\
&&- \hat g'_{1}&0&&\\
&&&&0&\hat g'_{2}\\
&&&&-\hat g'_{2}&0
\end{array}\right),
\ee
where as before we take both $\hat g'_{1} \to \infty$ and $\hat g'_{2} \to \infty$ at the end of computations.  For a general $\hat F_{5}$ on D5,  the discussion goes the same as what has been discussed for the $p = p' = 5$ case in the previous section.  An example of the following flux on D5 
\be\label{sampleflux5-4}
 (\hat F_{5})_{\alpha\beta} = \left(\begin{array}{cccccc}
0 & \hat f &&&&\\
- \hat f&0& &&&\\
&&0& \hat g_{1}&\\
&&- \hat g_{1}&0&&\\
&&&&0&\hat g_{2}\\
&&&&-\hat g_{2}&0
\end{array}\right),
\ee
along with the extension (\ref{extendedflux-1-4}) corresponds just to a special case of what has been discussed in great detail in \cite{Jia:2018mlr}.  So we refer there for detail and will not repeat it here.  For this, it is also essentially the same as the $\nu_{0}$ being the imaginary case of the D6-D2 system discussed above.\\

\noindent
{\bf The $p = 4$ case:} Here $p' = 0$.  The extension of no flux on D0 to $\hat F'_{4}$, following (\ref{extendedfluxp'}), as  
 \be\label{extendedflux0-4}
(\hat F'_{4})_{\alpha\beta} = \left(\begin{array}{ccccc}
0 &&&&\\
&0&\hat g'_{1}&&\\
&-\hat g'_{1}&0&&\\
&&&0&\hat g'_{2}\\
&&&-\hat g'_{2}&0
\end{array}\right),
\ee
where once again we take both $\hat g'_{1} \to \infty$ and $\hat g'_{2} \to \infty$ at the end of computations. For a general flux on D4, the discussion will go the same as that for the 
$p = p' = 4$ case given in the previous section. We could give some sample discussion for the present extended flux (\ref{extendedflux0-4}) and some special choice of flux on D4 but this will not give anything new. The closed string cylinder amplitude, the open string annulus one, the potential decay rate and the potential open string pair production rate can all be read from the corresponding from (\ref{t-amplitude-cylinder-pp'}),  (\ref{t-amplitude-annulus-pp'}), (\ref{decayrate-pp'}) and (\ref{pprate-pp'}), respectively, for the present consideration. So we omit to write each of them explicitly here.
 
 \subsection{The $p - p' = 6$ case\label{subsection4.3}}
This is the last case to be considered in this section.  For $p \le 6$, we have only one case to consider, namely, $p = 6, p' = 0$. The extension of no flux on D0 to $\hat F'_{6}$, following (\ref{extendedfluxp'}), as 
 \be\label{extendedflux0-6}
(\hat F'_{6})_{\alpha\beta} = \left(\begin{array}{ccccccc}
0 &&&&&&\\
&0&\hat g'_{0}&&&&\\
&-\hat g'_{0}&0&&&&\\
&&&0&\hat g'_{1}&&\\
&&&-\hat g'_{1}&0&&\\
&&&&&0&\hat g'_{2}\\
&&&&&-\hat g'_{2}&0
\end{array}\right),
\ee
where similarly we need to take $\hat g'_{0} \to \infty, \hat g'_{1} \to \infty$ and $\hat g'_{2} \to \infty$ at the end of computations. As before, for a general flux $\hat F_{6}$ on D6, the relevant discussion goes more or less the same as that for the $p = p' = 6$ case discussed in the previous section and we will not repeat it here.  We could give a sample discussion for the following flux on D6,
\be\label{flux6-6}
(\hat F_{6})_{\alpha\beta} = \left(\begin{array}{ccccccc}
0 &\hat f_{1}&\hat f_{2}&&&&\\
-\hat f_{1}&0&\hat g_{0}&&&&\\
-\hat f_{2}&-\hat g_{0}&0&&&&\\
&&&0&\hat g_{1}&&\\
&&&-\hat g_{1}&0&&\\
&&&&&0&\hat g_{2}\\
&&&&&-\hat g_{2}&0
\end{array}\right),
\ee
and this is still a rather general case of the more general discussion for the $p = p' = 6$ case mentioned above but now with
\be\label{nu0nu1nu2-60-6}
\tan\pi\nu_{0} = \frac{\sqrt{1 - \hat f^{2}_{1} - \hat f^{2}_{2}}}{\hat g_{0}}, \quad \tan\pi\nu_{1} = \frac{1}{\hat g_{1}}, \quad \tan\pi\nu_{2} = \frac{1}{\hat g_{2}},
\ee
where we have taken $\hat g'_{0} \to \infty, \hat g'_{1} \to \infty$ and $\hat g'_{2} \to \infty$.   If we further set $\hat g_{1} = \hat g_{2} = 0$,  we have $\nu_{1} = \nu_{2} = 1/2$. For this special case, the closed string cylinder amplitude can be read from (\ref{t-amplitude-cylinder-pp'}) as
\bea\label{t-amplitude-cylinder-60-6}
\Gamma_{6,0} &=& - \frac{ V_{1} \sqrt{1 + \hat g^{2}_{0} - \hat f^{2}_{1} - \hat f^{2}_{2}}\, \sin^{2}\pi\nu_{0} } { 2 (8 \pi^2 \alpha')^{\frac{1}{2}}}  \int_0^\infty \frac{d t}{t^{\frac{3}{2}}}  e^{- \frac{y^2}{2\pi\alpha' t}} \nn
 &\,& \times  \prod_{n=1}^{\infty} \frac{\left[(1 + |z|^{4n})^{2} - 4 |z|^{4n} \cos^{2}\pi\nu_{0}\right]^{2}}{(1 - |z|^{2n})^{2}
 (1 + |z|^{2n})^{4} (1 - 2 |z|^{2n} \cos 2\pi\nu_{0} + |z|^{4n})},
 \eea
where we have used (\ref{C}) for $C_{n}$. Except for the overall constant factor, this amplitude looks essentially  the same as the corresponding one for the $p = 6,\, p' = 2$ case discussed in subsection \ref{subsection4.2}. For real and non-vanishing $\nu_{0}$, which requires $\hat f^{2}_{1} + \hat f^{2}_{2} < 1$ from (\ref{nu0nu1nu2-60-6}),  the amplitude is negative and therefore the interaction is repulsive again since here $\nu_{0} + \nu_{1} > \nu_{2}$ if $\nu_{0} \le 1/2$ or $\nu_{1} + \nu_{2} > \nu_{0}$ if $\nu_{0} > 1/2$, i.e. the possible largest one is less than the sum of  the remaining two as discussed for the $p = p' = 6$ case in the previous section. When $\nu_{0} = 0$ for which $\hat g_{0} \neq 0$ and $\hat f^{2}_{1} + \hat f^{2}_{2} =  1$, the amplitude vanishes but this is different from the $p = 6, \, p' = 2$ case for which the the fluxes on D6 and D2 all vanish (or the fluxes on D6 are vanishing except for the ones along the D2 directions which are identical to those on the D2).  The explanation for the present vanishing interaction goes like this.  The interaction between a D0 and a D6 carrying no flux is repulsive.  The magnetic flux $\hat g_{0}$ stands for 
delocalized D4 within D6 which has no interaction with D0 since their dimensionality differs by four.  The electric flux $\hat f_{1}, \hat f_{2}$ stand for the delocalized fundamental strings within D6 which have attractive interaction with the D0. So the vanishing of this amplitude must imply the cancellation of the repulsive interaction between the D6 and the D0 with the attractive one between the D0 and the fundamental F-strings within the D6 when  $\hat f^{2}_{1} + \hat f^{2}_{2} =  1$.   This is also consistent with the general conclusion reached for the $p = p' = 6$ case in the previous section that $\nu_{0} + \nu_{1} = \nu_{2}$ for which the amplitude vanishes. 

For real non-vanishing $\nu_{0}$, given what we learned earlier in this paper, we expect no open string tachyonic instability. Let us check this explicitly by examining the corresponding 
open string one-loop annulus amplitude which can be read from (\ref{t-amplitude-annulus-pp'}) as
\be\label{t-amplitude-annulus-60-6}
\Gamma_{6, 0} = \frac{ V_{1} \sqrt{1 - \hat f^{2}_{1} - \hat f^{2}_{2}}}{ (8 \pi^2 \alpha')^{\frac{1}{2}}}   \int_0^\infty \frac{d t}{t^{\frac{3}{2}}}  e^{- \frac{y^2 t}{2\pi\alpha'}} \frac{\sinh\frac{\pi\nu_{0} t}{2} \left(\cosh^{2} \frac{\pi\nu_{0} t}{2} - \cosh^{2} \frac{\pi t}{2}\right)}{\cosh\frac{\pi \nu_{0} t}{2} \sinh^{2}\frac{\pi t}{2}} \nn \prod_{n=1}^{\infty} Z_{n}, 
 \ee    
where $Z_{n}$, read from (\ref{Z}), is
\bea\label{Z60-6}
Z_{n} &=& \frac{\left[1 - 2 |z|^{2n} \cosh\pi\nu_{0} t + |z|^{4n}\right]^{2}}{(1 - |z|^{2n})^{2} (1 - 2 |z|^{2n} \cosh\pi t + |z|^{4n})^{2} }\nn
&\,& \times \frac{ \left[(1 - 2 |z|^{2n} \cosh\pi\nu_{0} t\cosh\pi t + |z|^{4n})^{2} - 4 |z|^{4n} \sinh^{2}\pi \nu_{0} t \sinh^{2} \pi t \right]}{(1 -  2 |z|^{2n} \cosh 2 \pi \nu_{0} t + |z|^{4n})}.
\eea
For large $t$, $Z_{n} \approx 1$ and the integrand of the above amplitude behaves like 
\be\label{tachyon60-6}
\sim - e^{- \frac{y^2 t}{2\pi\alpha'}} ,
\ee
which vanishes for all $y \neq 0$, therefore no tachyonic divergence as expected.  

Let us consider $\nu_{0}$ to be imaginary which requires $\hat g_{0} \neq 0$ and $1 < \hat f^{2}_{1} + \hat f^{2}_{2} < 1 + \hat g^{2}_{0}$. We now set $\nu_{0} = i \bar\nu_{0}$ with 
$\bar \nu_{0} \in (0, \infty)$. We have now from (\ref{nu0nu1nu2-60-6})
\be\label{barnu0-60-6}
\tanh\pi\bar\nu_{0} = \frac{\sqrt{\hat f^{2}_{1} + \hat f^{2}_{2} - 1}}{|\hat g_{0}|},
\ee
in addition to $\nu_{1} = \nu_{2} = 1/2$ when we take $\hat g_{1} = \hat g_{2} = 0$. The open string one-loop annulus amplitude is now, from (\ref{t-amplitude-annulus-60-6}) with
 $\nu_{0} = i \bar\nu_{0}$, 
\be\label{t-amplitude-annulus-60-6-barnu0}
\Gamma_{6, 0} = \frac{ V_{1} \sqrt{ \hat f^{2}_{1} + \hat f^{2}_{2} - 1}}{ (8 \pi^2 \alpha')^{\frac{1}{2}}}   \int_0^\infty \frac{d t}{t^{\frac{3}{2}}}  e^{- \frac{y^2 t}{2\pi\alpha'}} \frac{\sin\frac{\pi\bar \nu_{0} t}{2} \left(\cosh^{2} \frac{\pi t}{2} - \cos^{2} \frac{\pi\bar\nu_{0} t}{2} \right)}{\cos\frac{\pi \bar \nu_{0} t}{2} \sinh^{2}\frac{\pi t}{2}}  \prod_{n=1}^{\infty} Z_{n}, 
\ee
where $Z_{n}$ continues to be given by (\ref{Z60-6}) but now with $\nu_{0} = i \bar\nu_{0}$. Now this amplitude has an infinite number of simples poles of its integrand occurring at
$t_{k} =  (2k - 1)/\bar\nu_{0}$ with $k = 1, 2, \cdots $, giving an imaginary part of the amplitude.  This further indicates the decay of the underlying system via the so-called open string pair production. The decay rate and the open string pair production rate are given, respectively, by (\ref{decayrate-pp'}) and (\ref{pprate-pp'}) for the present case with $\nu_{1} = \nu_{2} = 1/2$ as
\bea\label{decayrate-60}
{\cal W} &=& \frac{4\,\sqrt{ \hat f^{2}_{1} + \hat f^{2}_{2} - 1}} {\bar\nu_{0} (8 \pi^2 \alpha')^{\frac{1}{2}}}  \sum_{k = 1}^{\infty} \left(\frac{\bar\nu_{0}}{2k - 1}\right)^{\frac{3}{2}}  \frac{\cosh^{2} \frac{(2k - 1)\pi}{\bar\nu_{0}}}{\sinh^{2} \frac{(2k - 1)\pi}{\bar\nu_{0}}} e^{- \frac{(2k -1) y^2 }{2\pi\alpha' \bar\nu_{0}}}\, Z_{2k - 1} (\bar\nu_{0}, 1/2, 1/2),\qquad
\eea
where
\be\label{Zk-60-6}
Z_{k} (\bar\nu_{0}, 1/2, 1/2) = \prod_{n = 1}^{\infty} \frac{(1 + |z_{k}|^{2n})^{4} \left(1 + 2 |z_{k}|^{2n} \cosh \frac{k \pi}{\bar\nu_{0}} + |z_{k}|^{4n}\right)^{2}}
{(1 - |z_{k}|^{2n})^{4}  \left(1 - 2 |z_{k}|^{2n} \cosh \frac{k \pi}{\bar\nu_{0}} + |z_{k}|^{4n}\right)^{2}},
\ee
 with $|z_{k}| = e^{- k \pi/\bar\nu_{0}}$.  As before, the open string pair production rate is given by the $k = 1$ term of the above as
 \be\label{pprate-60-6}
 {\cal W}^{(1)} = \frac{4\,\sqrt{ \hat f^{2}_{1} + \hat f^{2}_{2} - 1}} { (8 \pi^2 \alpha')^{\frac{1}{2}}}  \bar\nu_{0}^{1/2}  \frac{\cosh^{2} \frac{\pi}{\bar\nu_{0}}}{\sinh^{2} \frac{\pi}{\bar\nu_{0}}} e^{- \frac{ y^2 }{2\pi\alpha' \bar\nu_{0}}}\, Z_{1} (\bar\nu_{0}, 1/2, 1/2).\qquad
 \ee
We would like to point out that with the choices of fluxes (\ref{extendedflux0-6}) and (\ref{flux6-6}), the above amplitudes and rates share qualitatively the same properties as their correspondences, respectively, in the case of $p = 2, p' =0$ discussed in subsection (\ref{subsection4.1}), even though the details are different.  For example,  both of the rates blow up when $\bar\nu_{0} \to \infty$ which occurs as $\hat f^{2}_{1} + \hat f^{2}_{2} - 1\to \hat g^{2}_{0}$, reaching the so-called critical field.  For small $\bar\nu_{0} \ll 1$, $Z_{1} \approx 1$ and the open string pair production rate above looks also like that of the $p = 2, p' =0$ case.  For a general $\bar \nu_{0}$,  these two rates are different. Note that we don't have the exponential enhancement of the rate for small $\bar\nu_{0}$, either.

\section{Discussion and conclusion}
We compute, in this paper, the closed string cylinder amplitude between a Dp and a Dp$'$, placed parallel at a separation along the directions transverse to the Dp, with each carrying their general constant worldvolume fluxes and with $p - p' =\kappa = 0, 2, 4, 6$ and $p \le 6$.  We find that the amplitude for each of the $p - p' = \kappa \neq 0$ cases can be obtained as just a special case of the corresponding amplitude for the $p = p'$ case based on the related physical consideration presented in the previous  sections. As such, we find a universal formula for this closed string cylinder amplitude, valid for all cases specified above, as
\bea\label{t-amplitude-cylinder-pp'-universal}
\Gamma_{p, p'} &=&  \frac{ 2^3 \, V_{p' + 1}\, \left[\det (\eta_{p'} + \hat F'_{p'})\det(\eta_{p} + \hat F_{p})\right]^{\frac{1}{2}}\prod_{\alpha =0}^{2} \sin \pi \nu_{\alpha} }{ 2^{\frac{\kappa}{2}} (8 \pi^2 \alpha')^{\frac{p' + 1}{2}}} \int_0^\infty \frac{d t} {t^{\frac{9 - p}{2}}} \frac{e^{- \frac{y^2}{2\pi\alpha' t}}}{\eta^{3} (it)} \nn 
&\,& \times \frac{ \theta_{1} \left(\left.\frac{\nu_{0} + \nu_{1} + \nu_{2}}{2}\right| it \right) \theta_{1} \left(\left.\frac{\nu_{0} - \nu_{1} + \nu_{2}}{2}\right| it \right)
 \theta_{1} \left(\left.\frac{\nu_{0} + \nu_{1} - \nu_{2}}{2}\right| it \right)\theta_{1} \left(\left.\frac{\nu_{0} - \nu_{1} - \nu_{2}}{2}\right| it \right)}{\theta_{1} (\nu_{0} | it)\theta_{1} (\nu_{1} | it)\theta_{1} (\nu_{2} | it)}\nn 
  &=&  \frac{ 2^2 \, V_{p'+ 1} \left[\det (\eta_{p'} + \hat F'_{p'})\det(\eta_{p} + \hat F_{p})\right]^{\frac{1}{2}}  \left[\sum_{\alpha = 0}^{2}\cos^{2}\pi \nu_{\alpha} - 2\prod_{\alpha =0}^{2} \cos\pi\nu_{\alpha} - 1\right]}{ 2^{\frac{\kappa}{2}}(8 \pi^2 \alpha')^{\frac{p' + 1}{2}}} 
\nn
 &\,& \times  \int_0^\infty \frac{d t}{t^{\frac{9 - p}{2}}}  e^{- \frac{y^2}{2\pi\alpha' t}}  \prod_{n=1}^{\infty} C_{n},
 \eea
where $C_{n}$ continues to be given by (\ref{C}).  The amplitude for each given pair of $p$ and $p'$ and the corresponding given worldvolume fluxes can be obtained from the above as a special case as prescribed in the previous two sections.  The corresponding open string one-loop annulus universal amplitude can be obtained from the above via the Jacobi transformation $t \to t' = 1/t$ along with the relations for the $\theta_{1}$-function and the Dedekind $\eta$-function given in (\ref{jacobi}) as

\bea\label{t-amplitude-annulus-pp'-universal}
\Gamma_{p,p'} &=& - \frac{ 2^3 \, i \, V_{p' + 1}\, \left[\det (\eta_{p'} + \hat F'_{p'})\det(\eta_{p} + \hat F_{p})\right]^{\frac{1}{2}}\prod_{\alpha =0}^{2} \sin \pi \nu_{\alpha} }{2^{\frac{\kappa}{2}}(8 \pi^2 \alpha')^{\frac{p' + 1}{2}}} \int_0^\infty \frac{d t} {t^{\frac{p - 3}{2}}} \frac{e^{- \frac{y^2 t}{2\pi\alpha' }}}{\eta^{3} (it)} \nn 
&\,& \times \frac{ \theta_{1} \left(\left. \frac{\nu_{0} + \nu_{1} + \nu_{2}}{2} i t \right| it \right) \theta_{1} \left(\left. \frac{\nu_{0} - \nu_{1} + \nu_{2}}{2} it \right| it \right)
 \theta_{1} \left(\left.  \frac{\nu_{0} + \nu_{1} - \nu_{2}}{2} it \right| it \right)\theta_{1} \left(\left.  \frac{\nu_{0} - \nu_{1} - \nu_{2}}{2} it \right| it \right)}{\theta_{1} (i \nu_{0} t | it)\theta_{1} ( i \nu_{1} t | it)\theta_{1} (i \nu_{2} t | it)}\nn
 &=&  \frac{ 2^2 \, V_{p'+ 1} \left[\det (\eta_{p'} + \hat F'_{p'})\det(\eta_{p} + \hat F_{p})\right]^{\frac{1}{2}} }{2^{\frac{\kappa}{2}} (8 \pi^2 \alpha')^{\frac{p' + 1}{2}}}   \int_0^\infty \frac{d t}{t^{\frac{p - 3}{2}}}  e^{- \frac{y^2 t}{2\pi\alpha'}} \prod_{\alpha =0}^{2} \frac{\sin \pi \nu_{\alpha}}{\sinh \pi \nu_{\alpha} t} \nn
 &\,& \times \left[\sum_{\alpha = 0}^{2}\cosh^{2}\pi \nu_{\alpha} t- 2 \prod_{\alpha = 0}^{2} \cosh\pi \nu_{\alpha} t - 1\right] \prod_{n=1}^{\infty} Z_{n},
 \eea     
where we have dropped the prime on the open string variable $t$ and $Z_{n}$ continues to be given by (\ref{Z}).  If one of three $\nu_{\alpha}$, say $\nu_{0}$,  is imaginary, the underlying system decays via the open string pair production. The general decay rate is 
\bea\label{decayrate-pp'-universal}
{\cal W}_{p, p'} &=& \frac{ 2^{3- \frac{\kappa}{2}} \, \left[\det (\eta_{p'} + \hat F'_{p'})\det(\eta_{p} + \hat F_{p})\right]^{\frac{1}{2}} \sinh\pi\bar\nu_{0} \sin\pi\nu_{1} \sin\pi\nu_{2}} {\bar\nu_{0} (8 \pi^2 \alpha')^{\frac{p' + 1}{2}}}  \sum_{k = 1}^{\infty} (-)^{k + 1} \left(\frac{\bar\nu_{0}}{k}\right)^{\frac{p - 3}{2}} \nn
&\,&\times \frac{\left(\cosh\frac{k \pi \nu_{1}}{\bar\nu_{0}} - (-)^{k} \cosh\frac{k\pi\nu_{2}}{\bar\nu_{0}}\right)^{2}}{\sinh\frac{k\pi\nu_{1}}{\bar\nu_{0}}\sinh\frac{k\pi\nu_{2}}{\bar\nu_{0}}} e^{- \frac{k y^2 }{2\pi\alpha' \bar\nu_{0}}}\, Z_{k} (\bar\nu_{0}, \nu_{1}, \nu_{2}),
\eea
where $Z_{k}$ is given by (\ref{Zk}).  The corresponding open string pair production rate is given by the leading $k = 1$ term of the above as
 \bea\label{pprate-pp'-universal}
 {\cal W}^{(1)}_{p, p'} &=&  \frac{ 2^{3- \frac{\kappa}{2}} \, \left[\det (\eta_{p'} + \hat F'_{p'})\det(\eta_{p} + \hat F_{p})\right]^{\frac{1}{2}} \sinh\pi\bar\nu_{0} \sin\pi\nu_{1} \sin\pi\nu_{2}} { (8 \pi^2 \alpha')^{\frac{p' + 1}{2}}} \bar\nu_{0}^{\frac{p - 5}{2}} \, e^{- \frac{y^2 }{2\pi\alpha' \bar\nu_{0}}}\nn
 &\,&\times \frac{\left(\cosh\frac{ \pi \nu_{1}}{\bar\nu_{0}} + \cosh\frac{\pi\nu_{2}}{\bar\nu_{0}}\right)^{2}}{\sinh\frac{\pi\nu_{1}}{\bar\nu_{0}}\sinh\frac{\pi\nu_{2}}{\bar\nu_{0}}} \, Z_{1} (\bar\nu_{0}, \nu_{1}, \nu_{2}).
 \eea
With the above, we have studied various properties of the amplitudes for each of the systems  considered such as the nature of the interaction, the open string tachyonic instability, and the open string pair production if it exists and the associated enhancement.  In particular, we find that the interaction can be repulsive for $p' \le p = 6$ with all three parameters $\nu_{0}, \nu_{1}, \nu_{2}$ being real and non-vanishing, i.e., $\nu_{\alpha} \in (0, 1)$ with $\alpha = 0, 1 , 2$. Since the amplitude is symmetric with respect to the three $\nu_{0}, \nu_{1}, \nu_{2}$, we can assume $\nu_{0} \le \nu_{1} \le \nu_{2}$ without loss of generality. The repulsive interaction occurs indeed when $\nu_{0} + \nu_{1} > \nu_{2}$ and $\nu_{0} + \nu_{1} + \nu_{2} < 2$.  In other words,  whenever the sum of two smaller $\nu$'s (here $\nu_{0}$ and $\nu_{1}$) is larger than the largest $\nu$
(here $\nu_{2}$) and $\nu_{0} + \nu_{1} + \nu_{2} < 2$, the underlying interaction is repulsive.  The reason for the above requirements is simple.  The repulsive inter-brane interaction occurs, in the absence of the worldvolume fluxes, only for the system of $p = 6$ and $p' =0$ for which we have $\nu_{0} = \nu_{1} = \nu_{2} = 1/2$ following the description given in the previous two sections (Here $\nu_{0} + \nu_{1} > \nu_{2}$ and $\nu_{0} + \nu_{1} + \nu_{2} = 3/2 < 2$ meet the above conditions for repulsive interaction). For all other choices of $p$ and $p'$, the inter-brane interaction, in the absence of the worldvolume fluxes, is either attractive or vanishing. So to have a potential repulsive-interaction, we first need to have the presence of D6  and secondly we need to have D0  which can be realized in general for $p' \le p = 6$ with all three $\nu_{0}, \nu_{1}, \nu_{2} \in (0, 1)$.  Note also that when $\nu_{0}, \nu_{1}, \nu_{2} \in (0, 1)$, the worldvolume fluxes give rise to not only D0 but also D2 and D4.  The latter branes give instead the attractive inter-brane interaction in addition to the repulsive one between D6 and D0.  So whether the net inter-interaction is repulsive, attractive or vanishes depends on the competition between the repulsive component and the attractive one mentioned above.  Using the above assumption $\nu_{0} \le \nu_{1} \le \nu_{2}$, we have shown in section 3 and checked for each case considered later that whenever $\nu_{0} +  \nu_{1} > \nu_{2}$ and $\nu_{0} + \nu_{1} + \nu_{2} < 2$, the net interaction is repulsive. The interaction vanishes whenever $\nu_{0} + \nu_{1} = \nu_{2}$ or $\nu_{0} + \nu_{1} + \nu_{2} = 2$. The interaction is attractive whenever $\nu_{0} + \nu_{1} < \nu_{2}$ and $\nu_{0} + \nu_{1} + \nu_{2} < 2$ or $\nu_{0} + \nu_{1} > \nu_{2}$ and $2 < \nu_{0} + \nu_{1} + \nu_{2} < 3$. 

We also find that there is a correlation between the nature of interaction and the existence of the open string tachyonic instability of the underlying system when the brane separation reaches the distance determined by the so-called tachyonic shift.  When the interaction is repulsive, there is no open string tachyonic instability, independent of the brane separation.   When the brane separation is attractive, we do have the onset of tachyonic instability when the brane separation reaches the distance set by the tachyonic shift. We analyze this from various means and confirm this correlation. 

When one of three parameters $\nu_{0}, \nu_{1}, \nu_{2}$ is imaginary, the underlying system is unstable and decays via the so-called open string pair production. This is reflected in that the open string one-loop amplitude has an imaginary part. Again without loss of generality, we choose $\nu_{0} = i \bar \nu_{0}$ with $\bar\nu_{0} \in (0, \infty)$. This is related to the applied electric flux(es). When the applied electric flux reaches its critical one, we have $\bar\nu_{0} \to \infty$,  the open string pair production rate diverges and the pair production cascades, giving rise also to the other instability of the system.  We have also studied the potential enhancement of the pair production rate in the presence of magnetic fluxes and our findings here are consistent with our previous studies on this.  The enhancement is determined by the so-called tachyonic shift which can be given in general as $|\nu_{2} - \nu_{1}|/2$ with $\nu_{1}, \nu_{2} \in [0, 1)$.  In practice, 
all $\bar\nu_{0}, \nu_{1}, \nu_{2}$ are small. We have that the larger the shift is, the larger the open string pair production enhancement.  For this purpose, we prefer to have the presence of the larger of $\nu_{1}$ and $\nu_{2}$ while turning off the smaller one such that the enhancement is larger.  For example, we keep $\nu_{1}$ while drop $\nu_{2}$. So the question is: can we realize the largest shift which is $\nu_{1}/2 = 1/2$?  This is one of the motivations for this paper as mentioned in the Introduction. Though we cannot realize $\nu_{1}/2 = 1/2$ but  we can have $\nu_{1}/2 = 1/4$, almost equally as good,  and it can come from the system of $p - p' = 2$ without adding any worldvolume magnetic fluxes given the above consideration.  This is due to that the Dp$'$ brane acts effectively as a magnetic field which can give rise to $\nu_{1} = 1/2$ as shown in subsection 4.1.  The largest pair production rate for practically given small $\bar\nu_{0}$ occurs for $p = 3, p' =1$ with purely added  electric fluxes along the D1-directions. 
This system has $\nu_{1} = 1/2$, giving the possible largest enhancement.  This may have a potential application in practice which we would like to pursue in the near future. 

One last thing, we have briefly mentioned in the introduction, is the relationship of the present discussion with that for a system of Dp and Dp$'$ ($p \ge p'$) with the two branes not at rest but with a constant relative motion transverse at least to the Dp$'$ and/or a rotation between certain transverse directions and the brane directions.  As discussed in \cite{Di Vecchia:1999fx}, a Dp brane carrying a constant electric flux along certain spatial direction  is equivalent to a boosted and delocalized D(p - 1) brane along this direction. They are related by a T-duality along this direction and the boost velocity is determined by the electric flux.   By a similar token, a Dp brane carrying a magnetic strength $F_{ij}$ with $i < j$ (also called magnetized brane), for example,  is equivalent to intersecting D(p - 1) branes at an angle between the spatial $i$-direction and  the spatial j-direction and determined by the magnetic flux, for example, by a T-duality along j-direction\cite{Berkooz:1996km, Balasubramanian:1996uc, Breckenridge:1996tt, Breckenridge:1997ar,Rabadan:2001mt, Di Vecchia:1999fx}.   Here the magnetized Dp brane and the intersecting D(p - 1) branes are related by a T-duality along the j-direction and the rotation is determined by the magnetic flux.  Since the resulting D(p - 1) brane(s) in either case is delocalized along certain directions transverse to the brane, it is probably much easier and much more straightforward to compute the same interaction between two such D branes using their equivalent ones carrying fluxes at rest as discussed in this paper even though computations of the interaction for localized such objects are known (for example, see \cite{Arfaei:1996rg, SheikhJabbari:1997cv, Di Vecchia:1999fx}). In this paper we only use the so-called no-force condition to discuss whether the underlying system preserves certain supersymmetry but for intersecting branes the underlying supersymmetry can be analyzed in detail following \cite{polbooktwo, Behrndt:1996pm, Berkooz:1996km, Balasubramanian:1996uc, Bergshoeff:1996rn, Breckenridge:1996tt, Breckenridge:1997ar, SheikhJabbari:1997cv, Rabadan:2001mt} (see \cite{Blumenhagen:2006ci} for a rather complete list of references on phenomenological applications to supersymmetry breaking for intersecting branes).   

\section*{Acknowledgments}
The authors would like to thank the anonymous referees for their suggestions and 
questions raised which help to improve the present manuscript and acknowledge support by grants from the NSF of China with Grant No: 11775212, 11947301 and 11235010. 

\section*{Appendix A\label{AA}}
In this Appendix, we first give a general discussion of the eigenvalues of the matrix $w$ (\ref{matrixlw}). Note that 
\be\label{ss'w}
s = (\eta - \hat F)(\eta + \hat F)^{-1}, \quad s' =  (\eta- \hat F')(\eta + \hat F')^{-1}, \quad w = s s'^{T},
\ee
and each of them satisfies the same relation, e.g.,
\be\label{lorentz}
w_{\alpha}\,^{\gamma} (w^{T})_{\gamma}\,^{\beta} = \delta_{\alpha}\,^{\beta}.
\ee
This is actually a relation satisfied by a Lorentz transformation in (1 + p) dimensions. In other words, either $s$ or $s'$ is a general Lorentz transformation since either flux $\hat F$ or $\hat F'$ counts the number of independent parameters of $SO(1, p)$ as $(p + 1)p/2$.  
This holds also true for $w$ since it is the product of $s$ and $s'$. In addition, we have $\det s = \det[(\eta - \hat F)(\eta + \hat F)^{-1}] = \det (\eta - \hat F) \det(\eta + \hat F)^{-1} = \det(\eta - \hat F) \det(\eta - \hat F)^{-1} = 1$ where  we have used $(\eta + \hat F)^{T} = (\eta - \hat F)$ in the third equality since $\hat F$ is antisymmetric.  This same holds for $s'$, too.  So we have $\det w = \det s \det s'^{T} = 1$ as well.  Note also here
\be\label{metric}
\eta_{\alpha\beta} = (-1, 1, \cdots 1),   \qquad \alpha, \beta = 0, 1, \cdots p.
\ee
Given the above, it is clear that a purely electric flux gives  a Lorentz boost while a purely magnetic one gives only a rotation of $SO(p)$. When both $\hat F$ and $\hat F'$ are each purely electric, the resulting $w$ is in general not a pure Lorentz boost unless the two electric fluxes are collinear.   However, the resulting $w$ is a rotation when both $\hat F$ and $\hat F'$ are each purely magnetic.   If $w$ is indeed a pure Lorentz boost, it can always be brought to the following form by a $SO(p)$ rotation $R$,
\be
w = r^{T} \tilde w r,
\ee
where 
\be 
\tilde w = \left(\begin{array}{ccc}
\gamma & \gamma v& \\
\gamma v & \gamma & \\
&&\mathbb{I}_{(p - 1)\times (p - 1)}\\
\end{array}\right), \qquad r = \left(\begin{array}{cc}
1 & \\
 & R_{p\times p} \\
 \end{array}\right),
\ee
with $\gamma = (1 - v^{2})^{- 1/2}$ and $v < 1$. The rotation just brings the velocity along the `1'-direction.  So it is clear that a general Lorentz boost has only one non-trivial pair real eigenvalues of $\lambda_{\pm} = \gamma (1 \pm v)$ with $\lambda_{+} \lambda_{-} = 1$ and the rest are all unity. If $w$ is instead a pure SO(p) rotation, e.g. $w = r$ with $r$ given above, we have its eigenvalues 1 and the rest in pairs\footnote{We choose here the same conventions as used in Section 3.} as $\lambda_{\alpha}, \lambda^{-1}_{\alpha}$ with $\lambda_{\alpha} = e^{2\pi i \nu_{\alpha}}$ and  $\alpha = 0,1, \cdots (p - 2)/2$ when $p =$ even or 1, 1, and the rest in pairs as before but now with $\alpha = 0, 1, \cdots, (p - 3)/2$ when $p =$ odd.  Here $\nu_{\alpha}$ are all real with each $\nu_{\alpha} \in [0, 1)$ for the amplitude as discussed in Section 3. 

For a general $w$, we need to put some extra efforts to figure out the nature of its eigenvalues.  For this, since $w$ is a Lorentz matrix with $\det w = 1$, we can set 
\be\label{lorentzw}
w = e^{K},
\ee
where from $(w^{-1})_{\alpha}\,^{\beta} = (w^{T})_{\alpha}\,^{\beta}$ we have
\be\label{K}
K_{\alpha}\,^{\beta} = - K^{\beta}\,_{\alpha}.
\ee
Now solving the eigenvalue problem of $w$ is transformed to that of $K$.  In other words, we have 
\be\label{eigenK}
f (\rho) = \det\left(\rho\, \delta_{\alpha}\,^{\beta} - K_{\alpha}\,^{\beta}\right) = 0.
\ee
Since $K^{T}$ has the same eigenvalues as $K$, we have also
\be\label{eigenKT}
f(\rho) = \det\left[\rho\,\delta_{\alpha}\,^{\beta} - (K^{T})_{\alpha}\,^{\beta}\right] = 0,
\ee
which implies whenever $\rho$ is an eigenvalue so is $- \rho$ from (\ref{K}). In other words, the eigenvalues appear in pairs. When $p =$ even, we have always one zero-eigenvalue since
$\det K_{\alpha}\,^{\beta} = - \det K_{\alpha \beta} = 0$, giving the unity eigenvalue of $w$ discussed in Section 3, and the rest are in pairs.  

Let us discuss the $p =$ even case first. For the zero-eigenvalue, we can choose the corresponding eigenvector as $e$ such that $K_{\alpha}\,^{\beta} e_{\beta} = 0$.  We have two sub-cases to consider:
$e\cdot e = - 1$ and $e\cdot e = 1$ where we have normalized each as indicated\footnote{For certain choice of the fluxes, we may have the eigenvector being light-like. This can always be taken as certain limit of the time-like or space-like limit as discussed.  When this happens, the corresponding $K$ matrix can still be diagonalized.  Here we use a simple example for $p = 2$ to illustrate this. Now the most general $K$ can be expressed as 
\be\label{fn-K}
K_{\alpha}\,^{\beta} = \left(\begin{array}{ccc}
0& f_{1} & f_{2}\\
f_{1}&0 & g\\
f_{2} & - g&0\\
\end{array}\right),
\ee
where we assume $f_{1}, f_{2}, g$ are all non-negative without loss of generality.  This matrix has three expected eigenvalues: $0, \rho_{0}, - \rho_{0}$ with $\rho_{0} = \sqrt{f_{1}^{2} + f_{2}^{2} - g^{2}}$.  It can be diagonalized as
\be
K = V \bar K_{0} V^{- 1},
\ee
where the diagonal matrix $({\bar K_{0}})_{\alpha}\,^{\beta} = (0, \rho_{0}, - \rho_{0})$ and the non-singular matrix 
\be
V = \frac{1}{\rho_{0}} \left(\begin{array}{ccc}
g & - f_{2} & f_{1} \\
\frac{f_{1} g - f_{2}\rho_{0}}{g^{2} - f_{2}^{2}} & \frac{g\rho_{0} - f_{1} f_{2}}{g^{2} - f_{2}^{2}}& 1\\
-f_{1} g - f_{2}\rho_{0} & g\rho_{0} + f_{1} f_{2} & f_{2}^{2} - g^{2}\\
\end{array}\right),
\ee
with its $\det V =  2$.  For the 0-eigenvalue, the corresponding eigenvector can be taken as $e = (g, f_{2}, - f_{1})^{T}$, giving $K_{\alpha}\,^{\beta} e_{\beta} = 0$.  Here $e^{2} = \eta^{\alpha\beta}
e_{\alpha} e_{\beta} = f_{1}^{2} + f^{2}_{2} - g^{2}$ which can be either time-like or space-like in general. However, it becomes null when $f_{1}^{2} + f_{2}^{2} =  g^{2}$, which can be taken as the corresponding limit of either time-like or space-like case.  So long this is taken as a limiting case, we don't have a problem to diagonalize the matrix $K$ since $\det V = 2$ is non-singular. 
}.  For the first subcase, we choose $e\,^{0} = e$ such that $\{e\,^{0}, e\,^{1}, \cdots, e\,^{p}\}$ forms a complete normalized basis of the eigenvector space, giving  
\be\label{basis}
 \eta^{\alpha\beta} e_{\alpha}\,^{\bar\alpha} e_{\beta}\,^{\bar\beta} = \eta^{\bar\alpha \bar\beta} \quad {\rm or}\quad \eta_{\alpha\beta} e^{\alpha}\,_{\bar\alpha} \,e^{\beta}\,_{\bar\beta} = \eta_{\bar\alpha\bar\beta},
 \ee
 where the $\alpha, \beta$ indices are raised or lowered using $\eta^{\alpha\beta}$ or $\eta_{\alpha\beta}$ and similarly for the $\bar\alpha, \bar\beta$ indices.  So  $e_{\alpha}\,^{\bar\alpha}$ or $e^{\alpha}\,_{\bar\alpha}$ is also a Lorentz transformation. 
We take $\alpha = (0, a)$ ($\bar \alpha = (0, \bar a)$) from now on with $a = 1, 2, \cdots p$ ($\bar a = 1, 2, \cdots, p$). We have now $K_{\alpha}\,^{\beta} e_{\beta}\,^{0} = 0$ and from (\ref{K}) we have also $(e^{T})^{0}\,_{\beta} K^{\beta}\,_{\alpha} = 0$. With these two, we have, from (\ref{basis}),
\be\label{Kbar}
\bar K_{\bar\alpha}\,^{\bar\beta} \equiv  (e^{T})_{\bar\alpha}\,^{\alpha} K_{\alpha}\,^{\beta} e_{\beta}\,^{\bar\beta} = \left(\begin{array}{cc}
0 & 0\\
0&\bar K_{\bar a}\,^{\bar b}\end{array}\right),
\ee
where $\bar K_{\bar a}\,^{\bar b} = \bar K_{\bar a \bar b} = - \bar K_{\bar b \bar a}$ is real and antisymmetric from the property of $K$ given in (\ref{K}).  So it can be diagonalized by a unitary matrix $u$ of the following form with its purely imaginary 
eigenvalues in pairs as ($\rho_{\bar a}, - \rho_{\bar a}$) with $\bar a = 1, \cdots, p/2$, 
\be\label{Kbardiagonal}
\bar K = U \bar K_{0} U^{+} = \left(\begin{array}{cc}
1& 0\\
0& u_{p\times p}\end{array}\right) \left(\begin{array}{cc}
0 & 0 \\
0& (\bar K_{0})_{p\times p}\end{array}\right) \left(\begin{array}{cc}
1& 0\\
0& u^{+}_{p\times p}\end{array}\right),
\ee
where 
\be\label{eigenbarK}
 (\bar K_{0})_{p\times p} = \left(\begin{array}{ccccc}
 \rho_{1}&0&\ldots&0&0\\
0& - \rho_{1}&\ldots&0&0\\
\vdots& \vdots &\ddots&\vdots &\vdots\\
0 &0 &\ldots &\rho_{p/2}&0\\
 0&0&\ldots& 0&- \rho_{p/2} \end{array}\right).
 \ee
With the above, the original $K$ is diagonalized as
\be\label{Kdiagonal}
K = V \bar K_{0} V^{-1},
\ee
where $V = e U$ with $(e^{-1})_{\bar \alpha}\,^{\beta} = (e^{T})_{\bar\alpha}\,^{\beta}$ and $U^{-1} = U^{+}$.  Here $\bar K_{0}$ gives the expected eigenvalues: one zero and the rest being purely imaginary in pairs as indicated in (\ref{eigenbarK}).

For the second subcase, i.e. $e \cdot e = 1$, we take now $e\,^{p} = e$ such that $\{e\,^{0}, e\,^{1}, \cdots, e\,^{p}\}$ forms a complete normalized basis of the eigenvector space. The following discussion goes exactly the same line as above and end up with now
\be\label{Kbarss}
\bar K_{\bar\alpha}\,^{\bar\beta} \equiv  (e^{T})_{\bar\alpha}\,^{\alpha} K_{\alpha}\,^{\beta} e_{\beta}\,^{\bar\beta} = \left(\begin{array}{cc}
\bar K_{\bar\alpha'}\,^{\bar \beta'}& 0\\
0&0\end{array}\right),
\ee
where $\bar\alpha', \bar\beta' = 0, 1, \cdots p - 1$. The diagonalisation of the matrix $\bar K_{\bar\alpha'}\,^{\bar\beta'}$ follows exactly the same as $K_{\alpha}\,^{\beta}$ for odd $p$. So we turn now to the odd $p$ case. 

For this case,  since the eigenvalues are in pairs as $(\rho_{\alpha}, - \rho_{\alpha})$ with $\alpha = 0, 1,\cdots (p - 1)/2$, the function $f(\rho)$ from (\ref{eigenK}) must be even in power of $\rho$ when we expand the determinant and is 
\be\label{eigenexp}
f(\rho) = \rho^{p + 1} + c_{1} \rho^{p - 1} + c_{3} \rho^{p - 3} + \cdots + c_{p -1} \rho^{2} + \det K_{\alpha}\,^{\beta}.
\ee 
Note that $\det K_{\alpha}\,^{\beta} = - \det K_{\alpha\beta} = - ({\rm pf} (K_{\alpha\beta}))^{2} < 0$ with $K_{\alpha\beta} = - K_{\beta\alpha}$ from (\ref{K}) and ${\rm pf (K)}$ denoting the Pfaffian of antisymmetric matrix $K_{\alpha\beta}$.
So we have $f(0) = \det K_{\alpha}\,^{\beta} < 0$.  For very large $\rho > 0$, the highest power of $\rho$ dominates and so we have $f (\rho) > 0$.  Therefore we must have at least  one pair $(\rho_{0}, -\rho_{0})$ with $\rho_{0}$ positive real
satisfying the eigenvalue equation $f (\pm \rho_{0}) = 0$.  If $\det K_{\alpha}\,^{\beta} =  0$, we will have a pair of zero eigenvalues unless the above $c_{p - 1} = 0$ and if this is case we can do what we have done for the above even $p$ case.  
If we have more zero eigenvalues, we just repeat this process until we have non-zero ones. For now, we assume $\det K_{\alpha}\,^{\beta} < 0$, so we have $\rho_{0} \neq 0$. The corresponding eigenvectors, denoting as $x^{0}$ and $x^{1}$, satisfy 
their respective equations as
\be\label{eigene}
K_{\alpha}\,^{\beta} x_{\beta}^{0} = \rho_{0}\, x_{\alpha}^{0}, \qquad K_{\alpha}\,^{\beta}  x_{\beta}^{1} = - \rho_{0}\, x_{\alpha}^{1}.
\ee
From the first one, we have $\rho_{0} x^{0} \cdot x^{0} = (x^{0})^{\alpha} K_{\alpha}\,^{\beta} x^{0}_{\beta} = x^{0}_{\alpha} K^{\alpha\beta} x^{0}_{\beta} = 0$ where we have used the property of $K$ from (\ref{K}).  This must imply $x^{0} \cdot x^{0} = 0$ since
$\rho_{0} \neq 0$. In other words, $x^{0}$ is a null vector. By the same token,  we have also $x^{1}$ as a null vector. We now show $x^{0} \cdot x^{1} \neq 0$. Since both are null vectors, without loss of generality, we can always choose $x^{0} = (1, 1, 0, \cdots, 0)$ and  $x^{1} = (|\vec a|, \vec a)$. If $x^{0} \cdot x^{1} = 0$, we must have $|\vec a| = a_{1} > 0$ and $x^{1} = (a_{1}, a_{1}, 0, \cdots, 0) = a_{1} x^{0}$. This contradicts the fact that the two eigenvectors are independent since they correspond to different eigenvalues. Therefore, we must have $x^{0} \cdot x^{1} \neq 0$.  For convenience, we choose to have $x^{0} \cdot x^{1} =  - 2$. With this, we define 
\be
e^{0} = \frac{1}{2} (x^{0} + x^{1}), \qquad  e^{1} = \frac{1}{2} (x^{0} - x^{1}),
\ee
such that $(e^{0})^{2} = - 1$, $(e^{1})^{2} = 1$ and $e^{0} \cdot e^{1} = 0$. We now construct an orthogonal basis $\{e^{0}, e^{1}, \cdots e^{p}\}$ satisfying the same relations as those given in (\ref{basis}). Note that 
$K_{\alpha}\,^{\beta} e_{\beta}\,^{0} = - (e^{T})^{0}\,_{\beta} K^{\beta}\,_{\alpha} = \rho_{0}\, e_{\alpha}\,^{1}$ and  $K_{\alpha}\,^{\beta} e_{\beta}\,^{1} = - (e^{T})^{1}\,_{\beta} K^{\beta}\,_{\alpha} = \rho_{0} \,e_{\alpha}\,^{0}$. So we have
\be\label{barKoddp}
\bar K_{\bar\alpha}\,^{\bar\beta} = (e^{T})_{\bar\alpha}\,^{\alpha} K_{\alpha}\,^{\beta} e_{\beta}\,^{\bar\beta} = \left(\begin{array}{ccc}
0&\rho_{0} &\\
\rho_{0}&0&\\
&&\bar K_{\bar a}\,^{\bar b}\end{array}\right),
\ee 
where $\bar K_{\bar a}\,^{\bar b}$  is a $(p - 1)\times (p - 1)$ antisymmetric matrix ($ \bar K_{\bar a}\,^{\bar b}= \bar K_{\bar a \bar b} = - \bar K_{\bar b \bar a}$) and can be diagonalized, as before, by a $(p - 1)\times (p -1)$ unitary matrix $u$, with its purely imaginary eigenvalues in pairs as $(\rho_{\bar c}, - \rho_{\bar c})$ with $\bar c = 1, 2, \cdots, (p - 1)/2$.  Note that the symmetric 
sub-matrix in $\bar K_{\bar\alpha}\,^{\bar \beta}$ can be diagonalized by a $2\times 2$  matrix $R$ as
\be
\left(\begin{array}{cc}
0&\rho_{0}\\
\rho_{0} & 0
\end{array}\right) = R  \left(\begin{array}{cc}
\rho_{0} & 0\\
0 & - \rho_{0}\end{array}\right) R^{-1},
\ee
where specifically 
\be
R  = \frac{1}{\sqrt{2}} \left(\begin{array}{cc}
 1& - 1\\
 1&1\end{array}\right),\qquad  R^{-1}  = \frac{1}{\sqrt{2}} \left(\begin{array}{cc}
 1& 1\\
 - 1&1\end{array}\right).
  \ee
In other words, the matrix $\bar K_{\bar\alpha}\,^{\bar\beta}$ can be diagonalized by the unitary matrix $r U$ as
\be\label{barKdiagonal}
\bar K = r \left(\begin{array}{ccc}
\rho_{0} & 0 &\\
0&- \rho_{0} &\\
&&\bar K_{(p - 1)\times (p -1)}
\end{array}\right) r^{-1} = r U \bar K_{0} U^{+} r^{T} = (r U) \bar K_{0} (r U)^{-1},
\ee
where 
\be
r = \left(\begin{array}{cc}
R & \\
& \mathbb{I}_{(p - 1)\times (p - 1)}\end{array}\right), \qquad U = \left(\begin{array}{cc}
\mathbb{I}_{2\times 2} &\\
& u \end{array}\right),
\ee
and the diagonal matrix $\bar K_{0} = (\rho_{0}, - \rho_{0}, \rho_{1}, - \rho_{1}, \cdots, \rho_{(p - 1)/2}, - \rho_{(p - 1)/2})$.  Given the above and from (\ref{barKoddp}), we have
\be\label{diagonalKoddp}
K = e \bar K e^{T} = e r U \bar K_{0} (r U)^{+} e^{T} = (erU) K_{0} (e r U)^{-1},
\ee
where we have used $e^{- 1} = e^{T}$.  So we prove that in general $K$ has a pair of real eigenvalue $(\rho_{0}, - \rho_{0})$ and the rest are all imaginary and given in pairs as
$(\rho_{c}, - \rho_{c})$ with $c = 1, 2, \cdots (p - 1)/2$.  

In summary, when $p =$ even,  we have two cases: 1) one zero eigenvalue and the rest are all imaginary and given in pairs as $(\rho_{c}, - \rho_{c})$ with $c = 1, 2, \cdots, p/2$; 2) one zero eigenvalue, a pair of real eigenvalues
$(\rho_{0}, \rho_{0})$ and the rest are all imaginary and given in pairs as $(\rho_{c}, - \rho_{c})$ with $c = 1, 2, \cdots, (p - 2)/2$. For $p = $ odd, we have in general a pair of real eigenvalues $(\rho_{0}, - \rho_{0})$ and the rest are all 
imaginary and given as $(\rho_{c}, - \rho_{c})$ with $c = 1, 2, \cdots, (p - 1)/2$.  

If we set the positive real eigenvalue $\rho_{0} = 2\pi \nu_{0}$  and the imaginary eigenvalues $\rho_{c} = 2\pi i \nu_{c}$,  from (\ref{lorentzw}) we then obtain the same eigenvalues of $w$ as discussed in Section 3. 

\section*{Appendix B\label{AB}} 
The  zero-mode contribution to the amplitude  in the RR sector for $p - p' = \nu = 0, 2, 4, 6$ and for $p \le 6$ can be computed, following the regularization given in \cite{Yost, Billo:1998vr}, to give 
 \bea\label{0mpp'}
{}_{\rm 0R}\langle B', \eta'| B, \eta\rangle_{\rm 0R} &\equiv& {}_{\rm 0R}\langle B_{\rm sgh}, \eta'| B_{\rm sgh}, \eta\rangle_{\rm 0R} \, {}_{\rm 0R}\langle B'_\psi, \eta'| B_\psi, \eta\rangle_{\rm 0R} \nn
&=& - \frac{2^{4}\, \delta_{\eta\eta', +} }{\sqrt{\det(\eta_{p} + \hat F)\det (\eta_{p'} + \hat F')}} \sum_{n = 0}^{[\frac{p' + 1}{2}]} \frac{ [2 (n + \frac{\nu}{2})]!}{ 2^{2n + \frac{\nu}{2}}\, n! (n + \frac{\nu}{2})!}\nn
&\,&\times \hat F^{[\alpha'_{1}\beta'_{1}\cdots}\hat F^{\alpha'_{n}\beta'_{n}}
\hat F^{(p' + 1)(p' + 2) \cdots} \hat F^{(p' + \nu - 1)(p' + \nu)]}\hat F'_{[\alpha'_{1}\beta'_{1}\cdots} \hat F'_{\alpha'_{n}\beta'_{n}]} ,
\eea
where the indices inside the square bracket denote their anti-symmetrization.  As indicated already in the previous sections, the above zero-mode matrix element for lower $p$ and $p'$ cases can be obtained from either $p = 6$ or $p = 5$ case       
depending on $p$ and $p'$ being even or odd if their worldvolume fluxes are extended in a specific way which we turn next. Let us take two explicit examples to demonstrate this.  The first one is for $p = 5$ and $p' = 3$ and we will show that the corresponding 
zero-mode matrix element can be obtained from $p = p' = 5$ case if we extend the $p' = 3$ worldvolume flux $\hat F'_{\alpha'\beta'}$ to the $p' = p = 5$ worldvolume flux $\hat F'_{\alpha\beta}$ the following way,
\be\label{35extension}
\hat F'_{\alpha\beta} = \left(\begin{array}{ccc}
\hat F'_{\alpha'\beta'}&  &\\
&0& \hat g'\\
&- \hat g' &0\end{array}\right),
\ee
where we take the magnetic flux $\hat g' > 0$ to be infinite at the end of computations.   From (\ref{0mpp'}), we have for $p = 5$ and $p' = 3$ 
\bea\label{53case}
{}_{\rm 0R}\langle B', \eta'| B, \eta\rangle_{\rm 0R} &=& - \frac{2^{4} \delta_{\eta\eta', +}}{\sqrt{\det(\eta_{5} + \hat F)\det (\eta_{3} + \hat F')}} 
\left( \hat F^{45} + \frac{3}{2} \hat F^{[\alpha'\beta'}\hat F^{45]} \hat F'_{\alpha'\beta'} \right.\qquad\qquad\qquad \qquad\nn
&\,& \qquad \qquad \qquad\qquad \left. + \frac{15}{8} \hat F^{[\alpha'_{1}\beta'_{1}}\hat F^{\alpha'_{2}\beta'_{2}}\hat F^{45]} \hat F'_{[\alpha'_{1}\beta'_{1}}\hat F'_{\alpha'_{2}\beta'_{2}]}\right).
 \eea
 We now show that this can also be obtained from the $p = p' = 5$ case but with $\hat F'_{\alpha\beta}$ given by (\ref{35extension}). 
  We have now from (\ref{0mpp'})
  \bea\label{0mp=p'}
{}_{\rm 0R}\langle B', \eta'| B, \eta\rangle_{\rm 0R}  
&=& - \frac{2^{4} \delta_{\eta\eta', +} }{\sqrt{\det(\eta_{5} + \hat F)\det (\eta_{5} + \hat F')}} \left(1 + \frac{1}{2} \hat F^{\alpha\beta}\hat F'_{\alpha\beta}\right. \nn
&\,&  \left. + \left(\frac{1}{2^{2} 2!}\right)^{2} 4! \hat F^{[\alpha_{1}\beta_{1}}\hat F^{\alpha_{2}\beta_{2}]} 
\hat F'_{[\alpha_{1}\beta_{1}}\hat F'_{
\alpha_{2}\beta_{2}]} \right. \nn
&\,& \left.+ \left(\frac{1}{2^{3} 3!}\right)^{2} 6! \hat F^{[\alpha_{1}\beta_{1}}\hat F^{\alpha_{2}\beta_{2}}\hat F^{\alpha_{3}\beta_{3}]} 
\hat F'_{[\alpha_{1}\beta_{1}}\hat F'_{
\alpha_{2}\beta_{2}}\hat F'_{\alpha_{3}\beta_{3}]}\right) .
\eea
Note that $\det (\eta_{\alpha\beta} + \hat F'_{\alpha\beta}) =  (1 + \hat g'^{2}) \det (\eta_{\alpha'\beta'} + \hat F'_{\alpha'\beta'})$ which gives $ \hat g'^{2} \det (\eta_{\alpha'\beta'} + \hat F'_{\alpha'\beta'})$ for $\hat g' \to \infty$.  
To have a finite contribution at $\hat g' \to \infty$, we need  each term in the bracket to have a factor $\hat F'_{45} = \hat g'$. For this,
\bea
&&\hat F^{\alpha\beta} \hat F'_{\alpha\beta} = 2 \hat g' \hat F^{45} + \cdots,\nn
&&\hat F^{[\alpha_{1}\beta_{1}} \hat F^{\alpha_{2} \beta_{2}]} \hat F'_{[\alpha_{1}\beta_{1}}\hat F'_{\alpha_{2}\beta_{2}]} = 4 \hat g' \hat F^{[\alpha_{1}\beta_{1}} \hat F^{45]} \hat F'_{\alpha_{1}\beta_{1}} + \cdots,\nn
&&\hat F^{[\alpha_{1}\beta_{1}}\hat F^{\alpha_{2}\beta_{2}}\hat F^{\alpha_{3}\beta_{3}]} 
\hat F'_{[\alpha_{1}\beta_{1}}\hat F'_{
\alpha_{2}\beta_{2}}\hat F'_{\alpha_{3}\beta_{3}]} = 6 \hat g'  \hat F^{[\alpha_{1}\beta_{1}}\hat F^{\alpha_{2}\beta_{2}}\hat F^{45]} 
\hat F'_{[\alpha_{1}\beta_{1}}\hat F'_{\alpha_{2}\beta_{2}]} + \cdots,
\eea
where the $\cdots$ terms are independent of $g'$. With these and taking $\hat g' \to \infty$, we can check easily that (\ref{0mp=p'}) is exactly the same as (\ref{53case}).  

The second example is to take $p = p' = 4$ and we will show that the zero-mode 
matrix element can be obtained from the $p = p' = 6$ case by taking the worldvolume fluxes, respectively, as
\be\label{46extension}
\hat F_{\alpha\beta} = \left(\begin{array}{ccc}
\hat F_{\alpha'\beta'}&  &\\
&0& \hat g\\
&- \hat g&0\end{array}\right), \qquad \hat F'_{\alpha\beta} = \left(\begin{array}{ccc}
\hat F'_{\alpha'\beta'}&  &\\
&0& \hat g'\\
&- \hat g'&0\end{array}\right),
\ee
where $\alpha, \beta = 0, 1, \cdots, 6$ and $\alpha', \beta' = 0, 1, \cdots 4$.  Note that we need to take $\hat g, 
\hat g' \to \infty$ at the end of computations. For $p = p' =4$, the zero-mode matrix element from 
from (\ref{0mpp'}) is
 \bea\label{0m44}
{}_{\rm 0R}\langle B', \eta'| B, \eta\rangle_{\rm 0R} 
&=& - \frac{2^{4}\, \delta_{\eta\eta', +} }{\sqrt{\det(\eta_{4} + \hat F)\det (\eta_{4} + \hat F')}} \left(1 + \frac{1}{2} \hat F^{\alpha'\beta'} \hat F'_{\alpha'\beta'} \right. \qquad \qquad \qquad\nn
&\,&\qquad\qquad \qquad \left. + \frac{3}{8} \hat F^{[\alpha'_{1}\beta_{1}}\hat F^{\alpha'_{2}\beta'_{2}]}
\hat F'_{[\alpha'_{1}\beta'_{2}}\hat F'_{\alpha'_{2}\beta'_{2}]} \right).
\eea
For $p = p' = 6$ with the respective worldvolume fluxes given in (\ref{46extension}),  we have the corresponding zero-mode matrix element from (\ref{0mpp'}) as
 \bea\label{0m66}
&&{}_{\rm 0R}\langle B', \eta'| B, \eta\rangle_{\rm 0R} 
= - \frac{2^{4}\, \delta_{\eta\eta', +} }{\sqrt{\det(\eta_{6} + \hat F)\det (\eta_{6} + \hat F')}} \left(1 + \frac{1}{2} \hat F^{\alpha\beta} \hat F'_{\alpha\beta} \right. \nn
&& \left.+ \frac{4!}{2^{6}} \hat F^{[\alpha_{1}\beta_{1}} \hat F^{\alpha_{2}\beta_{2}]}
 \hat F'_{[\alpha_{1}\beta_{1}} \hat F'_{\alpha_{2}\beta_{2}]} + \frac{6!}{2^{6} (3!)^{2}} \hat F^{[\alpha_{1}\beta_{1}} \hat F^{\alpha_{2}\beta_{2}} \hat F^{\alpha_{3}\beta_{3}]}
 \hat F'_{[\alpha_{1}\beta_{1}} \hat F'_{\alpha_{2}\beta_{2}}\hat F'_{\alpha_{3}\beta_{3}]}\right).
  \eea
We will show that the above is actually the same as that given in (\ref{0m44}) for $\hat g,  \hat g' \to \infty$. For this, note that 
\be
\sqrt{\det(\eta_{\alpha\beta} + \hat F_{\alpha\beta})\det (\eta_{\alpha\beta} + \hat F'_{\alpha\beta})} = \hat g \hat g' \sqrt{\det(\eta_{\alpha'\beta'} + \hat F_{\alpha'\beta'})\det (\eta_{\alpha'\beta'} + \hat F'_{\alpha'\beta'})},
\ee
where we have used (\ref{46extension}) and taken $\hat g, \hat g' \to \infty$.  To have a finite limit, only those terms in the bracket proportional to $\hat g \hat g'$ in (\ref{0m66}) survive. We have
\bea
&& \hat F^{\alpha\beta} \hat F'_{\alpha\beta} = 2 \hat g \hat g' + \cdots, \nn
&& \hat F^{[\alpha_{1}\beta_{1}} \hat F^{\alpha_{2}\beta_{2}]} \hat F'_{[\alpha_{1}\beta_{2}} \hat F'_{\alpha_{2}\beta_{2}]} = \frac{2^{5}}{4!} \hat g \hat g' \hat F^{\alpha'\beta'} \hat F'_{\alpha'\beta'} + \cdots,\nn
&&\hat F^{[\alpha_{1}\beta_{1}} \hat F^{\alpha_{2}\beta_{2}} \hat F^{\alpha_{3}\beta_{3}]}
 \hat F'_{[\alpha_{1}\beta_{1}} \hat F'_{\alpha_{2}\beta_{2}}\hat F'_{\alpha_{3}\beta_{3}]} = \frac{(3!)^{2} 4!}{6!} \hat g \hat g' \hat F^{[\alpha'_{1}\beta'_{2}}\hat F^{\alpha'_{2}\beta'_{2}]} 
 \hat F'_{[\alpha'_{1}\beta'_{1}}\hat F'_{\alpha'_{2}\beta'_{2}]} + \cdots,\qquad\qquad
 \eea
where the $\cdots$ terms are independent of $\hat g \hat g'$.  Plugging the above terms to (\ref{0m66}) and taking $\hat g, \hat g' \to \infty$, we get exactly (\ref{0m44}). 

In summary, for various RR zero-mode contributions given in (\ref{0mpp'}), they each can be  obtained from the $p = p' = 6$ or the $p = p' = 5$ case by choosing the corresponding worldvolume fluxes in a way as indicated in the above examples. 
So we only need to focus on these two cases.  They each can be given from (\ref{0mpp'}) as
\bea\label{0m5or6}
{}_{\rm 0R}\langle B', \eta'| B, \eta\rangle_{\rm 0R} 
&=& - \frac{2^{4}\, \delta_{\eta\eta', +} }{\sqrt{\det(\eta_{p} + \hat F)\det (\eta_{p} + \hat F')}} \nn
&\,& \times \sum_{n = 0}^{3} \frac{ (2 n)!}{ 2^{2n}\, (n!)^{2}}  \hat F^{[\alpha_{1}\beta_{1}\cdots}\hat F^{\alpha_{n}\beta_{n}]}
\hat F'_{[\alpha_{1}\beta_{1}\cdots} \hat F'_{\alpha_{n}\beta_{n}]} ,
\eea
where $p = 5$ or $6$. For either case, let us write the zero-mode contribution as
\be\label{0m5or6-new}
{}_{\rm 0R}\langle B', \eta'| B, \eta\rangle_{\rm 0R} =  - \frac{2^{4}\, \delta_{\eta\eta', +} }{\sqrt{\det(\eta_{p} + \hat F)\det (\eta_{p} + \hat F')}} S,
\ee
where 
\be\label{S}
S  =  \sum_{n = 0}^{3} \frac{ (2 n)!}{ 2^{2n}\, (n!)^{2}}  \hat F^{[\alpha_{1}\beta_{1}\cdots}\hat F^{\alpha_{n}\beta_{n}]}
\hat F'_{[\alpha_{1}\beta_{1}\cdots} \hat F'_{\alpha_{n}\beta_{n}]} .
\ee
In what follows, we will show 
\be\label{0msquare}
({}_{\rm 0R}\langle B', \eta'| B, \eta\rangle_{\rm 0R} )^{2}  =  2^{7 - p} \delta_{\eta\eta', +}\, \det (\mathbb{I} + w),
\ee 
where $w$ is given in (\ref{matrixlw}) and for convenience we rewrite it explicitly here
\be\label{lwmatrix}
w = (\mathbb{I} - \hat F) (\mathbb{I} + \hat F)^{-1} (\mathbb{I} + \hat F')(\mathbb{I} - \hat F')^{-1},
\ee
with $\mathbb{I}$ the $(p + 1) \times (p + 1)$ unity matrix.  Note that
\bea
\det (\mathbb{I} + w) &=& \frac{\det [(\mathbb{I} + \hat F) (\mathbb{I} + w) (\mathbb{I} - \hat F')]}{\det(\mathbb{I} + \hat F) \det(\mathbb{I} - \hat F')}\nn
&=& \frac{2^{p + 1} \det(\mathbb{I} - \hat F \hat F')}{\det(\mathbb{I} + \hat F)\det (\mathbb{I} + \hat F')}.
\eea
With this, for (\ref{0msquare}) to hold, we need to show
\be\label{Ssquare}
S^{2} = \det (\mathbb{I} - \hat F \hat F').
\ee
For this, we represent the S in terms of the following Grassmannian integration,
\be\label{grassmiS}
S = (- )^{p + 1} \int \prod_{\gamma = 0}^{p} \left[d\theta^{\gamma} d\theta'_{\gamma} (1 + \theta^{\gamma} \theta'_{\gamma})\right] e^{- \frac{1}{2} \hat F_{\alpha\beta} \theta^{\alpha}\theta^{\beta}}  e^{\frac{1}{2} \hat F'^{\bar\alpha\bar\beta} \theta'_{\bar\alpha} \theta'_{\bar\beta}},
\ee
where $\theta^{\gamma}$ and $\theta'_{\gamma}$ are all real Grassmannian variables. With this, we have
\be\label{Ssquare}
S^{2} = \int \prod_{\gamma = 0}^{p} d \theta^{\gamma} d\theta'_{\gamma} d\tilde\theta^{\gamma} d\tilde\theta'_{\gamma} (1 + \theta^{\gamma}\theta'_{\gamma})(1 + \tilde\theta^{\gamma}\tilde\theta'_{\gamma})\,
e^{- \frac{1}{2} \hat F_{\alpha\beta}(\theta^{\alpha}\theta^{\beta} + \tilde\theta^{\alpha}\tilde\theta^{\beta}) }e^{\frac{1}{2} \hat F'^{\bar\alpha\bar\beta} (\theta'_{\bar\alpha} \theta'_{\bar\beta} + \tilde\theta'_{\bar\alpha}\tilde\theta'_{\bar\beta})}.\quad
\ee
Now we change the integration variables as
\be\label{changev}
\theta^{\gamma} = \frac{\eta^{\gamma} + \eta^{* \gamma}}{\sqrt{2}}, \quad \tilde\theta^{\gamma} = \frac{\eta^{\gamma} - \eta^{*\gamma}}{i \sqrt{2}}, \quad
\theta'_{\gamma} =  \frac{\eta'_{\gamma} + \eta'^{*}_{\gamma}}{\sqrt{2}}, \quad \tilde\theta'_{\gamma} = \frac{\eta'_{\gamma} - \eta'^{*}_{\gamma}}{i \sqrt{2}}.
\ee
where $^{*}$ denotes the complex conjugate. In terms of $\eta^{\gamma}, \eta^{*\gamma}, \eta'_{\gamma}$ and $\eta'^{*}_{\gamma}$, we have
\be\label{S-eta}
S^{2} = \int \prod_{\gamma =0}^{p} d\eta'^{*}_{\gamma} d\eta^{\gamma} d\eta'_{\gamma} d\eta^{*\gamma}  (1 + \eta^{\gamma} \eta'^{*}_{\gamma})(1 + \eta^{*\gamma}\eta'_{\gamma}) 
e^{- \hat F_{\alpha\beta} \eta^{\alpha} \eta^{*\beta}} e^{\hat F'^{\bar\alpha\bar\beta}\eta'_{\bar\alpha} \eta'^{*}_{\bar\beta} }.
\ee
The evaluation of the integral can be simplified if we do the following integration first 
\bea\label{I}
I &=& \int \prod_{\gamma = 0}^{p} d\eta'_{\gamma} d \eta^{*\gamma} (1 + \eta^{*\gamma} \eta'_{\gamma}) e^{- \hat F_{\alpha\beta} \eta^{\alpha} \eta^{*\beta}} e^{\hat F'^{\bar\alpha\bar\beta}\eta'_{\bar\alpha} \eta'^{*}_{\bar\beta} }\nn
&=&  e^{- (\hat F \hat F')_{\alpha}\,^{\bar\alpha}\eta^{\alpha}\eta'^{*}_{\bar\alpha}}.
\eea  
 With this, we have
 \bea\label{S2}
 S^{2} &=& \int \prod_{\gamma =0}^{p} d\eta'^{*}_{\gamma} d\eta^{\gamma}    (1 + \eta^{\gamma} \eta'^{*}_{\gamma}) I \nn
 &=&  \int \prod_{\gamma =0}^{p} d\eta'^{*}_{\gamma} d\eta^{\gamma} e^{[\mathbb{I} - (\hat F \hat F')]_{\alpha}\,^{\bar\alpha} \eta^{\alpha} \eta'^{*}_{\bar\alpha} }\nn
 &=& \det (\mathbb{I} - \hat F\hat F').
 \eea
 In other words, (\ref{0msquare}) holds indeed.  In Appendix A, we have shown $w = V w_{0} V^{-1}$ with $w_{0}$ diagonal with eigenvalue $1$ and others in pairs as 
 $(\lambda_{\alpha}, \lambda_{\alpha})$ with $\alpha =0, 1, 2$ for $p = 6$ or with eigenvalues in pairs as   $(\lambda_{\alpha}, \lambda_{\alpha})$ with $\alpha =0, 1, 2$ for 
 $p = 5$. We then have from (\ref{0msquare}) for $p = 5$ or $6$ 
 \bea\label{0msquare1}
({}_{\rm 0R}\langle B', \eta'| B, \eta\rangle_{\rm 0R} )^{2}  &=&  2^{7 - p} \delta_{\eta\eta', +}\, \det (\mathbb{I} + w)\nn
&=& 2^{7 - p} \delta_{\eta\eta', +} \det (\mathbb{I} + w_{0})\nn
& =& 2^{7 - p}\delta_{\eta\eta', +}  (1 + \delta_{p, 6}) \prod_{\alpha = 0}^{2} (1 + \lambda_{\alpha})(1 + \lambda^{-1}_{\alpha})\nn
&=& 2^{8}\, \delta_{\eta\eta', +} \, \prod_{\alpha =0}^{2} \cos^{2}\pi \nu_{\alpha},
\eea 
where we have used $\lambda_{\alpha} = e^{2i\pi\nu_{\alpha}}$. This gives the expected result 
\be\label{0mexplicit}
{}_{\rm 0R}\langle B', \eta'| B, \eta\rangle_{\rm 0R} = - 2^{4}  \delta_{\eta\eta', +} \prod_{\alpha = 0}^{2} \cos\pi\nu_{\alpha},
\ee
where we have used the known result ${}_{\rm 0R}\langle B', \eta'| B, \eta\rangle_{\rm 0R} = - 2^{4}  \delta_{\eta\eta', +}$ in the absence of fluxes for which $\nu_{\alpha} = 0$ and $\nu_{\alpha} \in [0, 1)$ to cover all flux cases. Given what has been discussed, for a 
general $p \le 6$, we have 
\be\label{0mgeneralp}
{}_{\rm 0R}\langle B', \eta'| B, \eta\rangle_{\rm 0R} = - 2^{4}  \delta_{\eta\eta', +} \prod_{\alpha = 0}^{[\frac{p - 1}{2}]} \cos\pi\nu_{\alpha}.
\ee


\end{document}